\documentclass[12pt]{iopart}
\usepackage{iopams}
\usepackage{setstack}
\usepackage{graphicx}
\usepackage{bm}
\usepackage{epsfig}
\usepackage{hyperref} 
\usepackage{cite}
\usepackage{float}
\usepackage{placeins}

\newcommand{\diag}{{\rm diag\,}}
\newcommand{\sign}{{\rm sign\,}}

\newcommand{\U}{{\rm U\,}}
\newcommand{\SU}{{\rm SU\vspace*{0.05cm}}}

\newcommand{\eins}{\leavevmode\hbox{\small1\kern-3.8pt\normalsize1}}
\eqnobysec
\newcommand{\be}{\begin{eqnarray}}
\newcommand{\ee}{\end{eqnarray}}
\newcommand{\nn}{\nonumber}
\newcommand{\ha}{\widehat  a}

\renewcommand{\hm}{\widehat  m}
\newcommand{\hmu}{\widehat{\mu}_{\rm I}}
\newcommand{\whm}{\widehat m}
\newcommand{\whmu}{\widehat{\mu}_{\rm I}}
\newcommand{\wha}{\widehat a}
\newcommand{\KSTVZ}{Kogut:2000ek} 
\newcommand{\SoSt}{Son:2000xc} 
\newcommand\KST{Kogut:1999iv}

\begin{document}

 \newtheorem{definition}{Definition}[section]
\newtheorem{assumption}[definition]{Assumption}
\newtheorem{theorem}[definition]{Theorem}
\newtheorem{lemma}[definition]{Lemma}
\newtheorem{corollary}[definition]{Corollary}

\title[Phase Diagram of Twisted Mass Wilson Fermions]{Phase Diagram of Dynamical Twisted Mass Wilson Fermions at Finite Isospin Chemical Potential}
\author{Oliver Janssen$^{(1)}$, Mario Kieburg$^{(2)}$, K.~Splittorff$^{(3)}$, Jacobus J. M. Verbaarschot$^{(4)}$ and Savvas Zafeiropoulos$^{(5,6 )}$}
\address{
$^{(1)}$ Department of Physics, New York University, New York, NY 10003, USA\\
$^{(2)}$ Fakult\"at f\"ur Physik, Universit\"at Bielefeld, Postfach 100131, D-33501 Bielefeld, Germany\\
$^{(3)}$ Discovery Center, The Niels Bohr Institute, University of Copenhagen, Blegdamsvej 17, DK-2100, Copenhagen {\O}, Denmark\\
$^{(4)}$ Department of Physics and Astronomy, SUNY, Stony Brook, New York 11794, USA\\
$^{(5)}$ Laboratoire de Physique Corpusculaire, Universit\'e Blaise Pascal, CNRS/IN2P3 63177 Aubi\`ere Cedex, France\\
$^{(6)}$ Institut f\"ur Theoretische Physik, Goethe-Universit\"at Frankfurt, Max-von-Laue-Str.~1, 60438 Frankfurt am Main, Germany}
\eads{$^{(1)}$ \mailto{opj202@nyu.edu},
$^{(2)}$ \mailto{mkieburg@physik.uni-bielefeld.de}, $^{(3)}$ \mailto{split@nbi.ku.dk}, $^{(4)}$ \mailto{jacobus.verbaarschot@stonybrook.edu}, $^{(5)}$ \mailto{zafeiro@th.physik.uni-frankfurt.de}}

\date{today}

\begin{abstract}
We consider the phase diagram of twisted mass Wilson fermions of two-flavor QCD in the parameter space of 
the quark mass, the isospin chemical potential, the twist angle and the lattice spacing.
This work extends earlier studies in the continuum and those at zero chemical potential.
We evaluate the phase diagram as well as the spectrum of the (pseudo-)Goldstone bosons
using the chiral Lagrangian for twisted mass Wilson fermions at non-zero isospin chemical potential.
The phases are obtained from a mean field analysis.
At zero twist angle we find that already an infinitesimal isospin chemical potential destroys the Aoki phase. The reason
is that in this phase we have massless Goldstone bosons with a non-zero isospin charge. At finite twist angle only two different phases are present, one phase which is continuously connected to the Bose condensed phase at non-zero chemical
potential and another phase which is continuously connected to the normal phase. For either zero or maximal twist the
phase diagram is more complicated as the saddle point equations allow for more solutions.

\end{abstract}

\section{Introduction}\label{sec1}

In the past decade lattice QCD simulations with twisted mass Wilson 
fermions have attracted a great deal of attention \cite{Boucaud:2007uk, Boucaud:2008xu, Blossier:2009bx, Baron:2009wt, Baron:2010bv}. 
The advancement of algorithms as well as the increasing power 
of computers have allowed for simulations in the deep chiral regime.
Simulations with ordinary Wilson fermions in this regime can be severely 
hindered by the so-called exceptional configurations. These are configurations for which a  
Dirac eigenvalue is extremely close to minus the quark mass so that the Dirac operator
cannot be inverted. 
With a twisted mass \cite{Frezzotti:2000nk} on the contrary the determinant 
is regulated by the mass, see for example~(\ref{posDefiniteness}) below. The twisted mass is  a mass which 
comes with a $\gamma_5$ in Dirac space and a $\tau_3$ in flavor space 
(see \cite{Shindler:2007vp, Sint:2007ug} for excellent reviews of
 the twisted mass formulation). 
In the twisted mass basis
the fermionic part of the Lagrangian reads
\begin{equation}\label{TMaction}
\mathcal{L}_{\rm F}^{\rm tm}=\bar\psi(D_{\rm W}(m)+\imath u\gamma_5\tau_3)\psi=\bar{\psi}(D_{\rm W}+m+\imath u \gamma_5\tau_3)\psi,
\end{equation} 
where $D_{\rm W}$ is the Wilson Dirac operator, $m$ the quark mass ( i.e $D_{\rm W}(m)=D_{\rm W}+m$ the massive Wilson Dirac operator), and $u$
the twisted quark mass. The quark fields are denoted by $\chi$ and $ \bar \chi$.
One can perform the following change of variables
 $\chi=e^{\imath\omega\gamma_5\tau_3/2}\psi$ and  $\bar{\chi}=\bar{\psi}e^{\imath\omega\gamma_5\tau_3/2}$ where one introduces $\omega=\arctan(u/m)$ and $\mathcal{M}=\sqrt{m^2+u^2}$. This transformation defines a new basis, the physical basis, where the Dirac operator assumes the usual form
\begin{equation}\label{TMactionPB}
 \mathcal{L}_{\rm F}^{\rm tm}=\bar{\chi}(D'_{\rm W}+\mathcal{M})\chi.
\end{equation}
However as indicated by the prime, the Wilson term is now rotated in the Dirac operator. 
Since the Wilson term vanishes in the naive continuum limit,  twisted mass QCD should be equivalent to ordinary QCD. 
The proper mapping between twisted mass QCD and ordinary QCD was 
analyzed in \cite{Frezzotti:2000nk}. 
The addition of the twisted mass renders the fermion determinant positive definite, 
\begin{eqnarray}\label{posDefiniteness}
\det(D_{\rm W}(m)+\imath u \gamma_5\tau_3)&=&\det(D_{\rm W}(m)+\imath u \gamma_5)\det(D_{\rm W}(m)-\imath u \gamma_5)\nonumber\\
&=&\det(D_{\rm W}(m)+\imath u \gamma_5)\det(\gamma_5D_{\rm W}(m)\gamma_5-\imath u \gamma_5)\nonumber\\
&=&\det(D_{\rm W}(m)D^{\dagger}_{\rm W}(m)+ u^2),
\end{eqnarray}
where we used the $\gamma_5$-Hermiticity of $D_{\rm W}=\gamma_5D_{\rm W}^\dagger\gamma_5$.
The other crucial property of twisted mass Wilson fermions is the automatic $O(a)$ improvement of both the actions as well as of the operators which is achieved at maximal twist without significant additional computational cost compared to standard Wilson fermions \cite{Frezzotti:2003ni}.
It is noteworthy to point out that the case of intermediate twist is not of pure academic interest. 
In \cite{Frezzotti:2003ni,ren1,ren2} intermediate twist angles were used
to achieve $\mathcal{O}(a)$ improvement in the simulations dedicated to the generation of ensembles employed in  renormalization projects.
The simulations with very small PCAC fermion masses were numerically unstable and 
had very large autocorrelation times. This could be cured by 
averaging the results for opposite PCAC fermion masses 
over the twist angle, which also achieved an 
 $\mathcal{O}(a)$ improvement. This method was proposed in \cite{Frezzotti:2003ni}.
 
In order to make full use of the results of lattice simulations with Wilson 
fermions it is essential to understand the non-trivial phase diagram which 
is caused by the specific choice of discretization.  
The phase diagram of Wilson fermions with a non-zero twisted mass has been
studied in \cite{Aokiclassic, SharpeSingleton,RS,BRS,Aoki-spec,GSS,Shindler,BNSep,Creutz:1996wg, Sharpe-Wu,sharpe-prd,Munster:2004am, Scorzato:2004da, Farchioni:2004us, Farchioni:2004fs, Farchioni:2005tu,Kieburg:2012fw}. 
In the present work we extend these studies to include
the isospin chemical potential. Thus we generalize
finite density studies in continuum QCD and QCD like theories \cite{\KST,Kogut:2000ek, Son:2000xc, Splittorff:2000mm, Kogut:2002zg, Toublan:2003tt, Klein:2003fy, Kogut:2004zg, Barducci:2004tt}.
QCD at finite isospin chemical potential is of interest in 
different physical contexts. Two prominent examples are neutron stars 
and heavy ion collisions. Moreover, studying QCD at finite 
isospin chemical potential can give useful insights into 
the notorious sign problem in QCD at non-zero baryon chemical potential 
\cite{SVphase}. 

In section~\ref{sec:2} we recall some properties of the low energy 
effective theory, especially the $p$-regime, for twisted mass Wilson fermions. Thereby we present the three leading order terms of the chiral Lagrangian and introduce the order parameters. The thermodynamic limit is performed in section~\ref{sec:3}. We start the discussion of the phase diagram with an analysis of the case with vanishing twist ($\omega=0$) and  determine the different phases that one encounters in the cases of zero, real and imaginary isospin chemical potential. Then the phase diagram can be studied in the plane of the quark mass and the lattice spacing.  At zero twist we find that the Aoki phase  only occurs at 
zero isospin chemical potential. Thereafter we also discuss the structure of the phase diagram at finite twist.
 Finally, we consider the phase diagram of the physically important case of maximum twist. The detailed computations to this analysis are carried out in~\ref{app1}.

In section~\ref{sec:4} we discuss the dependence of the pion masses on the chemical potential and lattice spacing. The detailed calculation to this part are given in~\ref{MassDer}. In section~\ref{sec5} we summarize our results.

The present work was partly published in the proceedings~\cite{KSVZ}. Moreover we want to point out that we work with natural units ($c=\hbar=k_{\rm B}=1$).

\section{Chiral Lagrangian of Twisted Wilson Fermions}\label{sec:2}

We consider the partition function of two dynamical Wilson fermions in the
 fundamental representation of the gauge group $\SU(3)$ with degenerate quark masses $m$,
\begin{eqnarray}\label{partition}
 Z&=&\int D[U] \exp[-S_{\rm gauge}(U)] \det\left(D_{\omega}(m)+\frac{\mu_{\rm I}}{2} \gamma_0\tau_3\right),
\end{eqnarray}
where $S_{\rm gauge}(U)$ is a discretized version of the Yang-Mills action, e.g.~the Wilson gauge action. Here we follow the notation in~\cite{Son:2000xc,KSVZ} for the isospin chemical potential $\mu_{\rm I}$ which enters with a factor $1/2$.
We consider both real and purely imaginary isospin chemical potential $\mu_{\rm I}$.

 We emphasize that the ensuing discussion also applies to $\SU(N_{\rm c}>3)$ 
with the fermions in the fundamental representation. Those theories have the same
pattern of spontaneous symmetry breaking as the one of QCD,
 and hence, share the same chiral Lagrangian for the Goldstone bosons, 
though the low energy constants might be different.

The operator $D_{ \omega}$ is the Wilson Dirac operator with a twisted mass, defined in (\ref{twist-Dirac}) below. Its properties are briefly recalled in subsection~\ref{sec:2.1}. In subsection~\ref{sec:2.2} we introduce the chiral Lagrangian in the $p$-regime as 
well as its mean field limit which coincides with the leading order
Lagrangian in the $\epsilon$-regime also known as the microscopic limit~\cite{SV}. The order parameters which become important later on are presented in subsection~\ref{sec:order}. They are given in terms of the parametrization of the Goldstone manifold  ${\rm SU}_{\rm L}(2)\times{\rm SU}_{\rm R}(2)/{\rm SU}_{\rm V}(2)\cong{\rm SU}(2)$  in subsection~\ref{paramet}.

\subsection{Properties of the Euclidean Wilson Dirac operator with twisted mass and isospin chemical potential}\label{sec:2.1}

Let $\gamma_\alpha$ be the Euclidean Dirac matrices. Then the Euclidean Wilson-Dirac operator \cite{Wilson},
\begin{eqnarray}\label{Wilson-Dirac}
 D_{\rm W}(m)=\gamma^\kappa D_\kappa-a \vec\nabla^2+m,
\end{eqnarray}
is $\gamma_5$-Hermitian, $D_{\rm W}^\dagger=\gamma_5D_{\rm W}\gamma_5$. 
 The Dirac operator with isospin chemical potential is $\gamma_5=\diag(1,1,-1,-1)$ Hermitian if $\mu_{\rm I}$ is imaginary otherwise it becomes $\gamma_5\tau_{1/2}$ Hermitian, i.e.
\begin{eqnarray}
 \fl\left(D_{\rm W}(m)+\frac{\mu_{\rm I}}{2}\gamma_0\tau_3\right)^\dagger=\left\{\begin{array}{cl} \displaystyle\gamma_5\tau_1\left(D_{\rm W}(m)+\frac{\mu_{\rm I}}{2}\gamma_0\tau_3\right)\gamma_5\tau_1=\gamma_5\tau_2\left(D_{\rm W}(m)+\frac{\mu_{\rm I}}{2}\gamma_0\tau_3\right)\gamma_5\tau_2, & \\
 &\hspace*{-2cm}\mu_{\rm I}\in\mathbb{R}, \\
 \displaystyle\gamma_5\left(D_{\rm W}(m)+\frac{\mu_{\rm I}}{2}\gamma_0\tau_3\right)\gamma_5, & \\
 &\hspace*{-2cm} \mu_{\rm I}\in\imath\mathbb{R}. \end{array}\right.\nonumber\\
 \fl\label{Hermiticiy}
\end{eqnarray}
From both Hermiticities it follows that the spectrum of $D_{\rm W}(m)+(\mu_{\rm I} /2)\gamma_0\tau_3$ consists of complex conjugate pairs or real eigenvalues.
However the product of the real eigenvalues  is positive only if $\mu_{\rm I}$ is real since $(\gamma_5 D_{\rm W}(m)+(\mu_{\rm I}/2) \gamma_5\gamma_0)^\dagger=\gamma_5 D_{\rm W}(m)-(\mu_{\rm I}/2) \gamma_5\gamma_0$ so that $\lambda^*(\mu_{\rm I})=\lambda(-\mu_{\rm I})$. This guarantees the positive semi-definiteness of the statistical weight in Eq.~\eref{partition} if  $\mu_{\rm I}\in\mathbb{R}$ while for imaginary $\mu_{\rm I}$ only the reality of the statistical weight is ensured.

The twisted mass QCD partition function~\eref{partition} with the Dirac operator is given by
\begin{eqnarray}\label{twist-Dirac}
 D_{\omega}(m)=\gamma^\kappa D_\kappa-a \vec\nabla^2+m\cos\omega+\imath m\sin\omega\gamma_5\tau_3. 
\end{eqnarray}
Any kind of Hermiticity of $D_\omega$ is  lost for an arbitrary twist $\omega$ if the isospin chemical potential is imaginary. Thus the positive definiteness of the statistical weight will never be achieved in this case.
 However, the Dirac operator is still $\gamma_5\tau_{1/2}$ Hermitian if $\mu_{\rm I}$ is real such that the statistical weight is still positive semi-definite.
 
 In the ensuing discussions we restrict the twist to $\omega\in[0,\pi/2]$. Indeed the twists in the other three intervals $[-\pi,-\pi/2]$, $[-\pi/2,0]$, and $[\pi/2,\pi]$ can be traced back to the case $\omega\in[0,\pi/2]$ by relabelling the up and down quarks $u\leftrightarrow d$ and/or reflecting the quark mass $m\to-m$. Therefore this assumption does not restrict generality but it has the advantage that both $\cos\omega$ and $\sin\omega$ are non-negative.
`\subsection{Chiral perturbation theory of Twisted Mass Fermions}\label{sec:2.2}

The chiral Lagrangian of Wilson fermions with a twisted mass and a finite chemical potential follows from the symmetries of the fermionic action.
Classically it is invariant under the group $\SU_{\rm L}(N_{\rm f})\times\SU_{\rm R}(N_{\rm f})\times\U_{\rm V}(1)\times\U_{\rm A}(1)$. Due to the chiral anomaly the group $\U_{\rm A}(1)$ is  broken explicitly. 
The  chiral symmetry  is spontaneously broken 
by a non-zero chiral condensate 
 $\Sigma=|\langle\bar{\psi}\psi\rangle|$ with symmetry breaking pattern
 $\SU_{\rm L}(N_{\rm f})\times\SU_{\rm R}(N_{\rm f})\rightarrow\SU_{\rm V}(N_{\rm f})$.
 Therefore the corresponding Goldstone bosons $U$ are elements in $U=U_0\exp[\imath\Pi/F_{\pi}]\in\SU(N_{\rm f})$ with $\Pi$ the pion fields.

To construct an effective theory one needs two ingredients. The first 
ingredient is the thermodynamic and low-mass, low-momentum, low-chemical potential and low-temperature limit so that the QCD partition function is
dominated by  the pseudo-Goldstone bosons.
Second, we need a counting scheme to order the terms in the chiral Lagrangian 
allowed by symmetry.
We choose the $p$-counting scheme, 
\begin{eqnarray}\label{counting-scheme}
  m^2 \sim 1/ V,\ \mu_{\rm I}^4 \sim 1/V ,\ a^4\sim 1/V,\ p_\kappa^4\sim 1/V,\ \Pi^4\sim 1/V\ {\rm fixed},
\end{eqnarray}
where $a$ is the lattice spacing and $p_\kappa$ is the four momentum of the pion fields $\Pi$. In this counting scheme the kinetic modes ($p\neq0$) and the modes with zero momentum ($p=0$) are still coupled. To lowest order
the action is given by~\cite{SharpeSingleton,RS,BRS,Aoki-spec,GSS}
\begin{eqnarray}\label{action-p}
 S_{\rm p}=\int d^4x \; \mathcal{L}_{\rm p}(U(x))
\end{eqnarray}
with the chiral Lagrangian
\begin{eqnarray}
 \fl\mathcal{L}_{\rm p}(U)&=&\frac{F_\pi^2}{4}\tr\widehat{D}_\kappa U\widehat{D}^\kappa U^\dagger-\frac{m\Sigma \cos\omega}{2}\tr(U+U^\dagger)-\frac{\imath m\Sigma \sin\omega}{2}\tr\tau_3(U-U^\dagger)\nonumber\\
 \fl&&+a^2\left(W_6+\frac{W_8}{2}\right)\tr^2(U+U^\dagger).
\label{chiLagr-p-regime}
\end{eqnarray}
The covariant derivatives in flavor space are given by \cite{\KST,\KSTVZ}
\begin{eqnarray}
 \widehat{D}_\kappa U=\partial_\kappa U-\frac{\mu_{\rm I}}{2}[U,\tau_3]_-\delta_{\kappa0},\quad\widehat{D}^\kappa U^\dagger=\partial_\kappa U^\dagger-\frac{\mu_{\rm I}}{2}[U^\dagger,\tau_3]_-\delta_{\kappa0}\label{cov-der}
\end{eqnarray}
with $[.,.]_-$ the commutator. The symmetries also allow 
a term linear in the lattice spacing $a$. However, this term is 
proportional to $\tr(U+U^\dagger)$,  such that one 
can eliminate  it by renormalizing the quark mass $m$ and the twist 
angle $\omega$. Therefore we can omit this term without loss of generality. 

The low energy constants are the chiral condensate $\Sigma$,
 the pion decay constant $F_\pi$ and 
the two low energy constants $W_{6/8}$ whose sign convention is chosen as in 
Refs.~\cite{DSV,Akemann-Nagao,Kieburg,KVZ}.
 The third low energy constant associated with the $a^2$ terms, 
usually denoted by $W_7 \tr^2(U-U^\dagger)$ vanishes for the two flavor $\SU(2)$ 
partition function, 
see Ref.~\cite{sharpe-prd}. However this does not mean that this low energy constant does not play any role in two-flavor QCD. It still affects the eigenvalue spectrum of the Wilson Dirac operator, see~\cite{Akemann-Nagao,Kieburg,KVZ} for analytical discussions of this operator at vanishing isospin chemical potential.

The two low energy constants corresponding to the finite lattice spacing reduce to one constant $c_2=W_6+W_8/2$ because the trace and the squared trace of a matrix $U\in\SU(2)$ are related to each other, $\tr^2 U=\tr U^2+2$. While $W_8$ 
is positive definite, see the discussions in~\cite{DSV,Kieburg:2012fw,HanShar}, the sign of $W_6$ is negative \cite{Kieburg:2012fw} so 
that $c_2$ can have either sign. 

We will derive the phase diagram using this effective theory. In order to 
do so it is useful to introduce four additional source terms which probe the vacuum structure
\be
J=\imath\sum_{k=1}^3 j_k \bar\psi \gamma_5\tau_k \psi+m_{\rm v}\bar\psi \psi,
\ee
which in the chiral Lagrangian lead to the terms
\be
{\cal L}_{\rm source}(U) = -\imath \sum_{k=1}^3  \frac{  j_k\Sigma}{2} {\rm tr}\, \tau_k(U-U^\dagger)-\frac{  m_{\rm v}\Sigma}{2}\tr (U+U^\dagger) .
\label{Lsource}
\ee
Note that the $j_0$-term and the $m_{\rm v}$-terms are not really new source terms since they enter like the twisted mass terms $m\cos\omega\eins$ and  $\imath m\sin\omega\gamma_5\tau_3$. Thus they can be certainly absorbed by the quark mass $m$ and the twist angle $\omega$. Nonetheless we still include these two terms due to two reasons. First, we want to underline the ${\rm SO}(4)$ transformation property (the adjoint action of the $\SU_{\rm L}(2)\times\SU_{\rm R}(2)$ flavor group with $\psi\to \exp[\imath(\alpha^k\tau_k+\beta^k\gamma_5\tau_k)]\psi$ and $\bar{\psi}\to \bar{\psi}\exp[\imath(-\alpha^k\tau_k+\beta^k\gamma_5\tau_k)]$ where $\alpha_k$ and $\beta_k$ are real angles) of the four dimensional Euclidean real vector $j=(j_1,j_2,j_3,m_{\rm v})$. Second, the sources $j_0$ and $m_{\rm v}$ can be chosen independently of the quark mass $m$ and the twist angle $\omega$. This becomes particularly important when either $\omega\equiv0,\pi/2$ or $m\equiv0$ and the saddle point manifold is more than only a single element of $\SU(2)$ in the thermodynamical limit.

The phase diagram will be determined from the saddle point where the effective action $S_{\rm p}$ takes its minimum. Let $U_0$ be this saddle point. Then the field $U_0$ has to be constant over the whole four-dimensional box $V$ since each kinematic part will increase the action. Therefore we expand the field $U$ as
\begin{eqnarray}\label{expansion}
\fl U&=&U_0\exp\left[\frac {\imath}{ F_\pi}\sum_{p,E}\Pi(p,E)\frac{\exp\left(\imath (p_k x^k- Et)\right)}{4\pi^2}\right]\\
\fl&=&U_0\left[\eins_2+\frac{\imath}{ F_\pi}\sum_{p,E}\Pi(p,E)\frac{\exp\left(\imath (p_k x^k- Et)\right)}{4\pi^2}\right.\nonumber\\
\fl&&\left.-\frac{1}{2 F_\pi^2}\sum_{p,E,p',E'}\Pi(p,E) \Pi(p',E')\frac{\exp\left(\imath[ (p_k+p'_k) x^k-(E+E')t]\right)}{16\pi^4}\right]+O(V^{-3/4})\nonumber
\end{eqnarray}
and
\begin{eqnarray}
\fl U^\dagger&=&\left[\eins_2-\frac{\imath}{ F_\pi}\sum_{p,E}\Pi^\dagger(p,E)\frac{\exp\left(-\imath (p_k x^k- Et)\right)}{4\pi^2}\right.\label{expansiondag}\\
\fl&&\left.-\frac{1}{2 F_\pi^2}\sum_{p,E,p',E'}\Pi^\dagger(p,E) \Pi^\dagger(p',E')\frac{\exp\left(-\imath[ (p_k+p'_k) x^k-(E+E')t]\right)}{16\pi^4}\right]U_0^\dagger+O(V^{-3/4}),\nonumber
\end{eqnarray}
where we employ the Fourier transform of the pion fields. 
 Note that the
unitarity of $U$ enforces the relation
\begin{eqnarray}
 \Pi(p,E)=\Pi^\dagger(-p,-E).
\end{eqnarray}
The factors of $1/{2\pi}$ are the normalization factors of the plane waves such that we have
\begin{eqnarray}
\fl\int d^4x \frac{\exp\left(\imath[ (p_k-p'_k) x^k-(E-E')t]\right)}{16\pi^4}=V\delta(E-E')\prod_{k=1}^3\delta(p_k-p'_k).
\end{eqnarray}

To improve the notation we define the dimensionless quantities
\begin{eqnarray}
 \fl\widehat{m}=m\Sigma V,\quad \widehat{\mu}_{\rm I}^2=\mu_{\rm I}^2VF_\pi^2,\quad \widehat{a}^2=a^2Vc_2,\quad \widehat{p}^2=p^2VF_\pi^2,\quad \widehat{E}^2=E^2VF_\pi^2.\label{dimensionless-param}
\end{eqnarray}
Then the expansion~\eref{expansion} is plugged into the chiral Lagrangian~\eref{chiLagr-p-regime}. Expanding up to second order in $\Pi$ the action reads 
\begin{eqnarray}\label{chiLagexp}
  S_{\rm p}&=&V\mathcal{L}_{0}(U_0)+\int d^4x\;\mathcal{L}_1(U_0,\Pi)+\int d^4x\;\mathcal{L}_2(U_0,\Pi) \nn\\
&&+\int d^4x\; {\cal L}_{\rm source}(U_0)+O(V^{-1/4}),
\end{eqnarray}
where $\mathcal{L}_{0}(U_0)$ is the mean field chiral Lagrangian evaluated at 
the saddle point $U_0$
\begin{eqnarray}
 \fl V\mathcal{L}_{0}(U_0) = \frac{\widehat{\mu}_{\rm I}^2}{16}\tr [U_0,\tau_3]_-[U_0^\dagger,\tau_3]_--\frac{\widehat{m}}{2}\tr(e^{\imath\omega\tau_3}U_0+e^{-\imath\omega\tau_3}U_0^\dagger)+\widehat{a}^2\tr^2(U_0+U_0^\dagger). 
  \label{chiLagr}
\end{eqnarray}
To the same order we have  to evaluate the source terms ${\cal L}_{\rm source}(U_0)$.
The zeroth order terms  $\mathcal{L}_{0}(U_0)+\mathcal{L}_{\rm source}(U_0)$ are 
of order $O(V^{-1/2})$ while the first order terms in  $\Pi$ given by
\begin{eqnarray}
\int d^4x \; \mathcal{L}_1(U_0,\Pi)&=&\frac{4\imath \pi^2}{F_\pi}\tr\Pi(0)\left(\frac{\widehat{\mu}_{\rm I}^2}{8}[\tau_3, U_0^\dagger\tau_3 U_0]_--\frac{\widehat{m}}{2}(e^{\imath\omega\tau_3}U_0-U_0^\dagger e^{-\imath\omega\tau_3})\right.\nonumber\\
 &&\left.+2\widehat{a}^2(U_0-U_0^\dagger)\tr(U_0+U_0^\dagger)\right)\label{L1}
\end{eqnarray}
and the second order terms in $\Pi$ given by
\begin{eqnarray}
 \fl \int d^4x \; \mathcal{L}_2(U_0,\Pi)&=&\frac{2}{F^2_\pi}\sum_{\widehat{p},\widehat{E}>0}\left[ \frac{\widehat{p}_k\widehat{p}^k-\widehat{E}^2}{4}\tr\Pi(\widehat{p},\widehat{E})\Pi^\dagger(\widehat{p},\widehat{E})-\frac{\widehat{\mu}_{\rm I} \widehat{E}}{4}\tr[U_0\Pi(\widehat{p},\widehat{E}),\Pi^\dagger(\widehat{p},\widehat{E})U_0^\dagger]_-\tau_3\right.\nonumber\\
 \fl&&\hspace*{-2cm}\left.+\frac{\widehat{\mu}_{\rm I}^2}{16}{\rm Re}\,\tr[\Pi(\widehat{p},\widehat{E}),\tau_3]_-[\Pi^\dagger(\widehat{p},\widehat{E}),U_0^\dagger\tau_3 U_0]_-+\frac{\widehat{m}}{4}\tr(e^{\imath\omega\tau_3}U_0+U_0^\dagger e^{-\imath\omega\tau_3})\Pi(p,E)\Pi^\dagger(\widehat{p},\widehat{E})\right.\nonumber\\
 \fl&&\hspace*{-2cm}\left.+\widehat{a}^2\left|\tr(U_0-U_0^\dagger)\Pi(\widehat{p},\widehat{E})\right|^2-\widehat{a}^2\tr(U_0+U_0^\dagger)\Pi(\widehat{p},\widehat{E})\Pi^\dagger(\widehat{p},\widehat{E})\tr(U_0+U_0^\dagger)\right]\label{L2}
\end{eqnarray}
are of order $O(V^{-3/4})$ and $O(V^{-1})$, respectively.

The saddle point $U_0$ is calculated in \ref{app1} and will be discussed together with the corresponding phases in section~\ref{sec:3}. At the saddle point $U_0$ the first order Lagrangian $\mathcal{L}_1$ vanishes.  
 The Lagrangian $\mathcal{L}_2$ is the lowest non-vanishing order for the non-zero
momentum modes of the pseudo-scalar pions. In section~\ref{sec:4} we will  derive  the dispersion relations using $\mathcal{L}_2(U_0,\Pi)$ and subsequently read 
off the masses. This is also the reason why we Wick rotated from Euclidean to Minkowski space, i.e. $\partial_0\to-\imath\partial_t$.

Before we proceed with this discussion, let us note that the shifted Lagrangian $V\mathcal{L}_0(U_0)-8\widehat{a}^2$, see eq.~(\ref{chiLagr}), is invariant under the following transformation
\be
\label{symmetry-U}
\omega &\to &\frac \pi 2 - \omega,\nn \\
U_0 &\to& \tau_1 U_0 \tau_2,\nn \\
\wha^2 &\to& -\wha^2,\nn \\
\whmu^2 &\to& \whmu^2 +32 \wha^2.
\ee
To see this one has to use that any $U_0\in \SU(2)$ can be parametrized by
\be
 U_0=\alpha\eins_2+\imath\beta_k\tau_k\ {\rm with}\ \alpha^2+\beta_k\beta_k=1\ {\rm and}\ \alpha,\beta_k\in\mathbb{R}.
\ee 
Since this symmetry of $\mathcal{L}_0$  is valid for all $U_0\in \SU(2)$
it can be extended to $\mathcal{L}_1$ with the additional change $\Pi(0)\to\tau_2\Pi(0)\tau_2$ because $\mathcal{L}_1$ is a derivative of $\mathcal{L}_0$.
The symmetry is violated by the kinetic term (in particular the term proportional to $\widehat{\mu}_{\rm I}\widehat{E}$) and does not hold for
$\mathcal{L}_2$. Therefore the  pion masses do not reflect this
symmetry.
As we will see below, this symmetry relates the phase diagram for maximal
twist to the phase diagram for zero twist.

\subsection{Order Parameters}
\label{sec:order}

The phases of the full partition function are determined by
\begin{equation}
Z_0=\int_{{\rm SU}(2)}d\mu(U_0)\exp[-V(\mathcal{L}_0(U_0)+\mathcal{L}_{\rm source}(U_0))]
\end{equation}
and are characterized by order parameters. 
We consider the following ones:

\noindent
i) The chiral condensate can be introduced by introducing the auxiliary variable $m_{\rm v}$,
\be\label{chicond-def}
 \Sigma(\widehat{m}) &=& -\frac 12 \displaystyle{\lim_{V\to\infty}} \frac 1V \int d^4 x 
\langle\bar{\psi}(x)\psi(x)\rangle\nn\\
&=&
\lim_{m_{\rm v}\to 0}\lim_{V\to\infty}\frac{1}{2V}\frac{\partial}{\partial m_{\rm v}}
{\rm ln}Z_0\nn\\
&=&\frac \Sigma 4 \displaystyle{\lim_{V\to\infty}}\langle \tr(U_0+U_0^\dagger)  \rangle.
\ee
Note that only $m\langle \bar \psi\psi\rangle$ is a renormalization group invariant,
and the actual order parameter is given by this combination. In the case of 
the Sharpe-Singleton scenario, the quark mass has to be replaced by the
distance between the quark mass and the Dirac spectrum. Since it remains
finite for $m \to 0$, we obtain a first order phase transition. Strictly
speaking it is not a phase transition because the two phases have  the
same physical properties. In particular the system can lie in two different but physically equivalent ground states at vanishing quark mass.

\noindent
ii) The $\pi^0$ condensate can be generated by the source $j_0=j_3$,
\be 
C_{\pi^0}&=&\frac \imath 2   \lim_{V\to\infty}\frac{1}{V} 
\int d^4 x \langle\bar{\psi}(x)\gamma_5\tau_3\psi(x)\rangle\nn \\
&=&\lim_{j_0\to 0} \lim_{V\to\infty}\frac{1}{2V}
\frac{\partial}{\partial j_0}{\rm ln}Z_0\nn \\
&=& \imath  \frac \Sigma 4 \langle \tr \tau_3(U_0-  U_0^\dagger) \rangle.
 \ee

\noindent
iii) For the charged pion condensates  we employ the standard notation that $\pi^+\propto \bar{d}u$ and $\pi^-\propto \bar{u}d$ such that $\bar{\psi}=(\bar{u},\bar{d})$. Therefore when defining $\tau_\pm=\tau_1\pm\imath\tau_2$ and $j_\pm=j_1\pm\imath j_2$ these condensates read
\be 
 C_{\pi^\pm}&=&\frac \imath 2 \lim_{V\to\infty}\frac{1}{V} 
\int d^4 x \langle\bar{\psi}(x)\gamma_5\frac{\tau_\mp}{2}\psi(x)\rangle\nn \\
&=&\lim_{j_\pm\to 0} \lim_{V\to\infty}\frac{1}{2V} 
\frac{\partial}{\partial j_\pm}{\rm ln}Z_0 \nn\\
&=& \imath  \frac \Sigma 8 \langle \tr \tau_\mp(U_0-  U_0^\dagger) \rangle.
\label{charged-cond-def} 
\ee

\noindent
iv) The isospin charge density is given by
\be
\label{isospinden-def}
 n_{\rm I}&=&\lim_{V \to \infty}\frac 1{2V} \int d^4 x \langle\bar{\psi}(x)\gamma_0\tau_3\psi(x)\rangle
\nn\\
&=& \lim_{V\to\infty}\frac{1}{2V}\frac{\partial}{\partial \mu_{\rm I}}{\rm ln}Z_0
\nn \\
&=&- \frac {\mu_{\rm I} F_\pi^2}{16} \langle \tr [U_0,\tau_3]_-[U_0^\dagger,\tau_3]_-\rangle. 
 \ee
Since the phases are already  fully characterized by the first three condensates
we do not need the charge density as an additional order parameter.

Nonetheless other interesting observables such as the pion covariances and the chiral variance always play an important role when the angles of the Goldstone manifold ${\rm SU}(2)$ do not completely freeze out but the ground states are only fixed via the source terms. These three covariances are given by
\be
\label{suscep0-def}
 \fl\Delta C_{\pi^0}&=&\lim_{V \to \infty}\frac 1{4V^2} \left[\left\langle\left(\int d^4 x\bar{\psi}(x)\gamma_5\tau_3\psi(x)\right)^2\right\rangle-\left\langle\int d^4 x\bar{\psi}(x)\gamma_5\tau_3\psi(x)\right\rangle^2\right]
\nn\\
\fl&=& -\lim_{j_0\to 0}\lim_{V\to\infty}\frac{1}{4V^2}\frac{\partial^2}{\partial j_0^2}{\rm ln}Z_0
\nn \\
\fl&=& -\frac{ \Sigma^2 }{16}\left[\langle \tr^2 \tau_3(U_0-U_0^\dagger)\rangle-\langle\tr \tau_3(U_0-U_0^\dagger)\rangle^2\right]
 \ee
for the variance of the $\pi^0$ condensate,
\be
\label{susceppm-def}
 \fl\Delta C_{\pi^\pm}&=&\lim_{V \to \infty}\frac 1{4V^2} \left[\left\langle\left|\int d^4 x\bar{\psi}(x)\gamma_5\frac{\tau_+}{2}\psi(x)\right|^2\right\rangle-\left|\left\langle\int d^4 x\bar{\psi}(x)\gamma_5\frac{\tau_+}{2}\psi(x)\right\rangle\right|^2\right]
\nn\\
\fl&=& \lim_{j_\pm\to 0}\lim_{V\to\infty}\frac{1}{4V^2}\frac{\partial^2}{\partial j_+\partial j_-}{\rm ln}Z_0
\nn \\
\fl&=& \frac{ \Sigma^2 }{64}\left[\left\langle \left|\tr \tau_+(U_0-U_0^\dagger)\right|^2\right\rangle-\left|\left\langle\tr \tau_+(U_0-U_0^\dagger)\right\rangle\right|^2\right]
 \ee
 for the covariance between the two charged pion condensates and
\be
\label{suscepchi-def}
 \fl\Delta \Sigma(\widehat{m})&=&\lim_{V \to \infty}\frac 1{4V^2} \left[\left\langle\left|\int d^4 x\bar{\psi}(x)\psi(x)\right|^2\right\rangle-\left\langle\int d^4 x\bar{\psi}(x)\psi(x)\right\rangle^2\right]
\nn\\
\fl&=&\left.\lim_{V\to\infty}\frac{1}{4V^2}\frac{\partial^2}{\partial m_{\rm v}^2}{\rm ln}Z_0\right|_{m_{\rm v}=m\cos\omega}
\nn \\
\fl&=& \frac{ \Sigma^2 }{16}\left[\langle \tr^2 (U_0+U_0^\dagger)\rangle-\langle\tr (U_0+U_0^\dagger)\rangle^2\right]
 \ee
for the variance of the chiral condensate.  We want to emphasize that these covariances would be proportional to the susceptibilities at finite volume $V$, i.e. our definition incorporates an additional factor of $1/V$. However in the case that the average of a condensate vanishes the corresponding (co-)variance remains finite while its susceptibility is of the order $\mathcal{O}(V)$.

\subsection{Parameterization of $U_0$}\label{paramet}

The mean field phase diagram is determined by the zero momentum part of the chiral Lagrangian~\eref{chiLagr} with parameters occurring in the combination
\be
mV, \mu^2 V, a^2 V.
\ee

The terms of the zero momentum chiral Lagrangian coincide with the leading order chiral Lagrangian of the $\epsilon$-counting scheme as well as with the chiral
Lagrangian that is obtained from chiral random matrix theory in the $\epsilon$-domain of QCD. Therefore the mean field phase diagram can also be obtained
from the Wilson random matrix partition function~\cite{DSV}. However, to determine the pion masses we need the kinetic term and have to use the full lowest order chiral 
Lagrangian in the p-counting scheme.

To perform a saddle point analysis we need a parameterization of the group $\SU(2)$. Our choice is 
\begin{eqnarray}\label{parametSU(2)}
 \fl U_0&=&V\Phi V^{-1}\\
 \fl&=&\left[\begin{array}{cc} 
\displaystyle\cos\frac{\vartheta_1}{2} & -\displaystyle e^{\imath\vartheta_2} \sin\frac{\vartheta_1}{2} \\ \displaystyle e^{-\imath\vartheta_2}\sin\frac{\vartheta_1}{2} & \displaystyle \cos\frac{\vartheta_1}{2} \end{array}\right]\left[\begin{array}{cc} e^{\imath\varphi} & 0 \\ 0 & e^{-\imath\varphi}\end{array}\right]\left[\begin{array}{cc} \displaystyle\cos\frac{\vartheta_1}{2} & \displaystyle e^{\imath\vartheta_2}\sin\frac{\vartheta_1}{2} \\ -\displaystyle e^{-\imath\vartheta_2}\sin\frac{\vartheta_1}{2} & \displaystyle \cos\frac{\vartheta_1}{2} \end{array}\right]\nonumber\\
 \fl&=&\cos\varphi\eins_2+\imath\sin\varphi\left[\begin{array}{cc}  \displaystyle  e^{\imath\vartheta_2/2} & \displaystyle 0 \\ \displaystyle 0 & \displaystyle e^{-\imath\vartheta_2/2} \end{array}\right]\left[\begin{array}{cc} \cos\vartheta_1 & \sin\vartheta_1 \\ \sin\vartheta_1 & -\cos\vartheta_1 \end{array}\right]\left[\begin{array}{cc} \displaystyle e^{-\imath\vartheta_2/2} & \displaystyle 0 \\ \displaystyle 0 & \displaystyle e^{\imath\vartheta_2/2} \end{array}\right]\nonumber
\end{eqnarray}
with the angles $\varphi\in[0,\pi]$, $\vartheta_1\in[0,\pi[$, and $\vartheta_2\in[0,2\pi[$. The Haar measure $d\mu(U)$ in terms of these coordinates follows from the invariant length element,
\begin{eqnarray}\label{metric}
 \tr d\widehat{U}d\widehat{U}^{-1}&=&\tr d\Phi d\Phi^{-1}+\tr[\Phi,V^{-1}dV]_-[\Phi^{-1},V^{-1}dV]_-\\
 &=&2d\varphi^2+8\sin^2\varphi d\vartheta_1^2+8\sin^2\varphi\sin^2\vartheta_1 d\vartheta_2^2,\nonumber
\end{eqnarray}
such that we have the invariant measure
\begin{equation}\label{measure}
 d\mu(U)=16\sin^2\varphi\sin\vartheta_1d\vartheta_1d\vartheta_2d\varphi.
\end{equation}
Note that $\sin\varphi$ as well as $\sin\vartheta_1$ are positive on their support.

In the coordinates~\eref{parametSU(2)} the chiral Lagrangian~\eref{chiLagr} is given by
\begin{equation}\label{chiLagrcoor}
  \fl V\mathcal{L}_{0}(\varphi,\vartheta_1)=-2\widehat{m}\cos\omega\cos\varphi+2\widehat{m}\sin\omega\sin\varphi\cos\vartheta_1+ 16\widehat{a}^2\cos^2\varphi-\frac{\widehat{\mu}_{\rm I}^2}{2}\sin^2\varphi  \sin^2\vartheta_1, 
\end{equation}
and
\begin{eqnarray}
\fl \mathcal{L}_{\rm source}(\varphi,\vartheta_1)&=&\Sigma(
2j_0\sin\varphi\cos\vartheta_1
+j_+ \sin\varphi \sin\vartheta_1 e^{\imath\vartheta_2}
+ j_- \sin\varphi \sin\vartheta_1 e^{-\imath\vartheta_2})\nn\\
\fl&&-2\Sigma m_{\rm v}\cos\varphi.
\end{eqnarray}
Note that the dependence on the angle $\vartheta_2$ completely drops out
of the microscopic Lagrangian $\mathcal{L}_0$ which determines the phase diagram. 
 The $\pi^\pm$ condensates have a phase associated with an exact Goldstone mode corresponding to the angle $\vartheta_2$. The exact nature of this mode can be fixed by the introduction of an appropriate source term which are the two terms proportional to $j_\pm$.

In the thermodynamic limit the angles $\varphi$ and $\vartheta_1$ are fixed by the saddle point equations. 
To measure these angles and, thus, in which phase the system is, 
we use the order parameters discussed in section \ref{sec:order}. In terms of the
parameterization (\ref{parametSU(2)}) they are given by
\begin{eqnarray}\label{condendates}
 \Sigma(\widehat{m})
&=&\Sigma  \lim_{m_{\rm v} \to 0}\lim_{V\to\infty}\langle \cos\varphi\rangle,\nn\\
  C_{\pi^0}&=&
\Sigma \lim_{j_0 \to 0}\lim_{V\to\infty}\langle \sin\varphi\cos\vartheta_1\rangle,\label{pion0cond}
\nn\\
  C_{\pi^\pm}&=& -\frac{\Sigma}{2} 
\lim_{j_\pm \to 0}\lim_{V\to\infty}\langle \sin\varphi\sin\vartheta_1 e^{\pm\imath\vartheta_2}\rangle.
\end{eqnarray}
Note that the order of the derivatives and the thermodynamic limit is crucial since they do not commute.
The isospin charge density in terms of these variables is given by
\be
n_{\rm I} = \frac{\mu_{\rm I} F_\pi^2}{2} \lim_{V\to\infty}\langle \sin^2 \varphi \sin^2 \vartheta_1 \rangle.
\ee
Also the variances take simple forms like
\be
\Delta C_{\pi^0}= \Sigma^2\lim_{V\to\infty}\left(\langle \cos^2\vartheta_1 \sin^2 \varphi\rangle-\langle \cos\vartheta_1 \sin \varphi\rangle^2\right)
\label{suscep}
\ee
for the $\pi^0$ variance and
\be
\Delta \Sigma= \Sigma^2\lim_{V\to\infty}\left(\langle  \cos^2 \varphi\rangle-\langle  \cos \varphi\rangle^2\right)
\label{suscepchi}
\ee
for the variance of the chiral condensate. The covariance between the two charged pion condensates can be traced back to the isospin density,
\begin{equation}\label{pi-pm-n-I}
\Delta C_{\pi^\pm}=\frac{n_{\rm I}\Sigma^2}{2\mu_{\rm I} F_\pi^2}-|C_{\pi^\pm}|^2,
\end{equation}
and, hence, does not yield anything new.
 
In the case that the angles $\varphi$, $\vartheta_1$ and $\vartheta_2$ freeze out at only one saddle point $U_0$ due to the large volume limit $V\to\infty$ (including source terms $j\neq0$) all quantities reduce to two terms only, namely
\begin{equation}\label{red-terms}
 \Sigma(\widehat{m})=\Sigma\cos\varphi_0\quad{\rm and}\quad C_{\pi^0}=\Sigma\sin\varphi_0\cos\vartheta_{1,0},
\end{equation}
where the subscript $0$ indicates the value at the saddle point.
The other terms take the values
\begin{eqnarray}\label{reduction}
 C_{\pi^\pm}&=&-\frac{\Sigma j_\pm}{2|j_\pm|}\sqrt{1-\left(\frac{\Sigma(\widehat{m})}{\Sigma}\right)^2-\left(\frac{C_{\pi^0}}{\Sigma}\right)^2},\nn\\
 n_{\rm I}&=&\frac{\mu_{\rm I} F_\pi^2}{2}\left[1-\left(\frac{\Sigma(\widehat{m})}{\Sigma}\right)^2-\left(\frac{C_{\pi^0}}{\Sigma}\right)^2\right],\nn\\
 \Delta C_{\pi^0}&=& \Delta C_{\pi^\pm}=\Delta \Sigma=0.
\end{eqnarray}
Note that the variances and covariances only vanish because the angles take certain values when switching off the source terms they can be non-zero.

\section{Phase Diagram of Twisted Wilson Fermions}\label{sec:3}

To determine the phase diagram we have to minimize the effective potential
given by $V{\cal L}_{0}$ in Eq.~(\ref{chiLagrcoor}).
It is useful to rewrite the potential in  the form
\begin{eqnarray}
\label{Lagrangian}
 \fl V\mathcal{L}_{0}&=& \left(16\widehat{a}^2
+ \frac{\widehat{\mu}_{\rm I}^2}{2}\right)\left(\cos\varphi
-\frac{2\widehat{m}\cos\omega}{32\widehat{a}^2+ \widehat{\mu}_{\rm I}^2}\right)^2+ \frac{\widehat{\mu}_{\rm I}^2}{2}\left(\sin\varphi\cos\vartheta_1
+\frac{2\widehat{m}\sin\omega}{\widehat{\mu}_{\rm I}^2}\right)^2\\
 \fl&&
-\frac{\widehat{\mu}_{\rm I}^2}{2}-\frac{2\widehat{m}^2\sin^2\omega}{\widehat{\mu}^2}
-\frac{2\widehat{m}^2\cos^2\omega}{32\widehat{a}^2+ \widehat{\mu}_{\rm I}^2},\nonumber
\end{eqnarray}
and we have to determine its minima with respect to 
the variables $\varphi\in[0,\pi]$ and $\vartheta_1\in[0,\pi[$. 
The third angle $\vartheta_2$ parametrizing the group $\SU(2)$ can only be
fixed by the source terms $j_\pm$.

From Eq.~\eref{Lagrangian} one can easily read off  two cases determined by the solutions
of the saddle point equations for  $\vartheta_1$,
\be
 \sin\varphi\cos\vartheta_1=-\frac{2\widehat{m}\sin\omega}{\widehat{\mu}_{\rm I}^2}
\quad {\rm and}\quad \sin \vartheta_1 =0.
\label{saddletheta}
\ee

\noindent
i)  The first solution in Eq.~(\ref{saddletheta}) is assumed when
the isospin chemical potential is real ($\widehat{\mu}_{\rm I}^2>0$) and the modulus of the mass is bounded from above as $0\leq 2|\widehat{m}|\sin\omega\leq \widehat{\mu}_{\rm I}^2\sin\varphi$. The Lagrangian simplifies to
\begin{eqnarray}\label{Lagrangian-simp1}
 \fl V\mathcal{L}_{0}= \left(16\widehat{a}^2+ \frac{\widehat{\mu}_{\rm I}^2}{2}\right)\left(\cos\varphi-\frac{2\widehat{m}\cos\omega}{32\widehat{a}^2+ \widehat{\mu}_{\rm I}^2}\right)^2-\frac{\widehat{\mu}_{\rm I}^2}{2}-\frac{2\widehat{m}^2\sin^2\omega}{\widehat{\mu}_{\rm I}^2}-\frac{2\widehat{m}^2\cos^2\omega}{32\widehat{a}^2+ \widehat{\mu}_{\rm I}^2}.
\end{eqnarray}
Minimizing with respect to $\varphi$ we find two different solutions
\be
\cos\varphi = \frac{2\whm\cos\omega}{32\wha^2+\whmu^2}\quad {\rm or} \quad
\sin \varphi =0.
\ee

The phase with $0 <\cos \varphi < 1$ and $0 <\cos \vartheta_1 < 1$  will
be denoted by $I$. The solution $\sin \varphi =0$ can  only exist for $\hm \sin \omega = 0$
(denoted by $III^{(\omega=0)}$). At zero twist we also have the solution 
$\cos \vartheta_1 = 0$ for $\hmu^2 > 0$. This gives two possible phases, one with
$|\cos \varphi| = 1$ and one with $|\cos \varphi| < 1$.
Otherwise the first solution in Eq.~(\ref{saddletheta}) cannot be satisfied.
  For $\omega = \pi/2$ we find $\cos \varphi = 0$ and 
$ \cos\vartheta_1 = - 2\whm/\whmu^2$ which will be referred to as phase $I^{(\omega=\pi/2)}$.

Further analysis of  this case including the computation of the parameter domain of the
various phases is carried out in \ref{app1.1}.

\begin{figure}[!t]
\centerline{\includegraphics[width=0.8\textwidth]{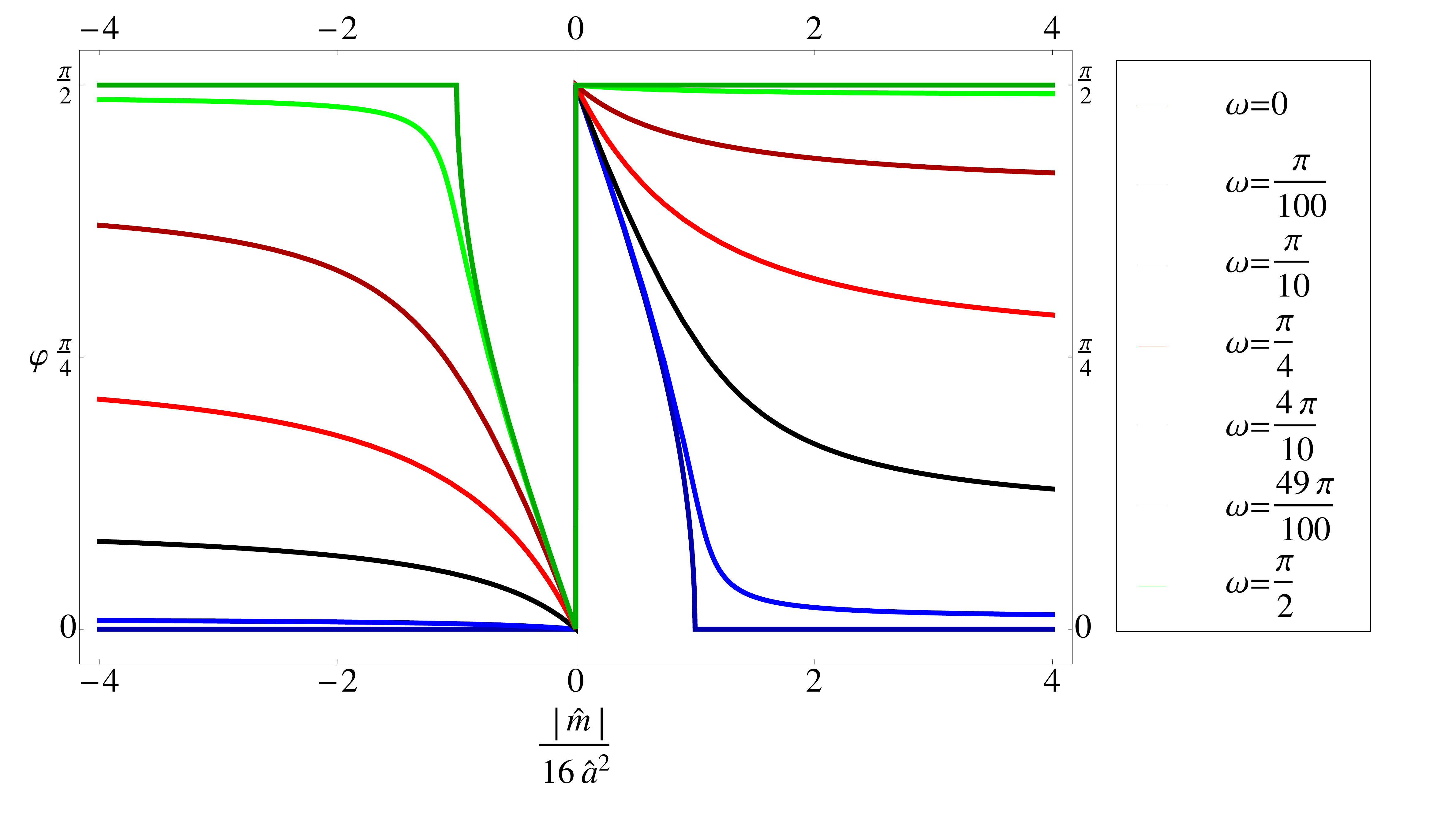}}
\caption{Solution of the transcendental equation~\eref{trans} for $\varphi \in[0,\pi/2]$, see Eq.~\eref{angles-P3} as an explicit  function of $|\whm|/16 \wha^2$. Note that the solution is not differentiable for $\omega=0,\pi/2$ which is the reason for the splitting of the phase $II$ into two phases at zero and maximal twist.}
\label{fig2c}
\end{figure}

ii) The second  solution in Eq.~(\ref{saddletheta}) minimizes the 
effective potential
when $ 2|\widehat{m}|\sin\omega>\widehat{\mu}_{\rm I}^2\sin\varphi$ which
includes the case $\hmu^2 < 0$.
Then $\cos\vartheta_1$ has to take one of its extremal values $\pm1$.
 Comparing both solutions we find that $\cos\vartheta_1=-\sign\widehat{m}$ is 
always the minimum in these cases. Then the Lagrangian 
reads 
\begin{eqnarray}\label{Lagrangian-simp2}
 V\mathcal{L}_{\rm 0}&=&
16\widehat{a}^2\cos^2\varphi-2|\widehat{m}|\sin\omega\sin\varphi
-2\widehat{m}\cos\omega\cos\varphi.
\end{eqnarray}
This Lagrangian satisfies up to a constant shift the following symmetry 
\be
\omega\to \frac \pi 2 -\omega, \quad
\varphi \to \frac \pi 2 {\rm sign}(\widehat{m}) -\varphi,\quad
\widehat{a}^2 \to - \widehat{a}^2, 
\ee 
which is reminiscent of the symmetry~(\ref{symmetry-U}). Note that no chemical potential is involved in this symmetry since the free energy is independent of it in this part of the phase diagram and in this order of the free energy. Therefore the system enters the ``Silver-blaze-property".

\begin{table}[t!] 
\begin{tabular}[c]{c||c}
 phase & region  \\ 
 \noalign{\vskip\doublerulesep\hrule height 2pt}
 $I$, $\omega\in[0,\pi/2]$ & 
 $\widehat{\mu}_{\rm I}^2>2|\widehat{m}|\sin\omega\geq0$ (if $\omega=0$ then $\widehat{\mu}_{\rm I}^2>0$) \\
 & and   $\displaystyle \overset{\ }{ 32\widehat{a}^2+\widehat{\mu}_{\rm I}^2}>\overset{\ }{\frac{2\widehat{\mu}_{\rm I}^2|\widehat{m}|\cos\omega}{\sqrt{\widehat{\mu}_{\rm I}^4-4\widehat{m}^2\sin^2\omega}}}>2|\widehat{m}|\cos\omega\geq0$
 \\ \hline 
 $II_{\pm}$, $\omega\in]0,\pi/2[$ & $\widehat{\mu}_{\rm I}^2\geq2|\widehat{m}|\sin\omega\geq0$ and $\displaystyle\overset{\ }{\frac{2|\widehat{m}||\widehat{\mu}_{\rm I}|^2\cos\omega}{\sqrt{\widehat{\mu}_{\rm I}^4-4\widehat{m}^2\sin^2\omega}}> 32\widehat{a}^2+\widehat{\mu}_{\rm I}^2}$;
 \\ & {\bf or} $2|\widehat{m}|\sin\omega> \widehat{\mu}_{\rm I}^2$
 \\ \noalign{\vskip\doublerulesep\hrule height 2pt} 
 $II_\pm^{(\omega=0)}$ &  $\widehat{\mu}_{\rm I}^2<0\ {\rm and}\ 16\widehat{a}^2>|\widehat{m}|>0$
 \\ \hline 
 $III_\pm^{(\omega=0)}$ & $\widehat{\mu}_{\rm I}^2>0\ {\rm and}\ 2|\widehat{m}|> 32\widehat{a}^2+\widehat{\mu}_{\rm I}^2$;
\\ & {\bf or} $0> \widehat{\mu}_{\rm I}^2\ {\rm and}\ |\widehat{m}|>16\widehat{a}^2$
 \\ \hline 
 Aoki$^{(\omega=0)}$ & $\widehat{\mu}_{\rm I}^2=0\ {\rm and}\ 16\widehat{a}^2>|\widehat{m}|>0$ 
\\ \noalign{\vskip\doublerulesep\hrule height 2pt} 
$II_\pm^{(\omega=\pi/2)}$ & $\widehat{\mu}_{\rm I}^2\leq0\ {\rm and}\ -16\widehat{a}^2>|\widehat{m}|>0$
 \\ & {\bf or} $\displaystyle -16\widehat{a}^2>|\widehat{m}|>0\ {\rm and}\ -32\widehat{a}^2>\widehat{\mu}_{\rm I}^2>0$ \\ \hline $III_\pm^{(\omega=\pi/2)}$ & $\widehat{\mu}_{\rm I}^2\leq0\ {\rm and}\ |\widehat{m}|>-16\widehat{a}^2$;
 \\ & {\bf or} $2|\widehat{m}|>\widehat{\mu}_{\rm I}^2>0\ {\rm and}\ |\widehat{m}|>-16\widehat{a}^2$
 \\ \hline
 Aoki$^{(\omega=\pi/2)}$ & $\widehat{\mu}_{\rm I}^2=-32\widehat{a}^2\ {\rm and}\ -16\widehat{a}^2>|\widehat{m}|>0$ 
\end{tabular}
\caption{ Regions of the phases. Note that the 
phase $III_\pm$ only occurs for $\omega\neq0,\pi/2$.  The Aoki phase only exists at $\widehat{\mu}_{\rm I}=\omega=0$ which connects the phases $I$ to $II_\pm^{(\omega=0)}$ along the chemical potential axis via a first order phase transition. It has an analogue at imaginary effective lattice spacing ($\widehat{a}^2<0$) and maximal twist ($\omega=\pi/2$) with real isospin chemical potential equal to $\widehat{\mu}_{\rm I}^2=-32\widehat{a}^2$. This phase exhibits a similar behavior of the corresponding condensates.}\label{t1}
\end{table}

The saddle point equation can be rewritten as
\be\label{trans}
8\wha^2 \sin2\varphi = -|\widehat{m}| \sin(\omega - \sign(\widehat{m})\varphi)
\ee
and can be solved numerically or analytically as we do in \ref{app1.2}.
In Fig.~\ref{fig2c} we show the solution for a range of twist angles.
For $0< \omega <\pi/2$ we always have that  $|\cos\varphi| < 1$ for
 $ |\cos\vartheta| =1$. This phase will be denoted by $II$. 

For $\omega = 0$ the solution 
of the transcendental equation~\eref{trans} splits into two branches
$ \cos \varphi =1 $ (denoted by phase $III^{(\omega=0)}$)  and $ \cos\varphi = \whm/16\wha^2$ (denoted by $II^{(\omega =0)}$). We underline that the  angle $\vartheta_1$ is not determined in the first case because the saddle point is $U_0=\eins_2$.
Also for $\omega=\pi/2$ the solution branches into $\cos\vartheta_1 = 1$
with $\cos\phi =0$ (denoted by phase $III^{(\omega=\pi/2)}$)
and $\cos\vartheta_1 = 1$
with $\sin\phi =\whm/16\wha^2$  (denoted by phase $II^{(\omega=\pi/2)}$). 

We further analyze  case ii) in \ref{app1.2}.

We find two phases at finite twist $\omega\neq0,\pi/2$ and three at each extremal twist $\omega=0,\pi/2$. Additionally there is an Aoki phase~\cite{Aokiclassic} at $\omega=0$ which has a counterpart at $\omega=\pi/2$ due to the symmetry~\eref{symmetry-U}.  The parameter domain  of the phases is summarized in Table~\ref{t1}. The corresponding order parameters are collectively presented in Table~\ref{t2}. Thereby we also show the covariances which are non-trivial for those condensates which are only aligned by the source terms $j$ and $m_{\rm v}$. We underline that at $j,m_{\rm v}\equiv0$ the condensates vanish while the covariances do not. However the condensates do not vanish if the thermodynamical limit
is taken before the limit $j,m_{\rm v}\to0$. This is a fundamental characteristic of spontaneous breaking of symmetries.

\subsection{Zero Twist}\label{sec:zero-twist}

First we concentrate on the case of zero twist ($\omega=0$). In this case
the mean field Lagrangian is given by
\be
V\mathcal{L}_0|_{\omega=0}=-2\widehat{m}\cos\varphi+ 16\widehat{a}^2\cos^2\varphi
-4\widehat{\mu}^2\sin^2\varphi \sin^2\vartheta_1 .
\ee
Before we discuss the general case of this kind of the Lagrangian  we first discuss the cases of zero lattice
spacing and zero chemical potential.

\begin{table}[t!] \centering
\rotatebox{90}{
\begin{tabular}[c]{c||c|c|c|c}
 phase &  $\displaystyle 
\frac{\Sigma(\widehat{m})}{\Sigma}=-\frac{\langle\bar{\psi}\psi\rangle}{2V\Sigma}$ & $\displaystyle \frac{C_{\pi^0}}{\Sigma}=\imath\frac{\langle\bar{\psi}\gamma_5\tau_3\psi\rangle}{2V\Sigma}$ & $\displaystyle\frac{\Delta\Sigma(\widehat{m})}{\Sigma^2}$ & $\displaystyle\frac{\Delta C_{\pi^0}}{\Sigma^2}$ \\ 
 \noalign{\vskip\doublerulesep\hrule height 2pt} 
 $I$ & 
$\overset{\ }{\frac{2\widehat{m}\cos\omega}{32\widehat{a}^2+ \widehat{\mu}_{\rm I}^2}}$
 & $-\frac{2\widehat{m}\sin\omega}{\widehat{\mu}_{\rm I}^2}$ & 0 &  
0 \\
 $\omega\in\left[0,\frac{\pi}{2}\right]$ &&&& \\
 \hline $II_{\pm}$ & $\frac{\widehat{m}}{|\widehat{m}|}\,F\left(\frac{|\widehat{m}|}{16\widehat{a}^2}\right)$ 
&\hspace*{-0.2cm}$-\frac{\widehat{m}}{|\widehat{m}|}\overset{\ }{\sqrt{1-F^2\left(\frac{|\widehat{m}|}{16\widehat{a}^2}\right)}}$\hspace*{-0.1cm} & $0$ & $0$ \\
 $\omega\in\left]0,\frac{\pi}{2}\right[$ &&&& \\ 
\noalign{\vskip\doublerulesep\hrule height 2pt} 
$II_\pm^{(\omega=0)}$ & $\overset{\ }{\frac{\widehat{m}}{16\widehat{a}^2}}$ 
& $-\frac{j_0}{|j_0|}\overset{\ }{\sqrt{1-\left(\frac{\widehat{m}}{16\widehat{a}^2}\right)^2}}$ & $0$ & \hspace*{-0.1cm}$\overset{\ }{\frac{1}{2}\left[1-\left(\frac{\widehat{m}}{16\widehat{a}^2}\right)^2\right]}$\hspace*{-0.1cm} \\ 
\hline  $\overset{\ }{III_\pm^{(\omega=0)}}$ & $\sign\widehat{m}$ & $0$ & $0$ & $0$ \\
  \hline  Aoki phase & $\overset{\ }{\frac{\widehat{m}}{16\widehat{a}^2}}$ 
& $-\frac{j_0}{|j_0|}\overset{\ }{\sqrt{1-\left(\frac{\widehat{m}}{16\widehat{a}^2}\right)^2}}$ & $ 0$ &   \hspace*{-0.1cm}$\overset{\ }{\frac{1}{3}\left[1-\left(\frac{\widehat{m}}{16\widehat{a}^2}\right)^2\right]}$\hspace*{-0.1cm}
\\  
$\widehat{\mu}_{\rm I}=\omega=0$ &&&& \\
 \noalign{\vskip\doublerulesep\hrule height 2pt} 
 $II_\pm^{(\omega=\pi/2)}$ 
& \hspace*{-0.1cm}$\frac{m_{\rm v}}{|m_{\rm v}|}\overset{\ }{\sqrt{1-\left(\frac{\widehat{m}}{16\widehat{a}^2}\right)^2}}$\hspace*{-0.1cm} 
& $\frac{\widehat{m}}{16\widehat{a}^2}$ 
& $\hspace*{-0.1cm}\overset{\ }{\frac{1}{2}\left[1-\left(\frac{\widehat{m}}{16\widehat{a}^2}\right)^2\right]}$\hspace*{-0.1cm} & $0$ \\ 
\hline  $\overset{\ }{III_\pm^{(\omega=\pi/2)}}$ & $0$ & $-\sign\widehat{m}$ & $0$ & $0$ \\
  \hline  Aoki-like phase & $\frac{m_{\rm v}}{|m_{\rm v}|}\overset{\ }{\sqrt{1-\left(\frac{\widehat{m}}{16\widehat{a}^2}\right)^2}}$ 
& $\overset{\ }{\frac{\widehat{m}}{16\widehat{a}^2}}$ & $\hspace*{-0.1cm}\overset{\ }{\frac{1}{3}\left[1-\left(\frac{\widehat{m}}{16\widehat{a}^2}\right)^2\right]}$\hspace*{-0.1cm} &   0
\\  
\hspace*{-0.1cm}$\widehat{\mu}_{\rm I}=-32\widehat{a}^2;\ \omega=\frac{\pi}{2}$\hspace*{-0.1cm} &&&& 
\end{tabular}}
\caption{\label{t2} List of order parameters in the corresponding phases. The function $\displaystyle F(|\widehat{m}|/16\widehat{a}^2)$ used for the phases $II_\pm$ is defined in Eq.~\eref{sol-app}. The subscript ``$\pm$" refers to the sign of the quark mass $\widehat{m}$. The charged pion condensates and the isospin charge density are $C_{\pi^\pm}=-j_\pm/(2|j_\pm|)\sqrt{\Sigma^2-\Sigma^2(\widehat{m})-C_{\pi^0}^2}$ and $n_{\rm I}=2\mu_{\rm I} F_\pi^2C_{\pi^{\pm}}^2/\Sigma^2$, respectively, cf. Eq.~\eref{reduction}. They vanish when $\Sigma^2(\widehat{m})+C_{\pi^0}^2=\Sigma^2$. The quantities proportional to the sign of the source terms $j$ and $m_{\rm v}$ would vanish if we would set $j,m_{\rm v}\equiv0$. Only due to the limit $j,m_{\rm v}\to0$, after the thermodynamical limit, those condensates acquire a non-vanishing value and reflect spontaneous breaking of the corresponding symmetries. This behavior is also reflected in non-vanishing variances. The one for the charged pion condensate is always $\Delta C_{\pi^\pm}=n_{\rm I}/(2\mu_{\rm I} F_\pi^2)$, see Eq.~\eref{pi-pm-n-I}. The prefactor $1/2$ in the variance of the phases $II_\pm^{(\omega=-,\pi/2)}$ results from averaging over the two minima while the factor $1/3$ in the Aoki and Aoki-like phase results from averaging over the angle $\vartheta_1$ (note the non-trivial measure~\eref{measure} and the symmetry~\eref{symmetry-U}).
}
\end{table}

\FloatBarrier

\subsubsection{Zero Twist and Zero Lattice Spacing.}

\begin{figure}[t!]
\centerline{\includegraphics[width=0.49\textwidth]{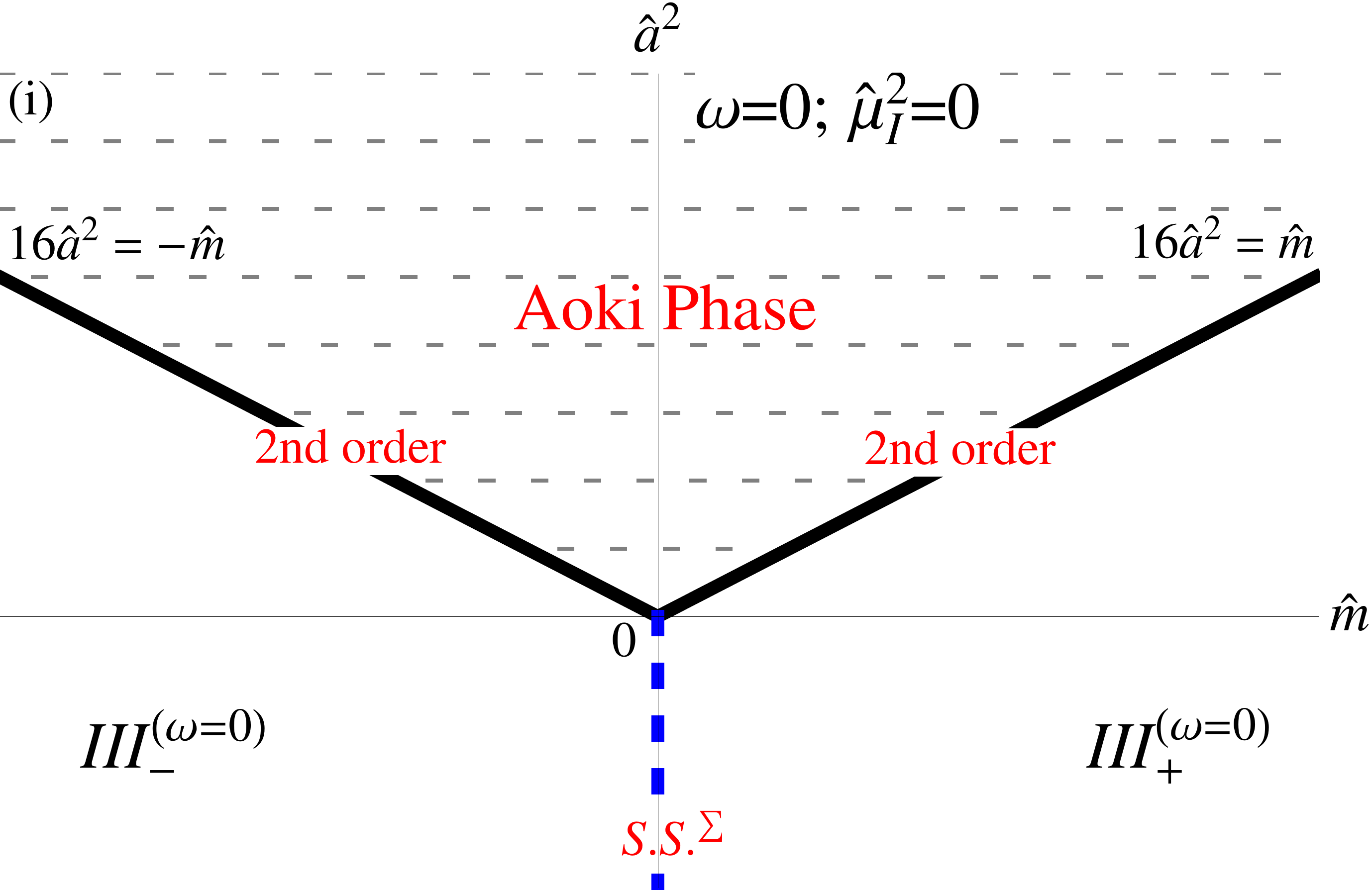}}
 \centerline{\includegraphics[width=0.49\textwidth]{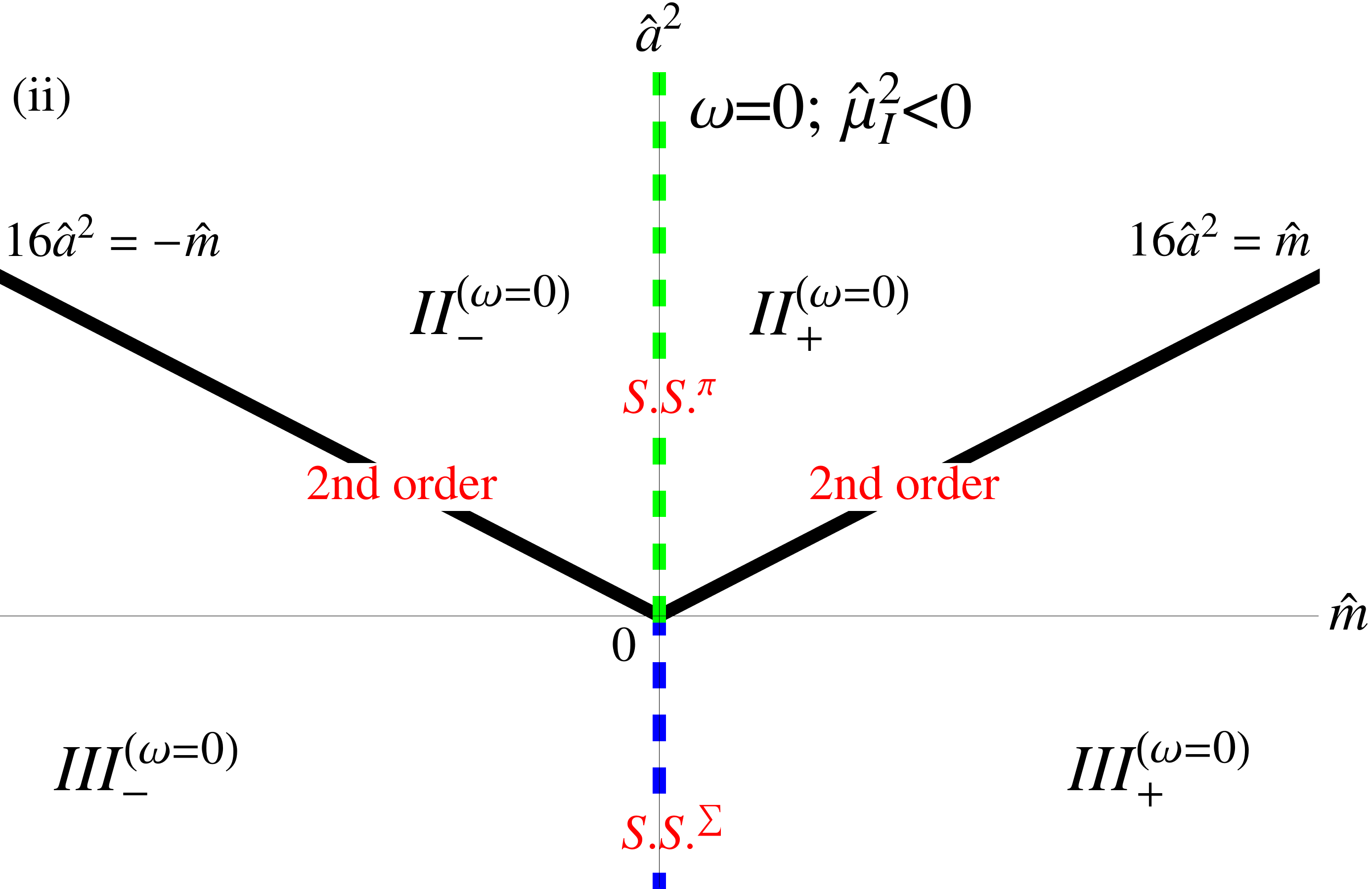}\hfill
\includegraphics[width=0.49\textwidth]{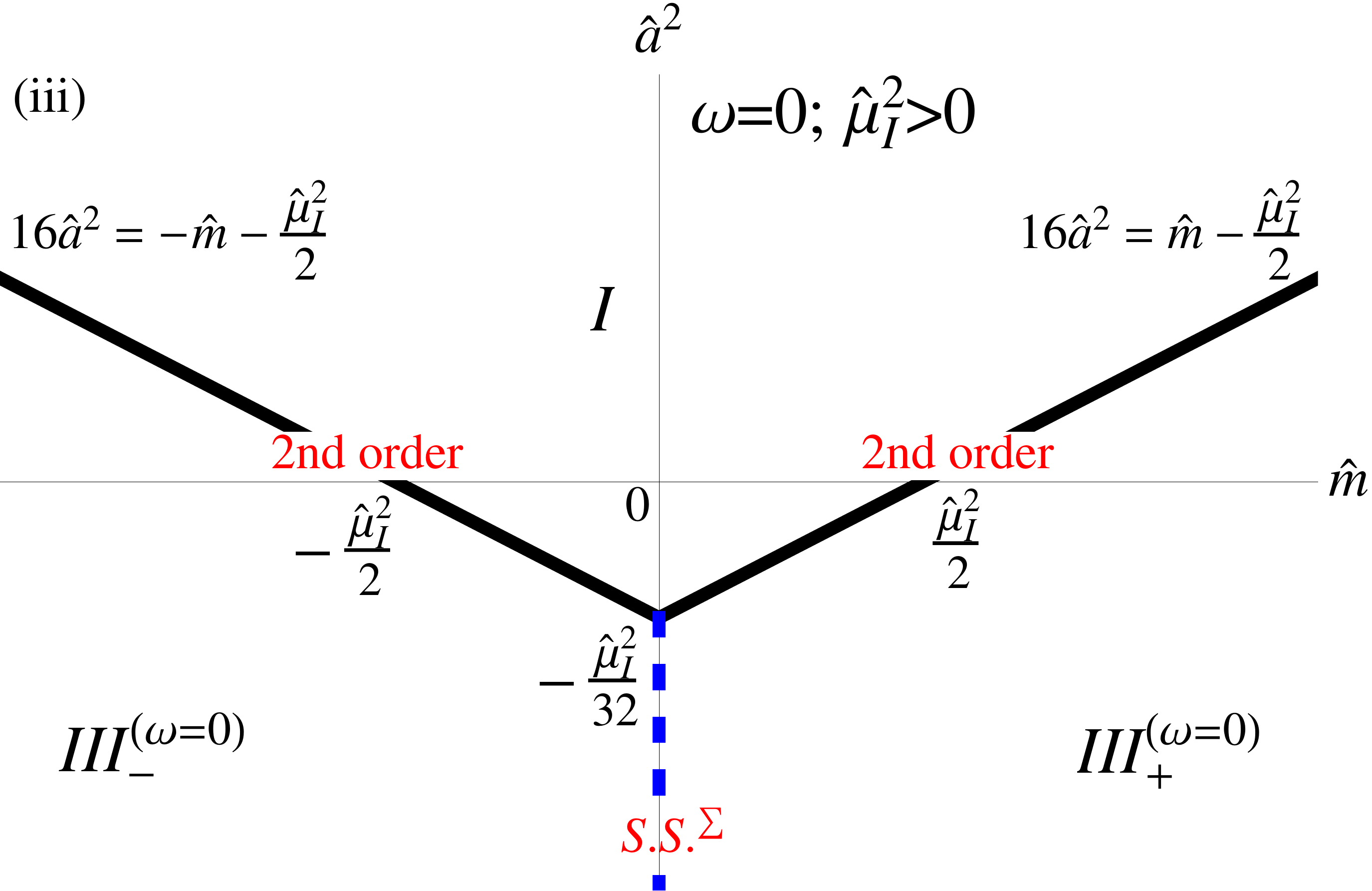}}
\centerline{\includegraphics[width=0.49\textwidth]{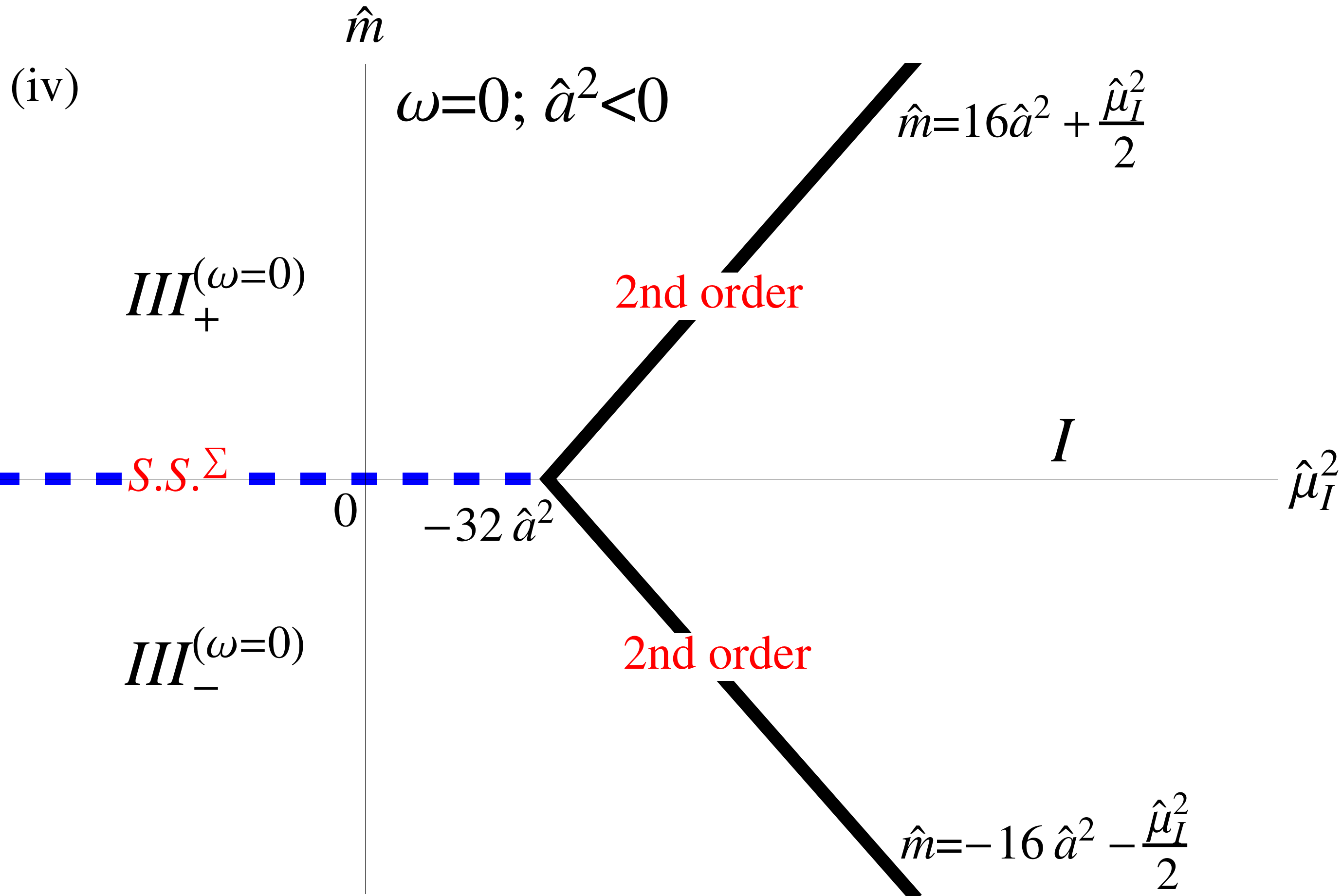}\hfill\includegraphics[width=0.49\textwidth]{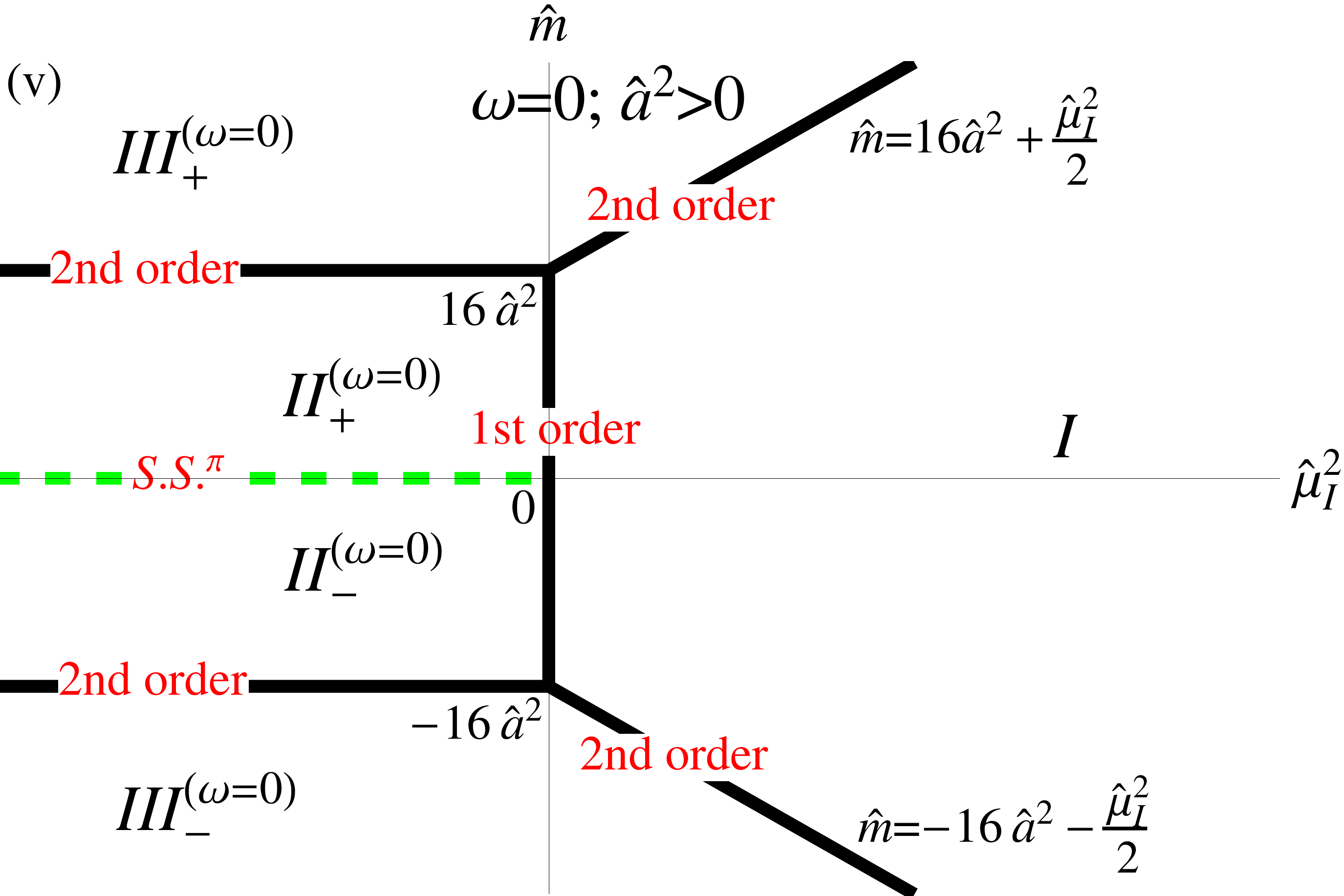}}
\caption{Phase diagram at vanishing twist ($\omega=0$) in the $\hm$-$\ha^2$-plane (Figs. (i), (ii) and (iii)) and in the $\widehat{\mu}_{\rm I}^2$-$\widehat{m}$-plane (Figs. (iv) and (v)).
 The generic cases are: 
 (i) zero isospin chemical potential ($\hmu^2=0$), 
 (ii) imaginary  isospin chemical potential ($\hmu^2<0$),  
 (iii) real  isospin chemical potential ($\hmu^2>0$),
 (iv) imaginary effective lattice spacing ($\widehat{a}^2<0$), and
 (v) real effective lattice spacing ($\widehat{a}^2>0$).
 The Aoki phase only exists at vanishing  isospin chemical potential.
 Recall that $\wha^2$ is a combination of two low energy constants. Therefore we can have $\wha^2<0$ although the physical lattice spacing $a$ is positive definite. The abbreviations ${\rm S.S.}^\Sigma$ and ${\rm S.S.}^\pi$ refer to the original Sharpe-Singleton scenario where the chiral condensate jumps when the quark mass crosses the origin and a Sharpe-Singleton-like scenario where the $\pi^0$ condensate jumps. The subscripts $\pm$ corresponds to the sign of the  quark mass. The $\pi^0$ condensate is only non-zero in the Aoki phase  if a twisted mass ($j_0$ or $\omega\searrow0$)
is introduced as a source term. This condensate spontaneously breaks parity and flavor symmetry.}
\label{fig1a}
\end{figure}

 In this case we want to recall the phase diagram in the low-temperature limit
as a function of the
isospin chemical potential and the quark mass. 
 The physics of this
problem is well understood, see \cite{\KSTVZ,\SoSt}. The phase diagram which is shown in Fig.~\ref{fig1a} for $\widehat{a}^2\to0$. At $\mu_{\rm I}  = m_\pi$ there is 
a phase transition to a Bose condensed phase of charged pions with
 one exactly massless Goldstone boson. For $\mu_{\rm I}<m_\pi$ the masses of the
pseudo-Goldstone modes are given by
\be
m_\pi = m_\pi(\mu_{\rm I}=0) +  q\mu_{\rm I}=\sqrt{\frac{2\Sigma m}{F_\pi^2}}+  q\mu_{\rm I},
 \ee
where $q$ is the isospin charge of the pions.

In Euclidean space-time the interpretation of the chemical potential is that
the retarded propagation 
is enhanced by $\exp(q\mu_{\rm I} x_0 )$ for particles with positive charge $q$ and
suppressed by a factor  $\exp(-q\mu_{\rm I}x_0 )$ for the anti-particles.
Asymptotically, the propagator in the imaginary time direction (temperature) behaves as
\be
e^{-(m_\pi-q\mu_{\rm I})\tau}.
\ee
This behavior is determined by the pole mass which can also be obtained from
the analytic continuation of the propagator to Minkowski space time.                        
At non-zero temperature this results in a net non-zero particle density while
at zero temperature, uninhibited propagation of the lightest pion 
only occurs for 
$\mu_{\rm I} > m_\pi$.

At imaginary chemical potential in Euclidean space time, retarded propagation
is modified by a factor $\exp(i q\mu_{\rm I} x_0)$ and a
factor $\exp(-iq \mu_{\rm I} x_0)$ for advanced propagation. This corresponds to
an imaginary pole mass. At imaginary chemical potential the partition function becomes
Roberge-Weiss periodic \cite{RW} with period $\mu_{\rm I}=2\pi  i T/3$ where $T$ is the temperature.
In the zero temperature mean field limit ($T=0$) we thus encounter this periodicity
immediately. In the microscopic domain, though, we have $\mu_{\rm I} \sim 1/\sqrt V$
which is well within the first Roberge-Weiss period \cite{svet-split}.
We will study the phase diagram in terms of the microscopic scaling variables, which
have to be taken much larger than unity so that the saddle point approximation
can be justified. In this way we can consider an imaginary chemical potential
while not running immediately in the Roberge-Weiss periodicity in the low
temperature limit.

\subsubsection{Zero Twist and Zero Isospin Chemical Potential.}
 At {\it vanishing chemical potential} the particular structure of the phase 
diagram was first predicted by Aoki \cite{Aokiclassic} and was  studied in detail
by Sharpe and Singleton \cite{SharpeSingleton} in the $p$-regime of chiral 
perturbation theory.

For {\it imaginary effective lattice spacing} $\widehat{a}$ the saddle point
in $\varphi$ is at $\varphi =0$ or at $\varphi = \pi$ depending
on the sign of the quark mass. The chiral condensate is independent of
the mass but flips sign when the quark mass traverses zero.
The two phases are physically identical but since the effective potential
is nonvanishing for $m=0$, they are not continuously connected
as a function of the quark mass.
This  is known as the 
Sharpe-Singleton scenario \cite{SharpeSingleton}. We denote these two phases by 
``$III_\pm^{(\omega=0)}$" where the subscript reflects the sign of the quark
 mass ($\sign\widehat{m}=\pm1$).  

At {\it real effective  lattice spacing } 
($\widehat{a}^2>0$) {\it and vanishing chemical potential}
the chiral condensate does not jump anymore. 
When decreasing the quark mass one  enters the  Aoki phase~\cite{Aokiclassic} through 
a second order 
phase transition. In this phase the saddle point in $\varphi$ is located at
\be
\cos \varphi = \frac \whm {16 \wha ^2}
\ee
while the angles $\vartheta_1$ and $\vartheta_2$ are not determined by the saddle point equations.
The chiral condensate decreases linearly as a function of $\whm$ until one reaches another 
second order phase transition for negative quark mass at the point where $\cos \varphi = -1$. 
The $\pi^0$ condensate can be calculated by switching on a small twisted
mass as source term and setting it to zero after differentiation and
taking the thermodynamic limit. Its value is
 non-zero in the Aoki phase. Thus parity and flavor symmetry are 
spontaneously broken. 
For $\widehat{\mu}_{\rm I}=\omega= 0$ the partition function is isospin
symmetric. A source term corresponding to a charged pion condensate $j_\pm\to0$ yields a hysteresis and polarizes this condensate due to spontaneous symmetry breaking.
The phase diagram at $\widehat{\mu}_{\rm I}=\omega=0$ is shown in Fig.~\ref{fig1a}.i).

\subsubsection{Non-zero Lattice Spacing and non-zero Chemical Potential.}

For {\it imaginary isospin chemical potential} and vanishing twist the 
phase diagram looks almost the same as the one at $\widehat{\mu}_{\rm I}=0$, cf. Figs.~\ref{fig1a}.i) and ii). 
The reason for this behavior is that $\sin \vartheta_1 = 0$ minimizes the saddle point action for $|\widehat{m}|<16\widehat{a}^2$. This fixes the charged pion condensates and their corresponding covariances at zero while
the $\pi^0$ condensate remains non-zero even in the absence of a source term.
The flavor symmetry is broken explicitly by the chemical potential
so that the would-be Goldstone bosons are massive in this phase 
which is  labeled by $II_\pm^{(\omega=0)}$. The subscript ``$\pm$" refers  to the sign of the quark mass since the non-vanishing $\pi^0$-condensate switches sign when $\widehat{m}$ crosses the origin. Because this behavior is reminiscent to the Sharpe Singleton scenario (denoted by ${\rm S.S.}^\Sigma$) where the mass dependent chiral condensate jumps  we refer to it as a Sharpe-Singleton-like scenario and abbreviate it  by ${\rm S.S.}^\pi$. Note that both parts, $II_+^{(\omega=0)}$ and $II_-^{(\omega=0)}$, refer to the same phase although they are connected by a discontinuity along the $\widehat{m}=0$ axis.
 This part of the phase diagram is shown in Figs.~\ref{fig1a}.ii), iv) and v). The details are summarized in Tables~\ref{t1} and \ref{t2}.
 
 At $\widehat{\mu}_{\rm I}= 0$ we are in the Aoki phase with two exactly massless Goldstone bosons if $|\widehat{m}|<16\widehat{a}^2$. At a {\it non-zero real isospin chemical potential} in the related regime $|\widehat{m}|<16\widehat{a}^2+\widehat{\mu}_{\rm I}^2/2$ we immediately enter a Bose condensed phase with an isospin density
that initially increases linearly with $\widehat{\mu}_{\rm I}$. The reason is that an exactly massless Goldstone boson with
non-vanishing isospin charge exists in this phase.
This phase is denoted by $I$. Note that for non-zero imaginary chemical potential, no Bose condensation
takes place (phase $II_\pm$), but the pions become massive because of the symmetry breaking by the chemical potential. The free
energy in the phase $I$ is quadratic in $\whmu$ for small chemical potential so that we have a second order phase transition for 
$|\whm| < 16 \wha^2$ at $\whmu ^2=2|\widehat{m}|-32\widehat{a}^2$.

In the regime $|\widehat{m}|>16\widehat{a}^2+\widehat{\mu}_{\rm I}^2/2$ we find the normal phase $III_\pm^{(\omega=0)}$ where the mass dependent chiral condensate $\Sigma(\widehat{m})$
is a constant  ($\cos \varphi =1 $) and all the other order parameters vanish. 
In contrast, for $|\whm| < 16 \wha^2 + \whmu^2/2$ (phase $I$)   we  have
\be
|\cos \varphi | < 1 \quad {\rm and} \quad \cos\vartheta_1 =0,
 \ee
so that the $\pi^0$ condensate always vanishes and parity is not
spontaneously broken. The flavor symmetry is broken explicitly by the
chemical potential though.
In the phase $I$ the chiral condensate drops linearly 
 in the quark mass  when approaching $\widehat{m} \to0$ as in the Aoki-phase. 
The phase boundaries are shifted down by the value $-\widehat{\mu}_{\rm I}^2/32$, 
in contrast to the case $\widehat{\mu}_{\rm I}^2\le 0$
cf. Figs.~\ref{fig1a}.ii) and iii). 
In the presence
of a charged pion source the corresponding condensate is nonvanishing in the phase $I$.
 This part of the phase diagram is shown in Figs.~\ref{fig1a}.iii), iv) and v).

\begin{figure}[t!]
\centerline{\includegraphics[width=0.49\textwidth]{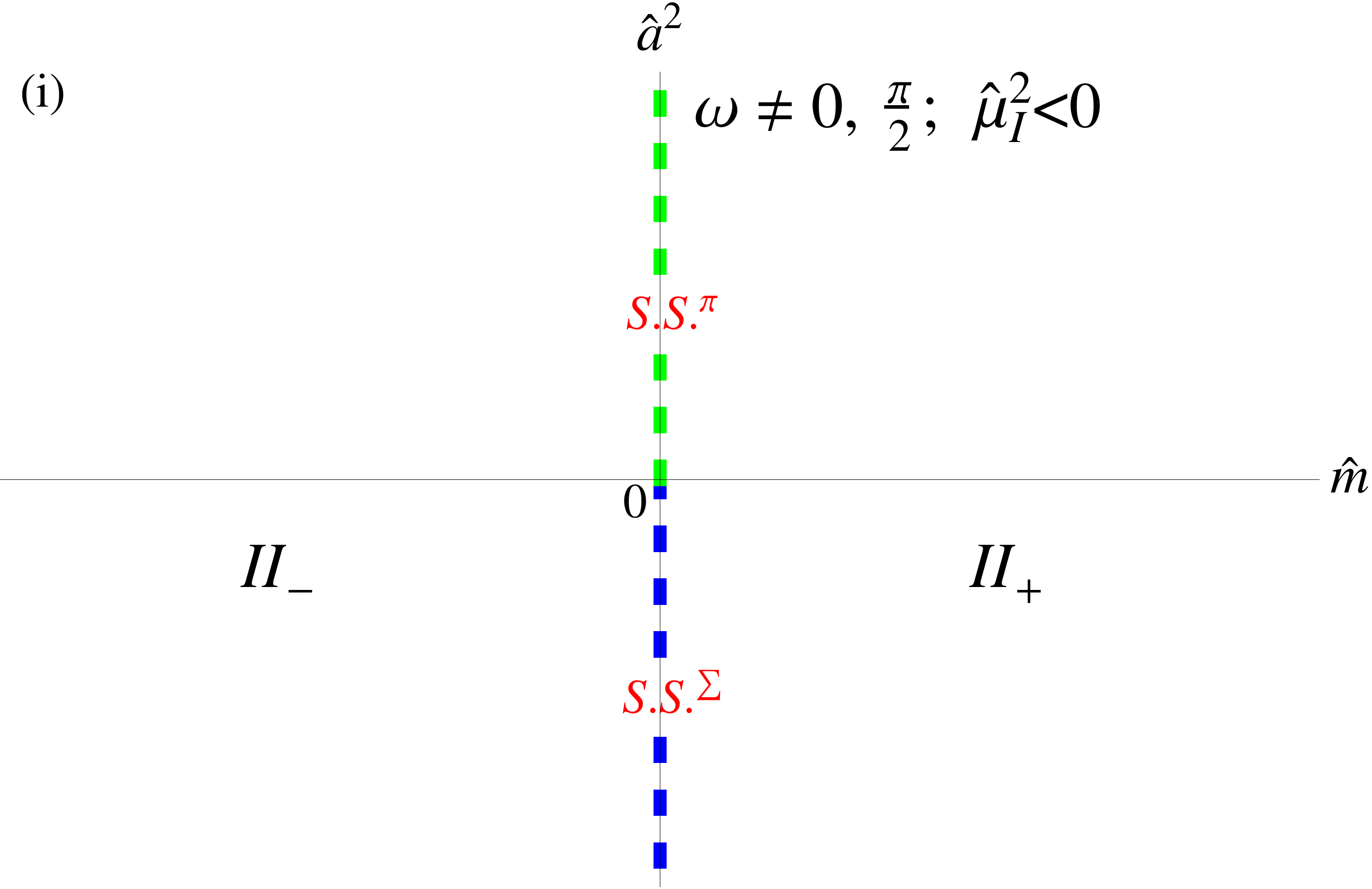}\hfill \includegraphics[width=0.49\textwidth]{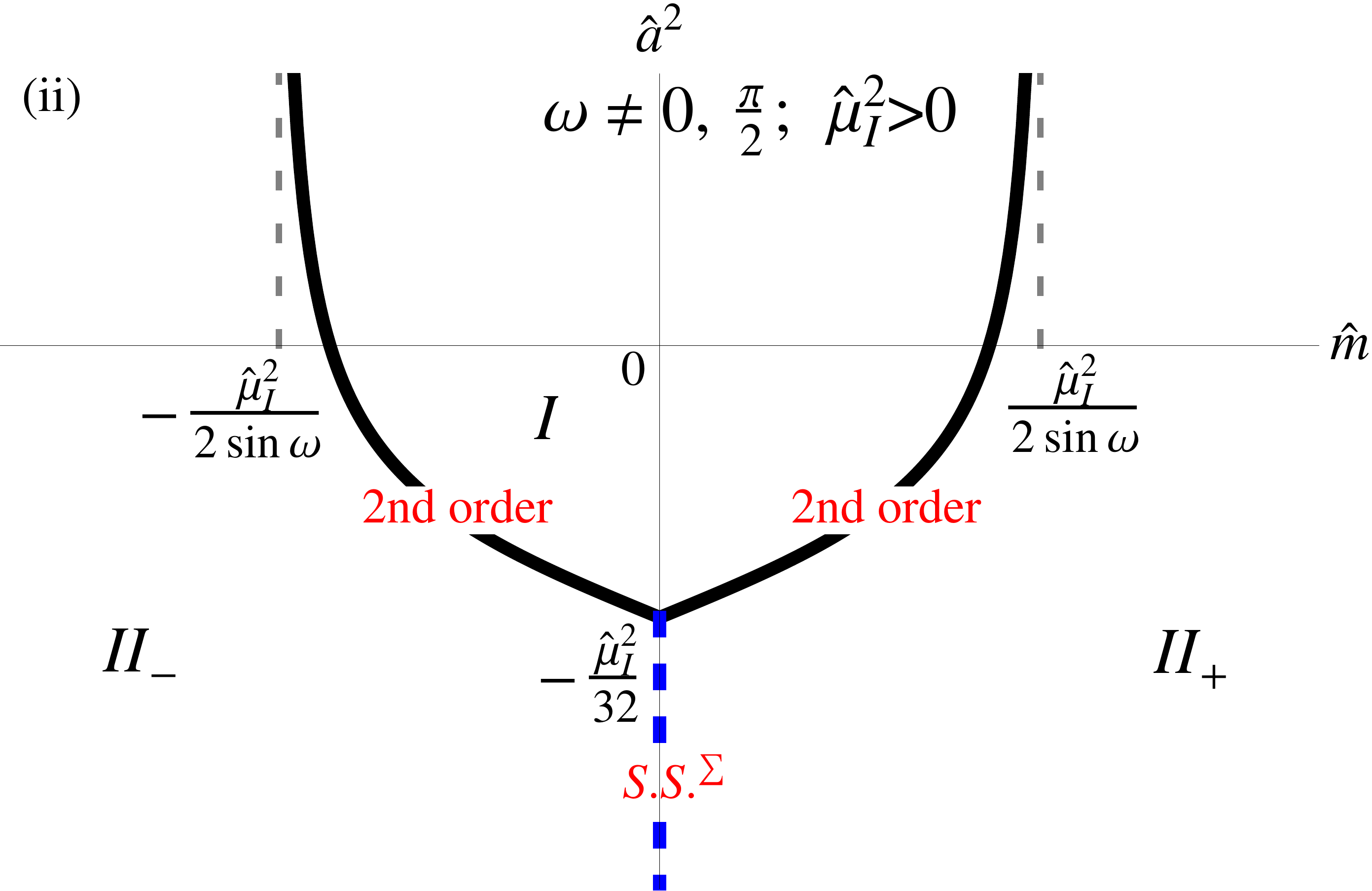}}
\centerline{\includegraphics[width=0.49\textwidth]{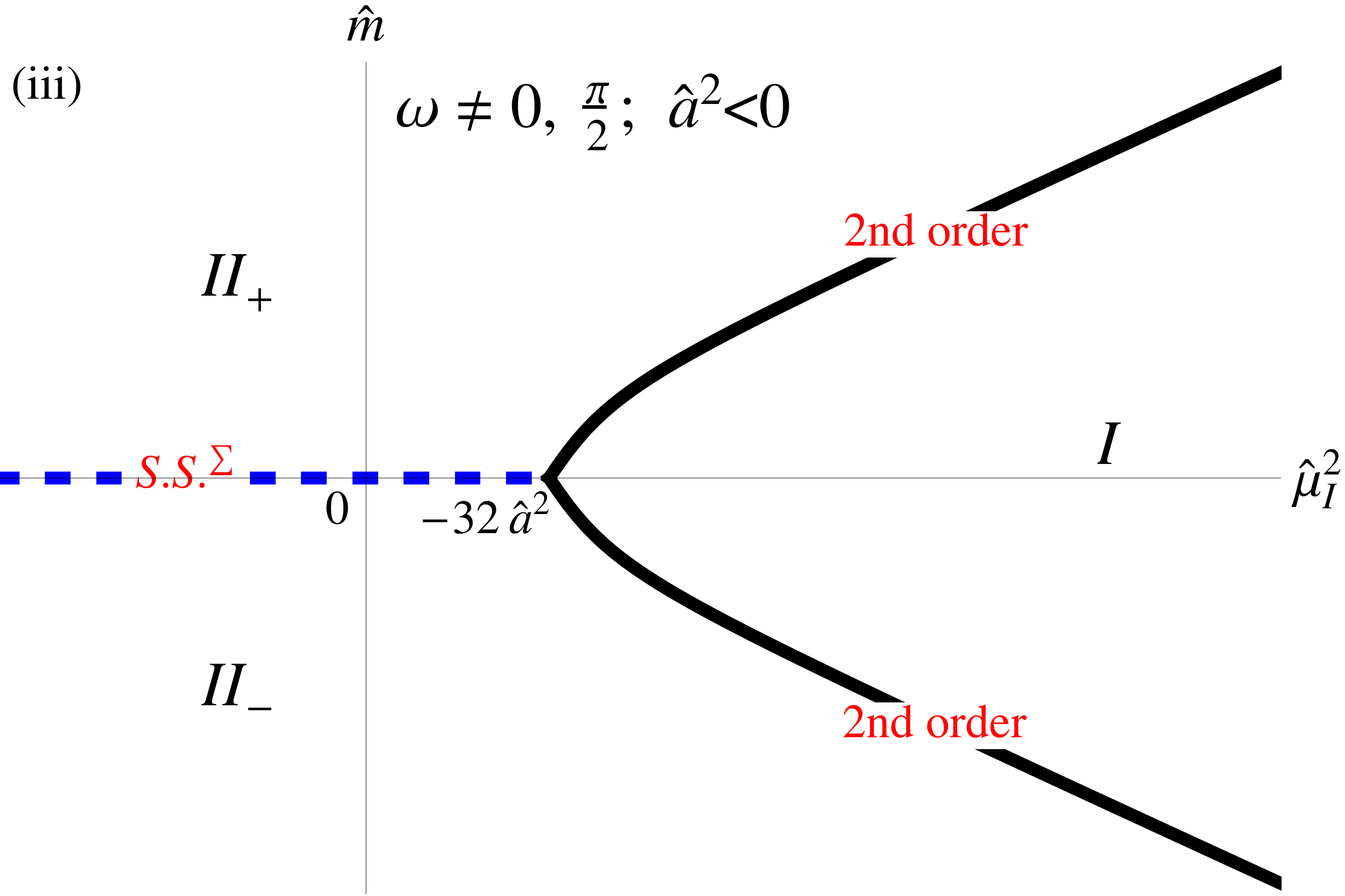}\hfill \includegraphics[width=0.49\textwidth]{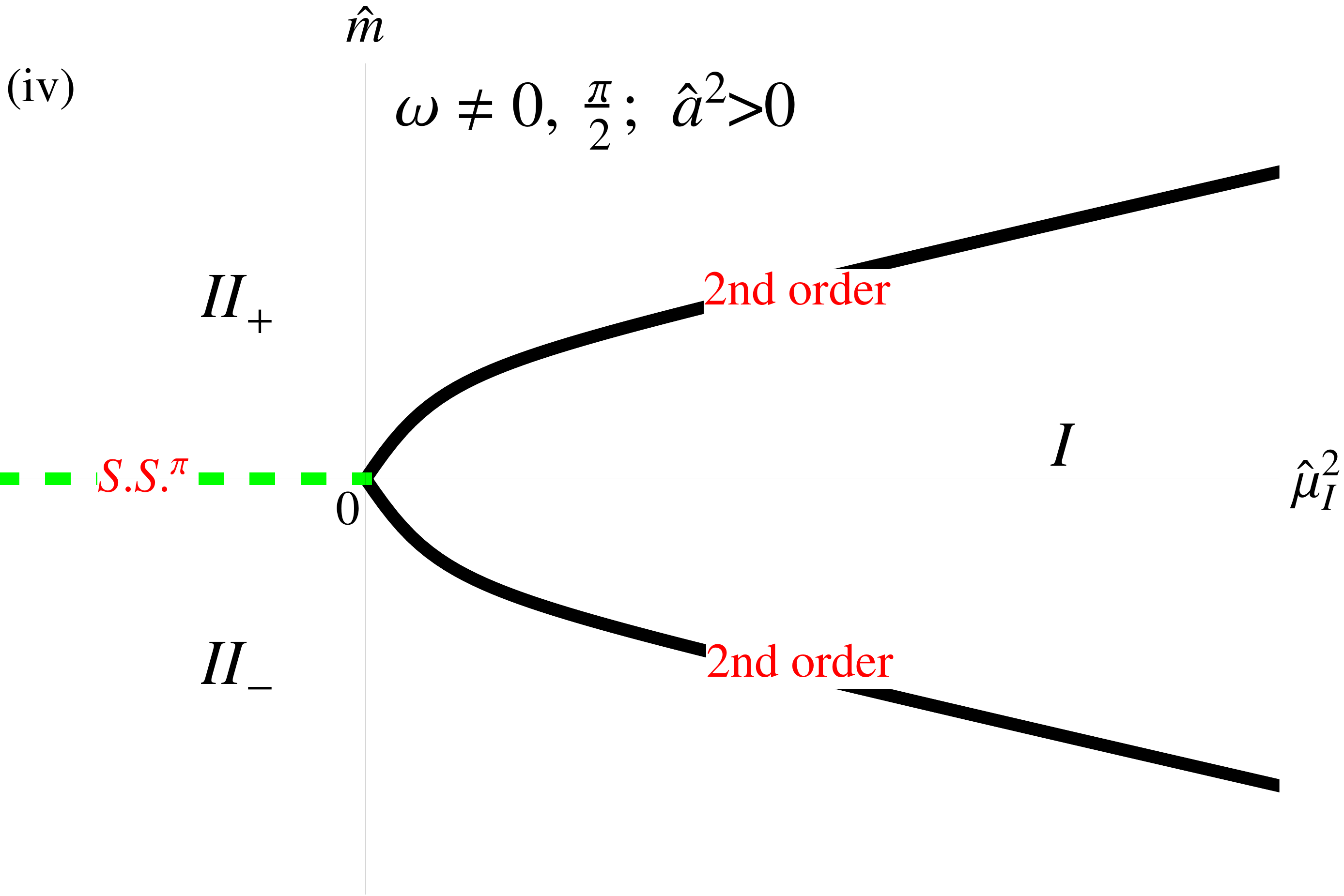}}
\caption{
Phase diagram in the $\hm-\ha^2$-plane  (Figs. (i) and (ii)) and in the $\hmu^2$-$\hm$-plane (Figs. (iii) and (iv)) for $0<\omega<\pi/2$. In particular we choose
(i) $\hmu^2<0$, (ii) $\hmu^2> 0$, (iii) $\ha^2<0$ and (iv) $\ha^2\geq 0$. The phase $I$ only exists for real isospin chemical potential.  
For $\hmu^2<0$ and $\wha^2 < 0$, 
the chiral condensate jumps when the mass 
crosses the origin and while the $\pi^0$-condensate jumps for $\wha^2 > 0$. This case is related to the Sharpe-Singleton scenario. The curves of the second order phase  
transitions at $\hmu^2>0$ in figures (ii)  extend to infinity  at $\whm=\pm\whmu^2/2\sin\omega$. 
There is no phase transition beyond these masses in the leading order expansion of the chiral Lagrangian. A finite, imaginary effective lattice spacing  $\ha^2<0$ shifts the bifurcation of the phase transition curves away from the origin (see Fig. (iii)).
}
\label{fig2a}
\end{figure}

\subsection{Phase Diagram at finite Twist}

At {\it imaginary isospin chemical potential} the saddle point equations dictate 
$\cos \vartheta_1 = -\sign \whm$ and $\cos\varphi=\sign\widehat{m}\cos\widetilde{\varphi}$ ($\widetilde{\varphi}\in[0,\pi/2]$) so that the   
phases are determined by the minimum of
\be
V\mathcal{L}_0|_{\cos \vartheta_1 = -\sign \whm}=-2|\whm|\cos(\widetilde{\varphi}-\omega)+16\wha^2\cos^2\widetilde{\varphi}.
\ee
In  Fig.~\ref{fig2c} we plot the 
solution of this equation for various values of $\omega\in[0,\pi/2]$ as a function of $|\whm|/16 \wha^2$.

The Sharpe-Singleton(-like) scenario extends to all
$\wha^2$ as well as masses since the twisted mass wipes out the second order phase transition
to the normal phase. In fact the Aoki phase as well as $II_\pm^{(\omega=0)}$ is absent for $0<\omega <\pi/2$ as can
be seen from the $\omega \to 0$ limit of the curves in Fig.~\ref{fig2c}. 
For $\wha^2 > 0$ the $\pi^0$ condensate
changes from $\Sigma$ to $-\Sigma$ when the quark mass traverses zero while  the mass dependent chiral condensate $\Sigma(\widehat{m})$ jumps  for $\wha^2 < 0$. The two minima are 
separated by a potential barrier  so that we have a first order phase 
transition when the mass crosses zero.
The corresponding two parts of the phase are  denoted by $II_\pm$
 (again $\pm$ refers to $\sign\widehat{m}=\pm1$).
In this phase the chiral condensate is given by $\cos \varphi $ and the $\pi^0$
condensate by $\sin \varphi$ so that
 the combination  $\Sigma^2(\widehat{m})+C_{\pi^0}^2=\Sigma^2$ is constant, 
cf. Fig.~\ref{fig2a}.i) and Table~\ref{t2}. 
The   chiral condensate and the $\pi^0$ 
condensate only depend on the angle $\omega$ and on the 
ratio $\widehat{m}/\widehat{a}^2$ such that we have some kind of  a ``Silver-blaze property" in the isospin chemical potential. 

The situation changes when {\it the isospin chemical potential is real}. 
For $2|\whm| \sin \omega < \whmu^2$ 
the effective potential also has a minimum at 
\be
\cos  \vartheta_1 = -\frac {\hm\sin\omega}{4\hmu^2 \sin\varphi },
\ee
which yields the  phase $I$. It shows up above the value
 $\widehat{a}^2>-\widehat{\mu}_{\rm I}^2/32$ and between the masses $-\widehat{\mu}_{\rm I}^2/2\sin\omega<\widehat{m}<\widehat{\mu}_{\rm I}^2/2\sin\omega$. The region where $\widehat{a}^2<-\widehat{\mu}_{\rm I}^2/32$ is reminiscent of the Sharpe-Singleton scenario since the chiral condensate is jumping at $\widehat{m}=0$ while the $\pi^0$ condensate does not because $\sin\varphi$ vanishes for $\widehat{a}^2<0$.
 The two second order phase transition lines asymptote to 
$\widehat{m}=\pm\widehat{\mu}_{\rm I}^2/2\sin\omega$. 
Hence we do not have the situation as for $\hmu=\omega=0$ where
 one can increase the effective lattice spacing, and independent of the 
value of the quark mass,
one  always enters
the Aoki phase.
For sufficiently large quark masses the system will always stay in the
 phase $II_\pm$ and will never enter the phase $I$ if $\omega>0$, cf. Fig.~\ref{fig2a}.ii). 
However when increasing the  isospin chemical potential $\widehat{\mu}_{\rm I}$ the system will eventually enter the phase $I$, see Fig.~\ref{fig2a}.iv). We underline that the described behavior can be caused by the fact that we only took the leading order of the chiral Lagrangian into account.

\begin{figure}[t!]
\centerline{\includegraphics[width=0.49\textwidth]{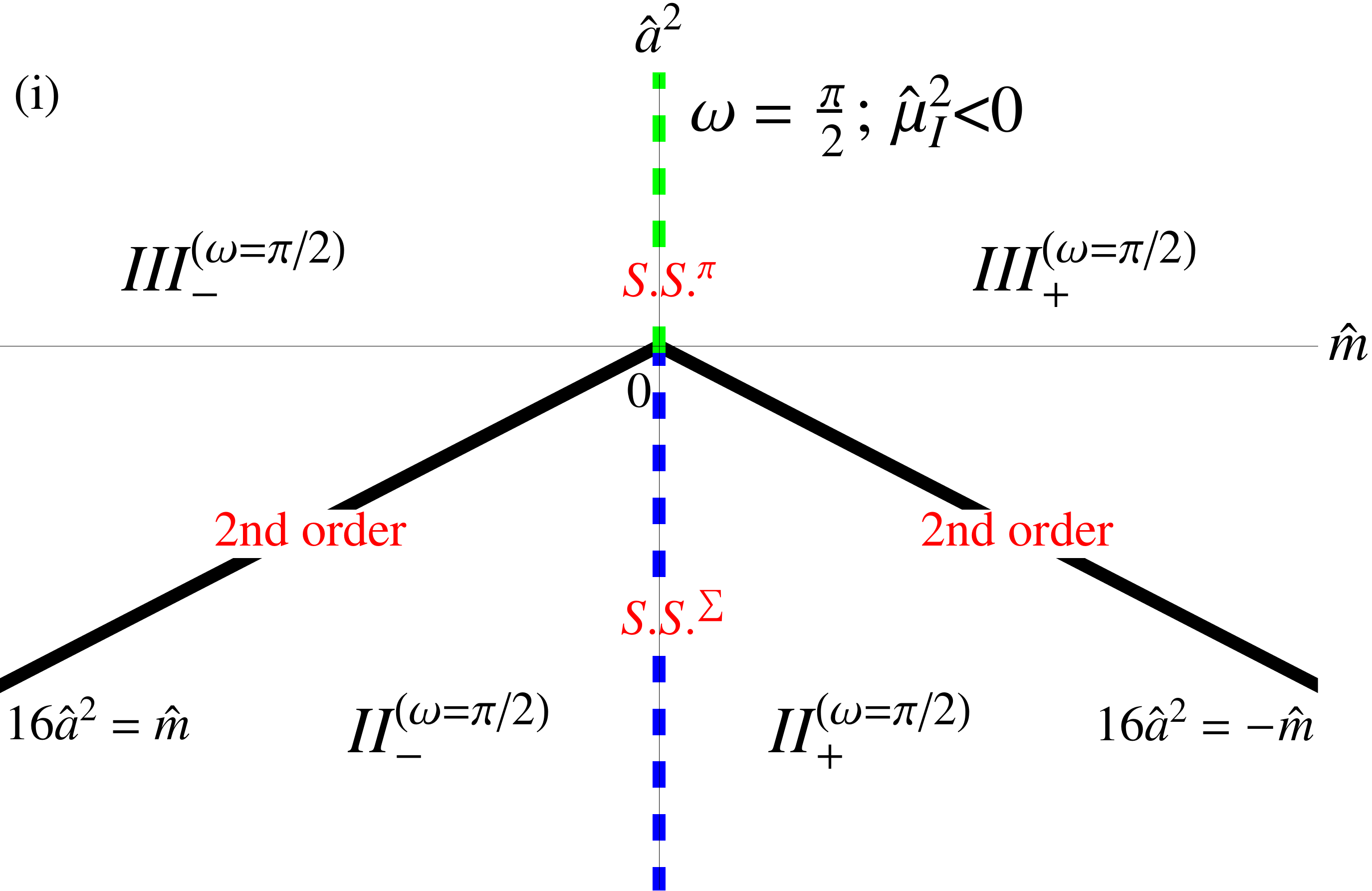}\hfill \includegraphics[width=0.49\textwidth]{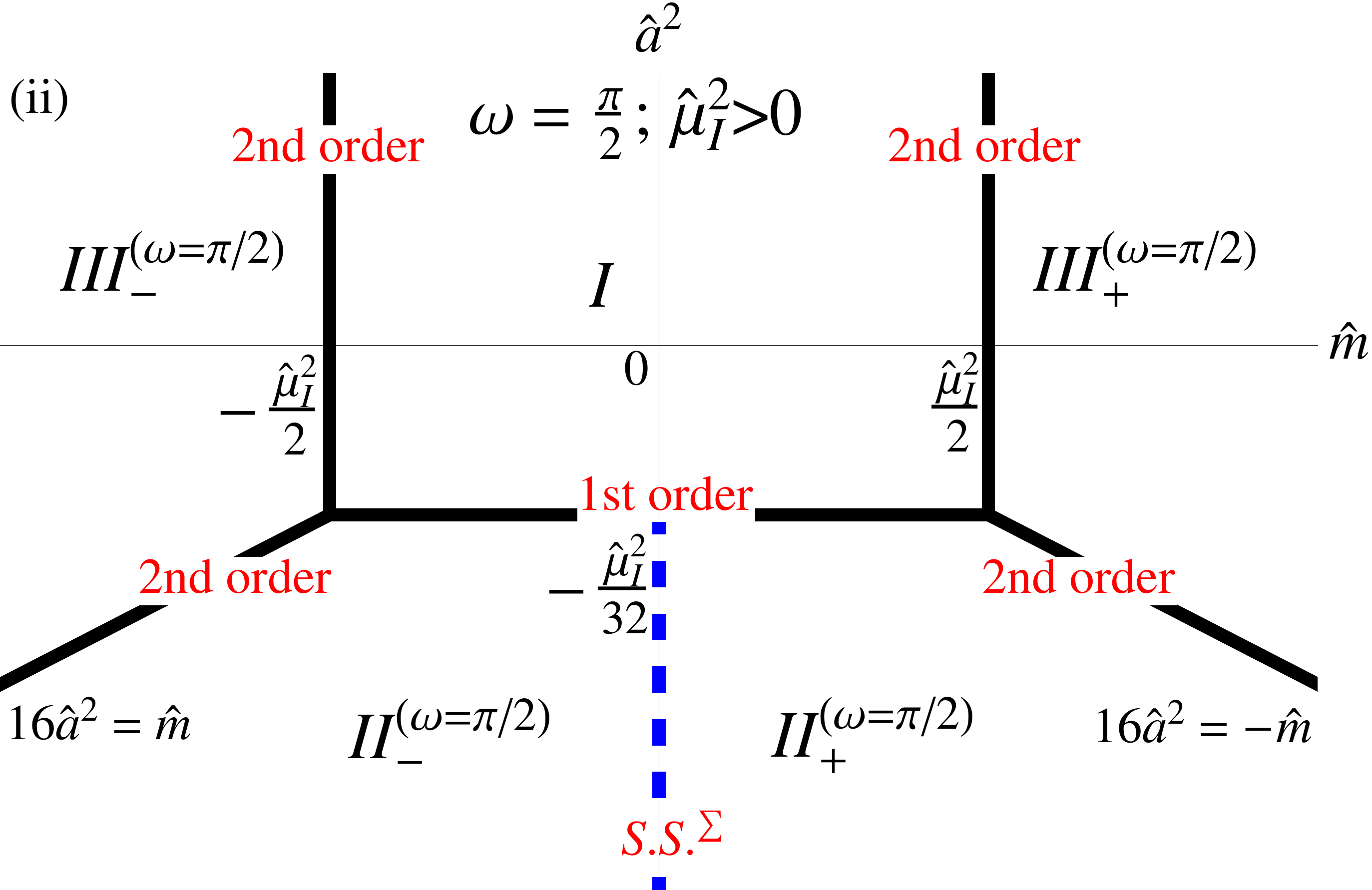}}
\centerline{\includegraphics[width=0.49\textwidth]{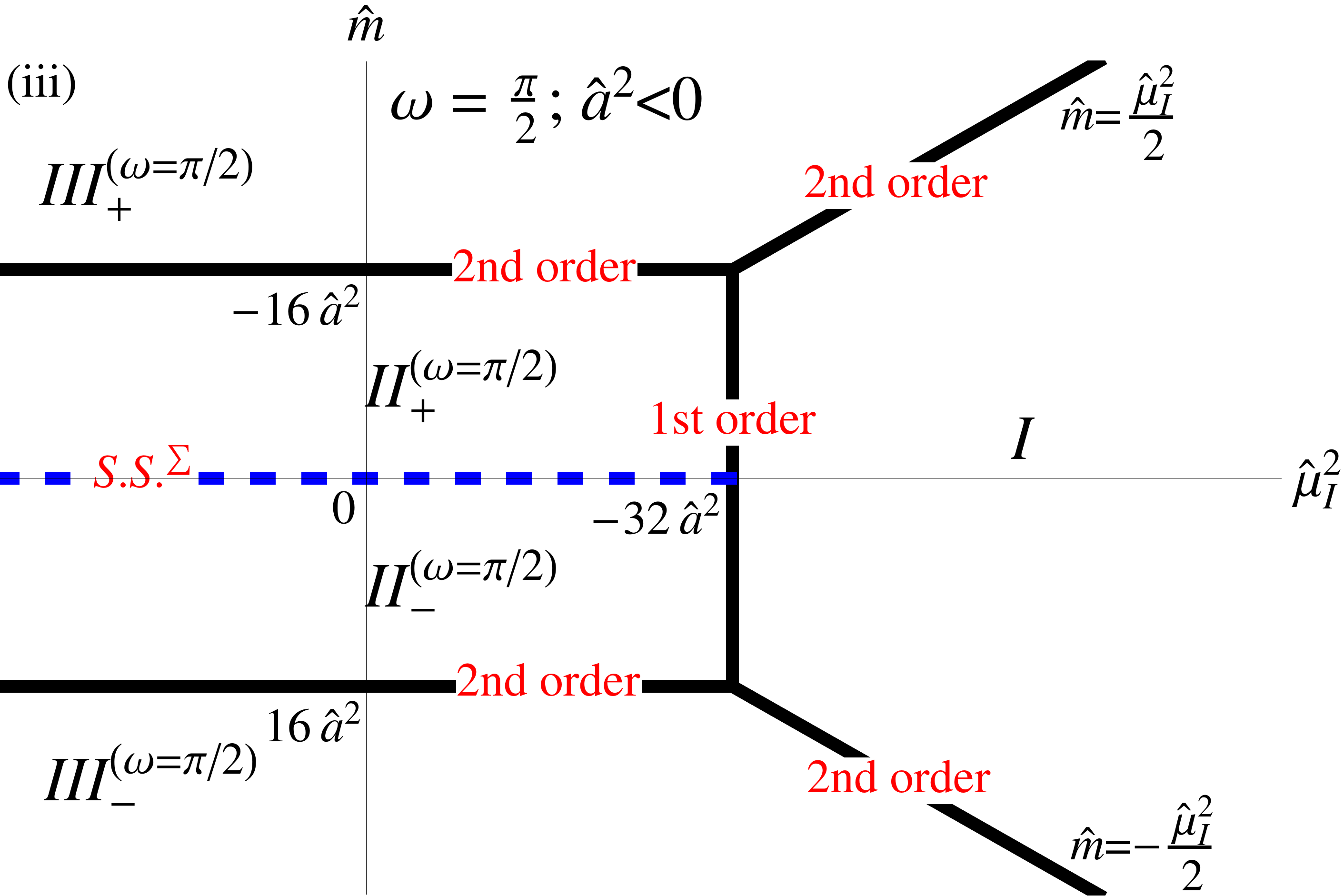}\hfill 
\includegraphics[width=0.49\textwidth]{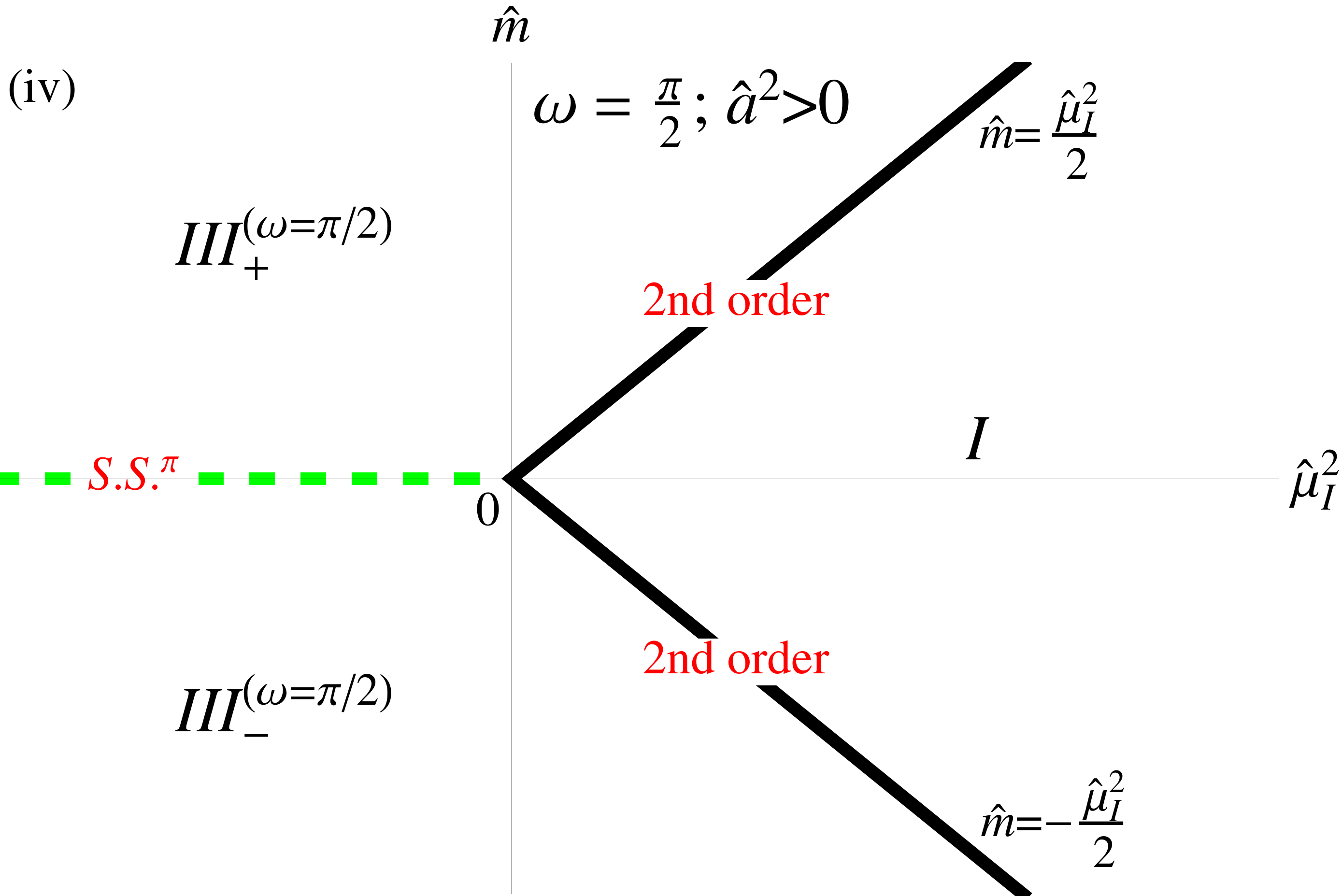}}
\caption{Phase diagram at maximal twist ($\omega=\pi/2$) in the $\hm-\ha^2$-plane (Figs. (i) and (ii)) and in the 
$\hmu^2-\hm$-plane (Figs. (iii) and (iv)). 
We draw the phase diagram at imaginary isospin chemical potential 
((i) $\hmu^2<0$), at real isospin chemical potential 
((ii) $\hmu^2>0$), at imaginary effective lattice spacing ((iii) $\widehat{a}^2<0$), and at real effective lattice spacing ((iv) $\widehat{a}^2>0$). 
Phase $I$ only appears if $\hmu^2>0$. 
The parts $II_+^{(\omega=\pi/2)}$ and $II_-^{(\omega=\pi/2)}$ of the phase $II_{\pm}^{(\omega=\pi/2)}$  are separated by a Sharpe-Singleton transition (chiral condensate jumps when $\widehat{m}$ crosses the origin) and extends to the phase $III_{\pm}^{(\omega=\pi/2)}$ where  the $\pi^0$ condensate jumps.
The case of real effective lattice spacing 
((iii) $a^2>0$) and maximal twist is the analogue of the case at imaginary 
effective lattice spacing and no twist, cf. Fig.~\ref{fig1a}.(iv) and vice versa. 
Comparing the phase diagrams at no and maximal twist  we observe the symmetry 
$(\omega,\whm,\whmu^2,\wha^2)\to(\pi/2-\omega,\whm,\whmu^2+32\wha^2,-\wha^2)$ 
of the static Lagrangian, cf. the symmetry transformation~\eref{symmetry-U}.
}
\label{fig3a}
\end{figure}

\subsection{Maximal Twist}

At {\it real isospin chemical potential} the cup-like structure of the phase transition 
lines becomes more and more rectangular shaped when increasing the twist angle.
 Eventually the smooth curves describing the phase transition lines develop a kink 
in the limit of maximal twist ($\omega=\pi/2$), see Fig.~\ref{fig3a}.ii) and iii). 
At the  corners a new second order phase transition   to the 
 ``normal" phase  $III_\pm^{(\omega=\pi/2)}$ appears, which corresponds to the solution
$\varphi=\pi/2$ that develops in the limit $\omega \to \pi/2$ for
$ \wha^2 < 0$.

In the regime $|\widehat{m}|<16\widehat{a}^2$ we find another phase $II_\pm^{(\omega=\pi/2)}$, see Figs.~\ref{fig3a}.ii) and iii).
 In both kinds of phases we have a non-vanishing $\pi^0$ condensate. However only in
 $III_\pm^{(\omega=\pi/2)}$ this condensate becomes independent of the quark mass and is 
 some kind of a ``Silver-Blaze property" in all parameters $|\widehat{\mu}_{\rm I}|^2$, $|\widehat{m}|$, and $|\widehat{a}|^2$ while in $II_\pm^{(\omega=\pi/2)}$ 
we have a linear dependence on $\widehat{m}/16\widehat{a}^2$. Furthermore, the mass dependent chiral condensate $\Sigma(\widehat{m})$ is also non-zero in the phase $II_\pm^{(\omega=\pi/2)}$ and becomes $\pm\Sigma$ at very small quark masses exhibiting the original Sharpe-Singleton scenario~\cite{SharpeSingleton}.
 This is not the case in the phase $III_\pm^{(\omega=\pi/2)}$. If the effective lattice spacing $\widehat{a}^2/V$ satisfies
 $\widehat{a}^2>-\widehat{\mu}_{\rm I}^2/4$, one does not cross the phase $II_\pm^{(\omega=\pi/2)}$ when varying the quark mass but the phase $I$ instead.

Both phases $II_\pm^{(\omega=\pi/2)}$ and $III_\pm^{(\omega=\pi/2)}$ carry over to {\it imaginary isospin chemical potential} while the phase $I$ does not exist in this region. The latter behavior was already the case for vanishing and finite twist. Note that the whole phase diagram at maximal twist is a flipped version of the phase diagram at vanishing twist reflecting the symmetry~\eref{symmetry-U},
 cf. Figs.~\ref{fig1a} and \ref{fig3a}. The Sharpe-Singleton transition lines inside the phases $II_\pm^{(\omega=\pi/2)}$ and $III_\pm^{(\omega=\pi/2)}$ along the $\widehat{m}=0$ axis can be also found at vanishing twist. In the region $\widehat{a}^2>0$ and $\widehat{\mu}_{\rm I}^2<0$ the $\pi^0$ condensate flips its sign when crossing with the quark mass the origin. At imaginary effective lattice
 spacing $\ha^2<0$ the $\pi^0$ condensate gets a kink at a second order phase transition and then drops linearly off while the mass dependent chiral condensate $\Sigma(\widehat{m})$ discontinuously switches the sign with the quark mass.
 The phase diagram is shown in Fig.~\ref{fig3a} and the details are summarized in Tables~\ref{t1} and \ref{t2}.

The special case $\whmu^2= -32\wha^2 > 0$ is related to the Aoki phase by the symmetry
transformation (\ref{symmetry-U}). Then the parameters of $U_0$ are only constrained by
\be
\sin \varphi \cos \vartheta_1 = -\frac {2\whm}{\whmu^2}.
\ee
Hence we have two massless Goldstone bosons when taking the limit $\widehat{\mu}_{\rm I}^2=\displaystyle{\lim_{\epsilon\searrow0}(32|\widehat{a}|^2+\epsilon)}$. Note that this limit is sometimes different from the limit $\widehat{\mu}_{\rm I}^2=\displaystyle{\lim_{\epsilon\searrow0}(32|\widehat{a}|^2-\epsilon)}$, see section~\ref{sec:4}. This is the crucial difference between the original Aoki phase at $\omega=0$ and the one identified at $\omega=\pi/2$.

\begin{figure}[t!]
 \centerline{\includegraphics[width=0.49\textwidth]{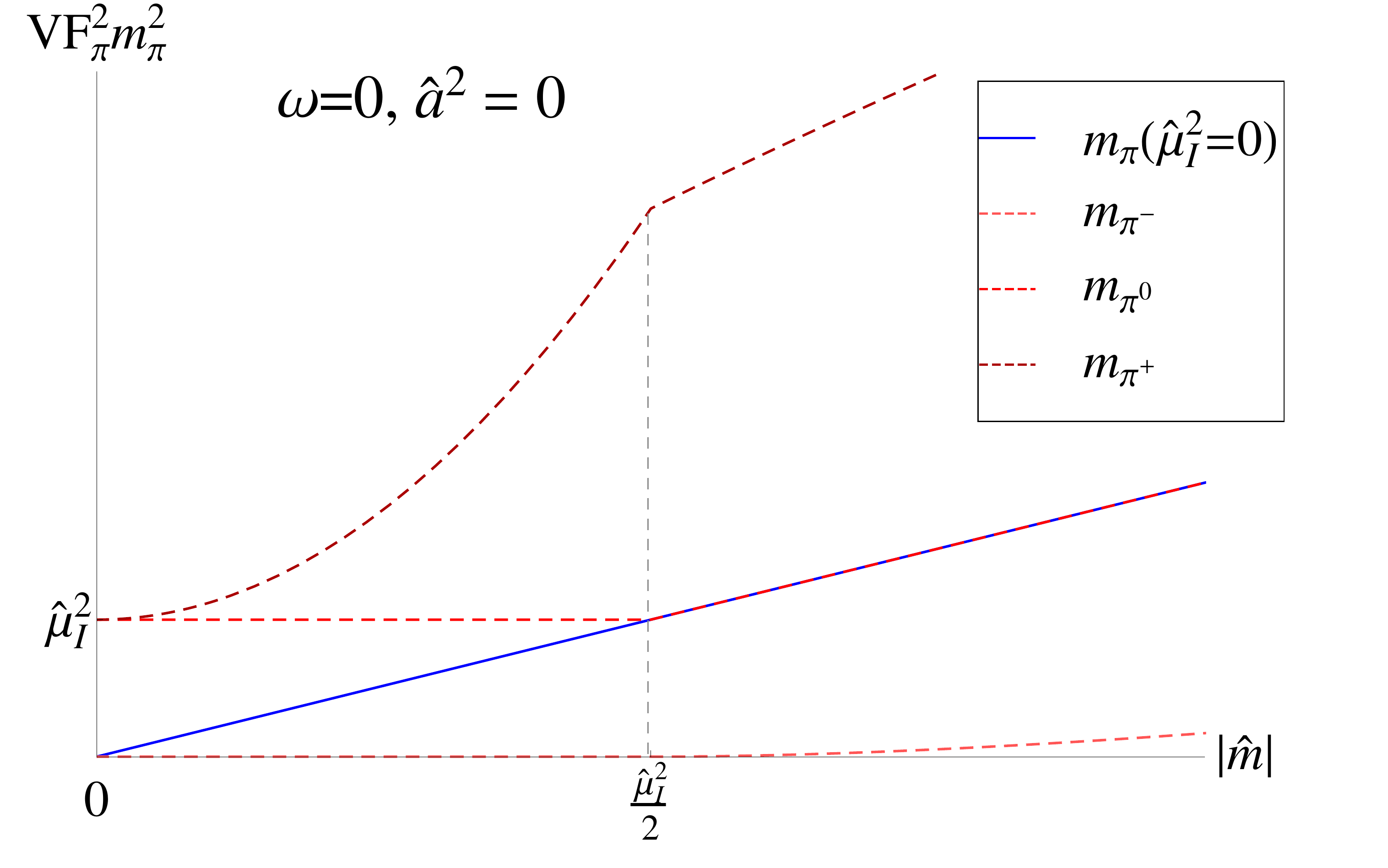}\hfill\includegraphics[width=0.49\textwidth]{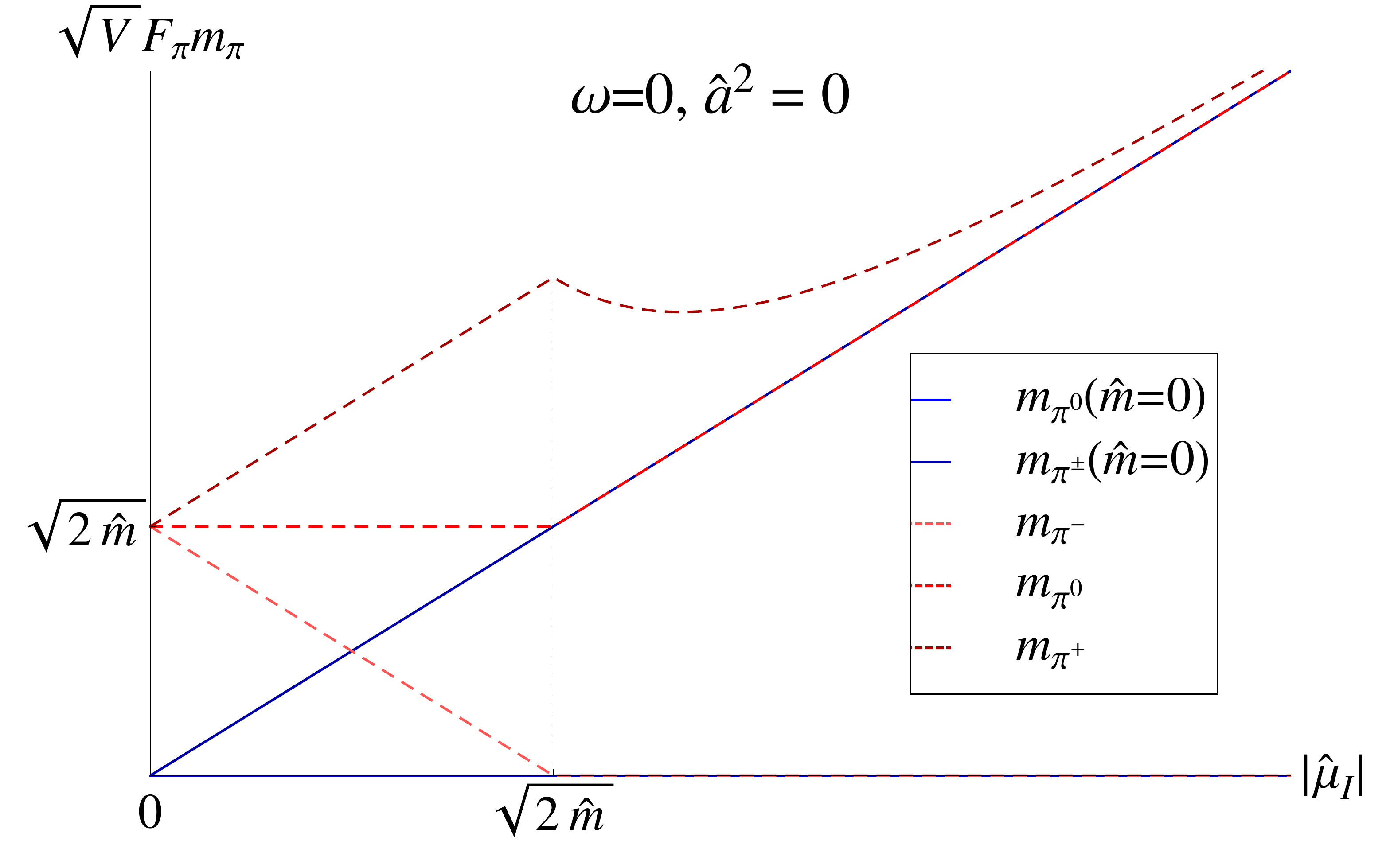}}
\caption{The three pion masses at vanishing effective lattice spacing ($\widehat{a}=0$) and at real isospin chemical potential ($\widehat{\mu}_{\rm I}^2\geq0$) as a function of
$|\widehat{m}|$ (left plot) and of  $|\widehat{\mu}_{\rm I}|$ (right plot). 
Notice that the squared pion mass is plotted in the right figure to underline the mass dependence. In this figure we show the functional behavior of the pion masses at 
zero chemical potential (blue curve in right figure) and at zero mass as a function of the
chemical potential (blue curves in left figure). The neutral pion mass remains massless as a function
of the chemical potential for zero quark mass. This diagram was already derived in \cite{Kogut:2000ek,Son:2000xc}.}
\label{fig8}
\end{figure}
\section{Masses of the pions}\label{sec:4}

The masses of the pions are obtained by expanding the chiral Lagrangian to 
second order in the pion fields, as it is done in $\mathcal{L}_2$, cf.~(\ref{L2}). Thereby we only consider a real isospin chemical potential.

The pole masses are defined by the energies $\widehat{E}$ when the quadratic
form containing the pion fields,
\be
\Pi_k H_{kl}(\widehat{E},\widehat{p}) \Pi_l
\ee
see\ Eq.~\eref{Hmatrix},
where we have used Minkowski metric, becomes singular. In particular we have to determine $\widehat{E}$ for which \cite{\KSTVZ}
\be
\det H(\widehat{E},\widehat{p}\equiv0)= 0.
\ee
This calculation is worked out in \ref{MassDer}. The results at vanishing and maximal twist are summarized in Table~\ref{t3}.

\begin{figure}[t!]
 \centerline{\includegraphics[width=1\textwidth]{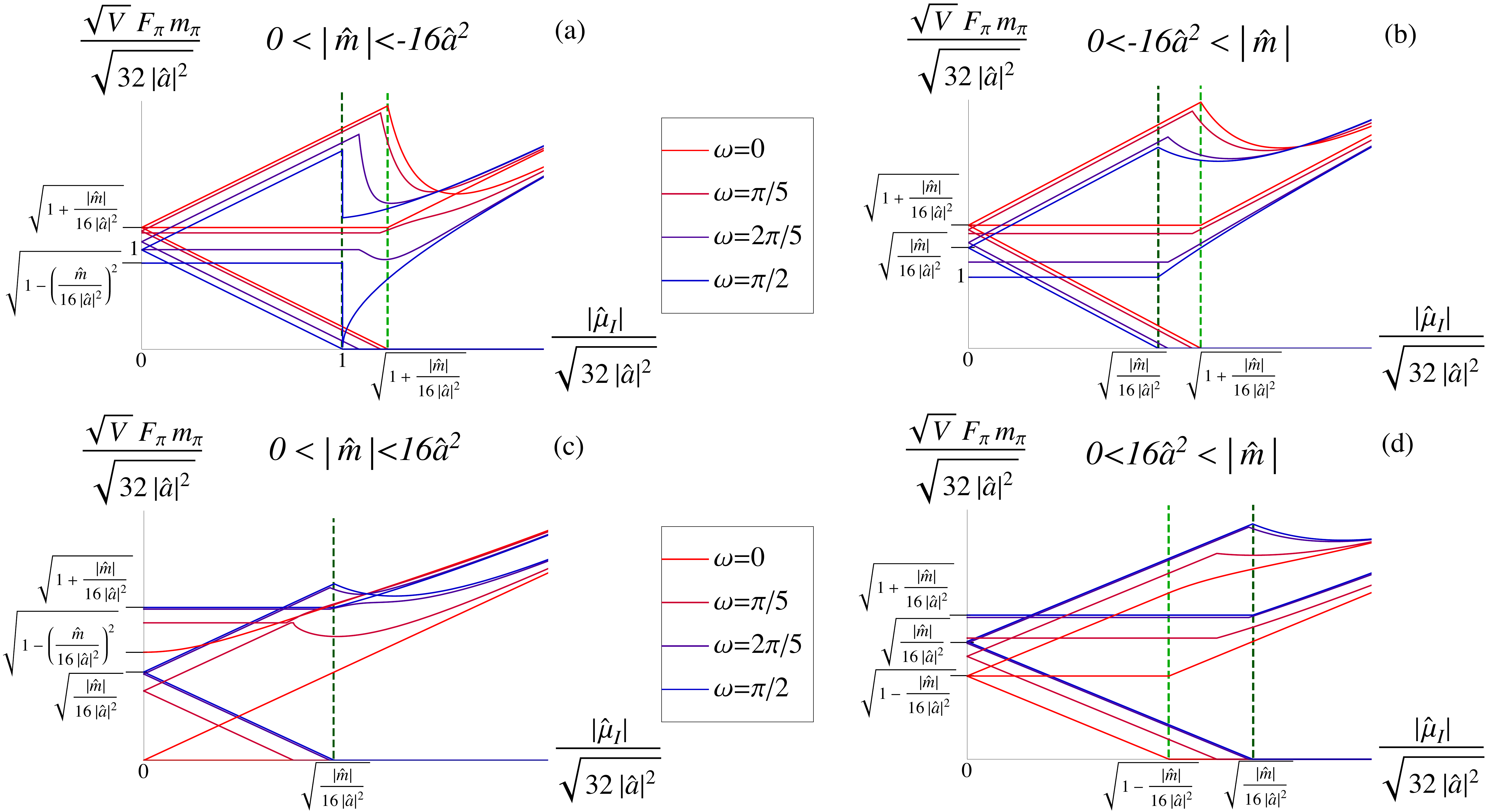}}
\caption{The behavior of all three pion masses drawn in the $m_\pi$-$|\widehat{\mu}|$-plane for various twisting angles $\omega$ and real isospin chemical potential ($\widehat{\mu}_{\rm I}^2\geq0$). In this figure we show the situations: (a) $0<|\widehat{m}|<-16\widehat{a}^2$; (b) $|\widehat{m}|>-16\widehat{a}^2>0$; (c) $0<|\widehat{m}|<16\widehat{a}^2$; and (d) $|\widehat{m}|>16\widehat{a}^2>0$. The vertical light green and dark green dashed lines indicate the positions of the phase transitions between the phases $I$ and $III_\pm$ at no and maximal twist, respectively. The jump of the masses in figure (a) for $\omega=\pi/2$ reflects the first order phase transition between the phases $I$ and $II_\pm^{(\omega=\pi/2)}$. Moreover there is no phase transition in figure (c) for $\omega=0$. The horizontal marks on the $y$-axis for $\omega=0,\pi/2$ are for orientation in the diagrams. In particular it should underline the generic form of these pion mass curves.}
\label{fig11}
\end{figure}

Physically it is clear  that for $\ha =0$ and sufficiently small $\widehat{\mu}_{\rm I}$ (such that the system is in one of the phases $II_\pm$ and $III_\pm$) the chemical potential dependence of
the pion masses is given by
\be\label{mass-relation}
m_{\pi^q} \approx m_{\pi^q}(\mu_{\rm I}=0) +  q\mu_{\rm I},
\ee
where $q$ is the isospin charge of the pions, compare with the linear dependence of the curves in Fig.~\ref{fig11}.
The pions condense for 
$ \mu_{\rm I} > m _\pi(\mu_{\rm I}=0)$ resulting in one exactly massless Goldstone boson (phase I), since
one of the angles in $U_0$ is not determined
by the saddle point equations.

At vanishing twist and lattice spacing  the pion masses are well known \cite{Kogut:2000ek,Son:2000xc}. We recall the detailed dependence of the pion masses on the 
quark mass and the isospin chemical potential in Fig.~\ref{fig8}. In the ``normal'' phase $III_\pm^{(\omega=0)}$ the mass relation~\eref{mass-relation} is exact with $VF_\pi^2m_{\pi^q}^2(\mu_{\rm I}=0)=2|\widehat{m}|$ while in the pion condensed phase we have one massless charged pion and two massive pions which are the other charged pion and the neutral one.

\begin{figure}[t!]
 \centerline{\includegraphics[width=0.49\textwidth]{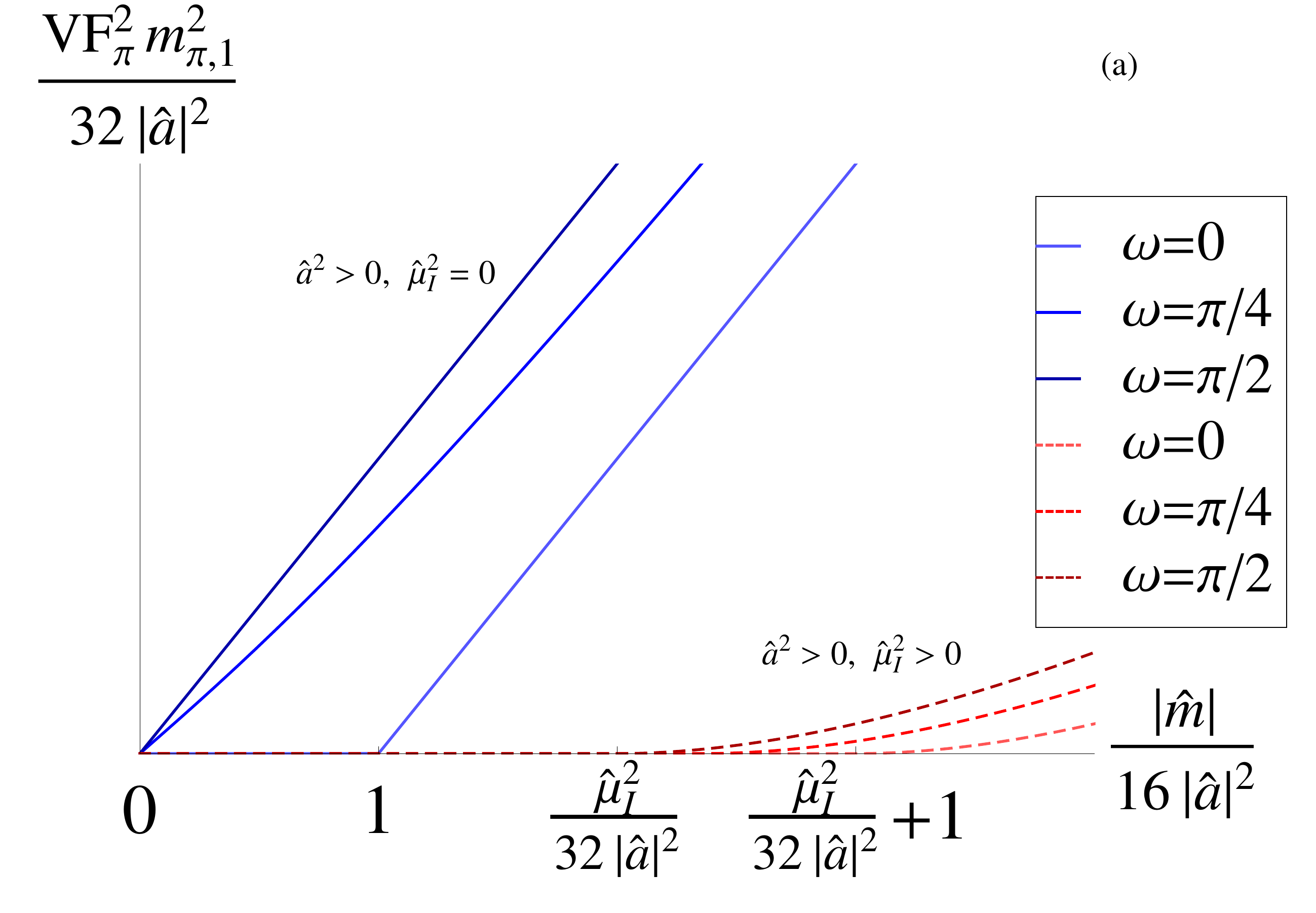}\hfill\includegraphics[width=0.49\textwidth]{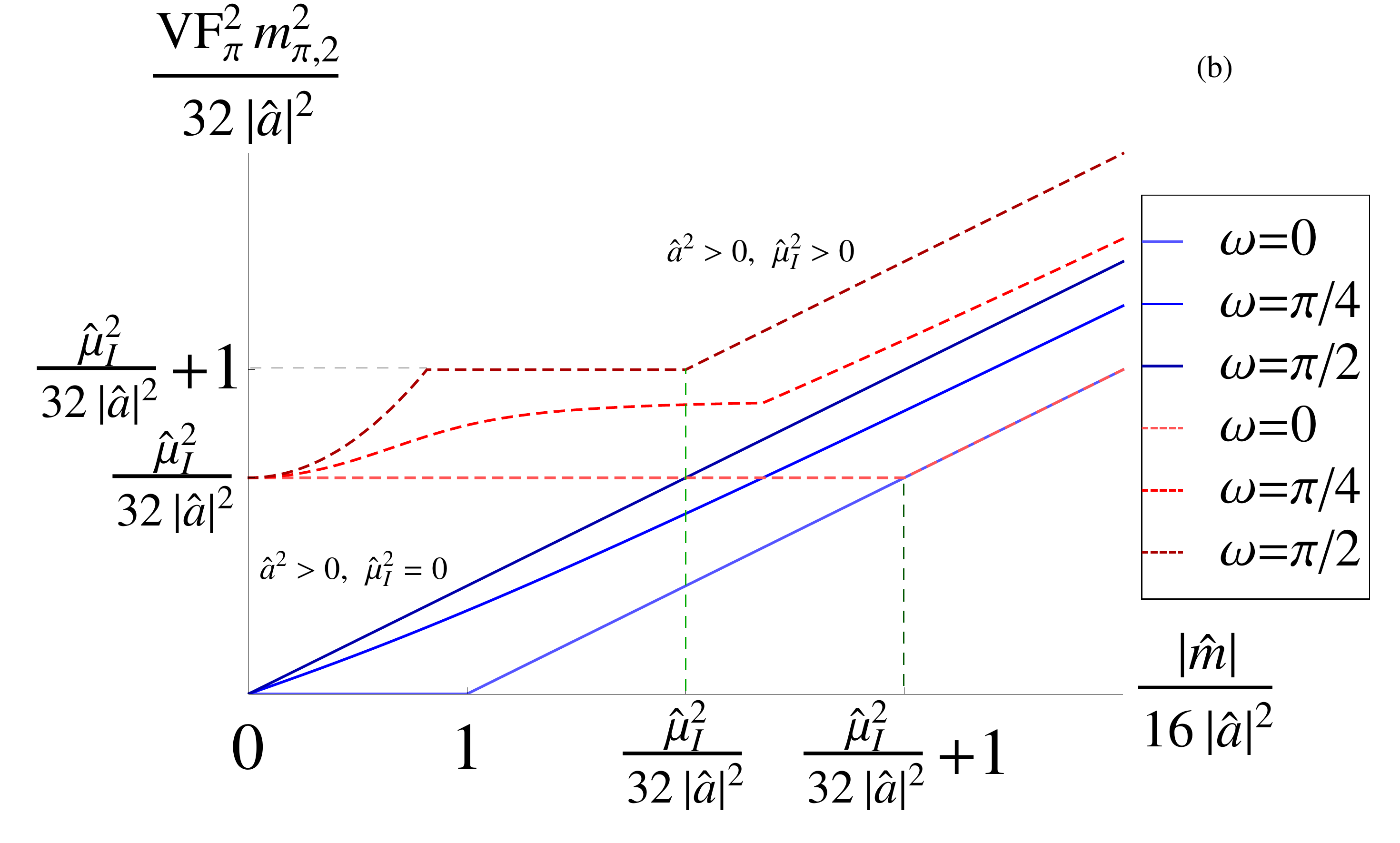}}
 \centerline{\includegraphics[width=0.49\textwidth]{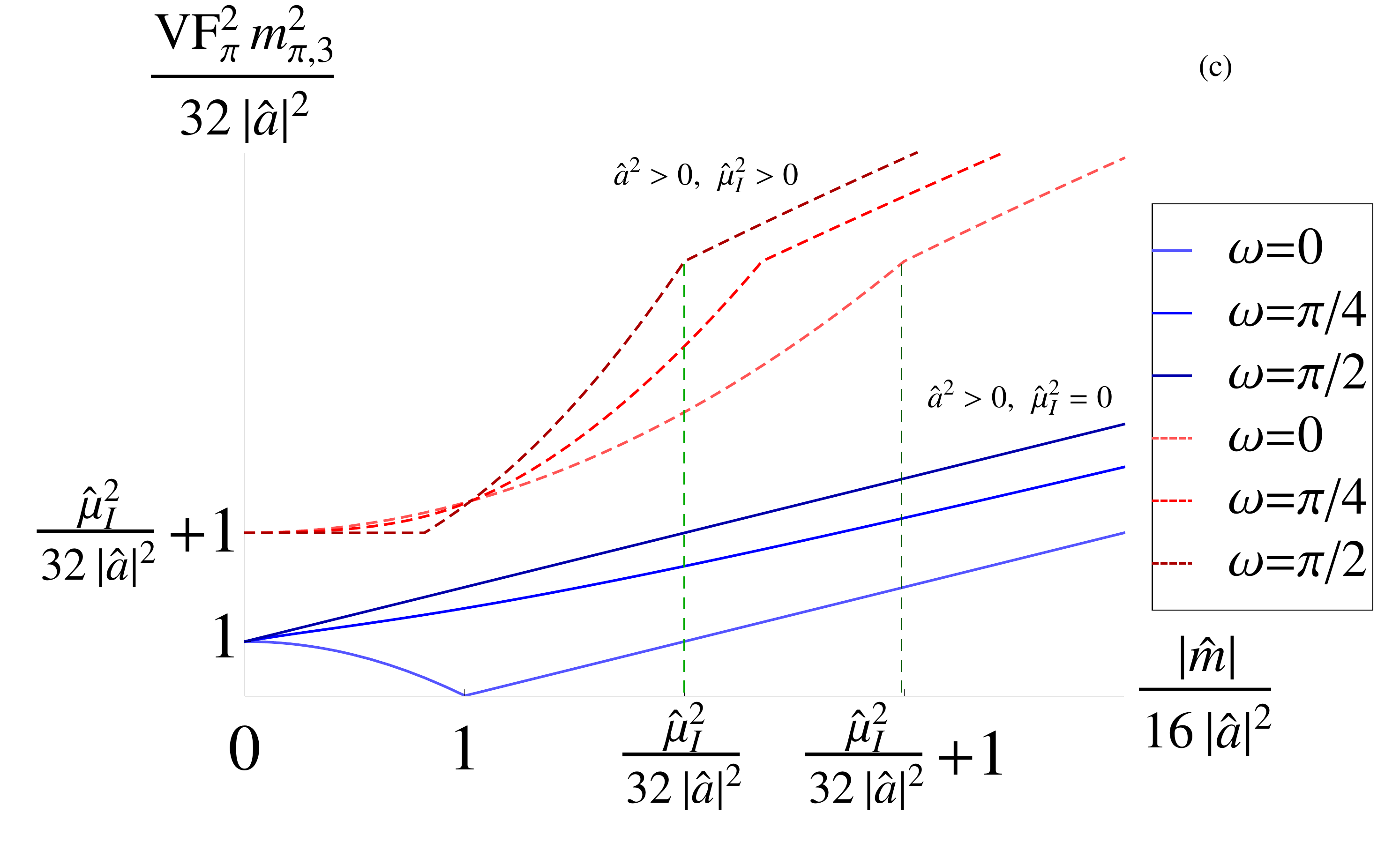}}
\caption{Pion masses at real effective lattice spacing ($\widehat{a}^2>0$) in the $m^2_\pi$-$|\widehat{m}|$-plane for various twisting angles $\omega$.  Shown is the generic behavior at vanishing isospin chemical potential (blue, solid curves) and at finite real isospin chemical potential (red, dashed curves). The vertical dashed lines indicate the positions of the phase transitions between the phases $I$ and $III_\pm$ at zero (dark green) and maximal twist (light green). The transition between the phases $I$ and $II_\pm$ at finite twisting angle is always between those two points. The kinks of the masses below these positions result from exact crossings of the masses and the ordering $m_{\pi,1}\leq m_{\pi,2}\leq m_{\pi,3}$. The phase diagram at $\widehat{\mu}_{\rm I}=0$ was discussed in \cite{Sharpe-Wu}.}
\label{fig9}
\end{figure}

The next simplest case is $\widehat{\mu}_{\rm I}=\omega = 0$ and $ \ha^2 > 0$. This case was analyzed in\cite{Aokiclassic,SharpeSingleton}. In the Aoki phase, 
$\hm < 16 \ha^2$, the saddle point equations only fix the angle $\varphi$, with
a nontrivial dependence of $U_0$ on $\vartheta_1$ and $\vartheta_2$ remaining.
This is consistent with having 
 two massless pions in this phase at  zero twist. At the phase
transition point also the third pion becomes massless (see Fig. \ref{fig9} for $\widehat{\mu}_{\rm I}^2=0$). At a finite twisting angle $\omega\neq0$ all three pions become massive when the quarks are massive ($\widehat{m}\neq0$) . The reason for this is that we never enter the phase $I$ for $\omega\neq0$, $\widehat{\mu}_{\rm I}^2=0$ and $\widehat{a}^2>0$.

\begin{figure}[!ht]
 \centerline{\includegraphics[width=1\textwidth]{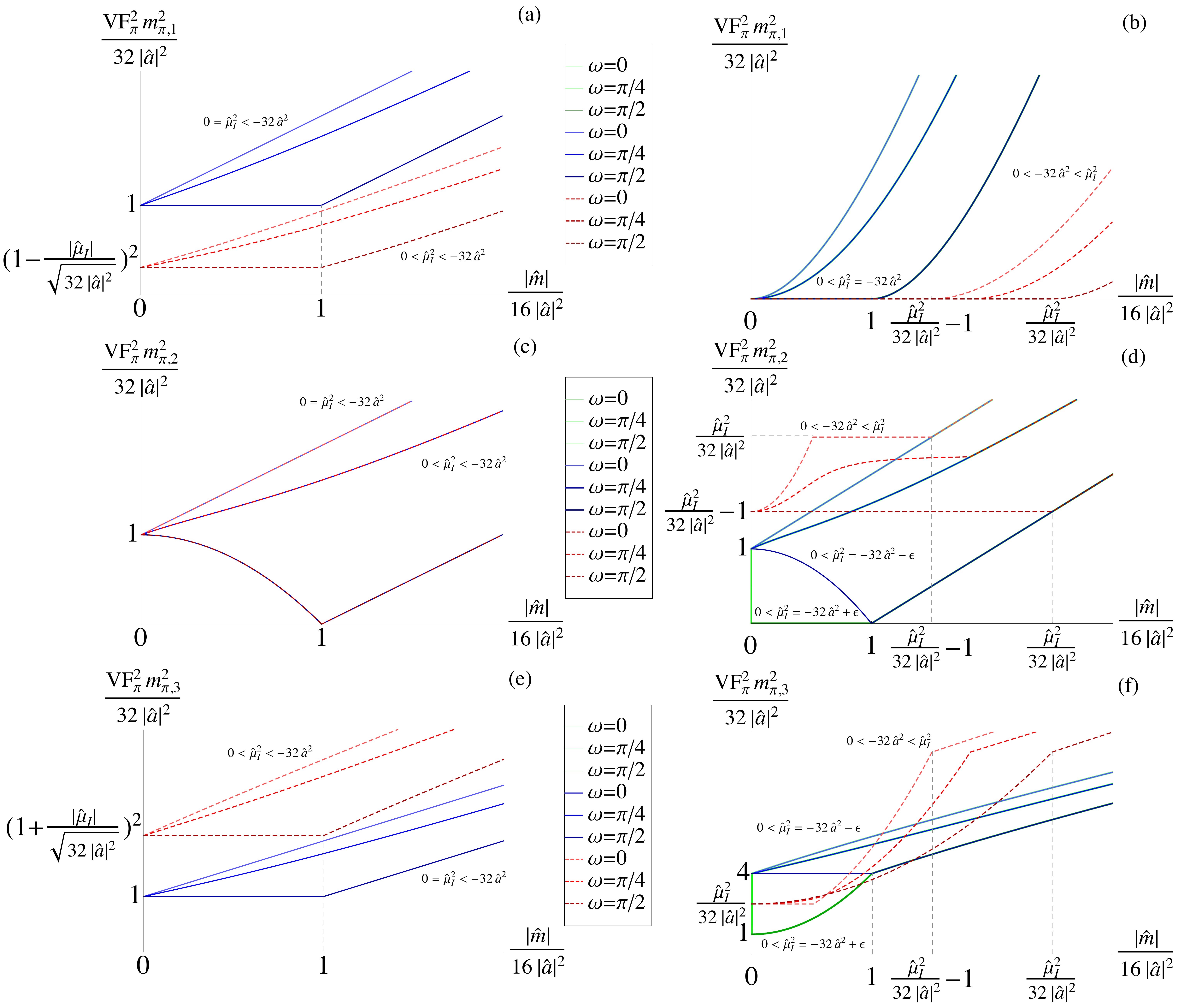}}
\caption{Pion masses at imaginary effective lattice spacing ($\widehat{a}^2<0$)  and at real isospin chemical potential ($\widehat{\mu}_{\rm I}^2\geq0$) as a function of the quark mass $|\widehat{m}|$ for various twisting angles $\omega$. Shown is the generic behavior for the following isospin chemical potentials:  $\widehat{\mu}_{\rm I}^2=0$ (blue, solid curves in (a), (c), and (e)),   $\widehat{\mu}_{\rm I}^2<-32\widehat{a}^2$ (red dashed curves in (a), (c), and (e)),  $\widehat{\mu}_{\rm I}^2=\displaystyle{\lim_{\epsilon\searrow0}(-32\widehat{a}^2-\epsilon)}$ (blue, solid curves in (b), (d), and (f)),  $\widehat{\mu}_{\rm I}^2=\displaystyle{\lim_{\epsilon\searrow0}(-32\widehat{a}^2+\epsilon)}$ (green, solid curves in  (b), (d), and (f)), and  $\widehat{\mu}_{\rm I}^2>-32\widehat{a}^2>0$ (red dashed curves in (b), (d), and (f)). Note, the mass curve for $m_{\pi,1}$ is the same for $\widehat{\mu}_{\rm l}^2=32\widehat{a}^2\pm\epsilon$. The vertical grey, dashed lines indicate the positions of the phase transitions between the phases $I$ and  $III_\pm^{(\omega=0,\pi/2}$ at no and maximal twist. Note that there is no phase transition for $\widehat{\mu}_{\rm I}^2\leq 32\widehat{a}^2 $ and $\omega\neq\pi/2$. The kinks of the masses below the positions of the phase transitions  for $\widehat{\mu}_{\rm I}^2>-32\widehat{a}^2>0$ result from exact crossings of the pion masses and an ordering $m_{\pi,1}\leq m_{\pi,2}\leq m_{\pi,3}$. The jump of the masses $m_{\pi,2}$ and $m_{\pi,3}$ at $\widehat{\mu}^2= -32\widehat{a}^2 $ reflects the first order phase transition between the phases $I$ and $II_{\pm}^{\omega=\pi/2}$. The curves for $\widehat{\mu}_{\rm I}=0$ were already derived in \cite{Sharpe-Wu}.}
\label{fig10}
\end{figure}

When the Aoki phase is perturbed by a real isospin chemical potential, one of the charged pions condenses. Despite the symmetry breaking by the chemical potential,
one
of the two massless pions therefore remains massless in the phase $I$ for all values of the twist angle, cf.  Fig. \ref{fig11}.c). When we are outside the Aoki phase for $\widehat{\mu}_{\rm I} =0$,
there are no massless pions at $\widehat{\mu}_{\rm I}=0$.
However one massless pion always appears for sufficiently large isospin chemical potential  when entering the phase $I$
(see Fig. \ref{fig11}.d).
 Also for $\ha^2 < 0 $ we find a transition to a pion condensed phase
for $\widehat{\mu}_{\rm I} > m_\pi(\widehat{\mu}_{\rm I}=0)$, see Figs.~\ref{fig11}.a) and b), \ref{fig9}, and \ref{fig10}.

At maximum twist we enter an Aoki like phase at $\hmu^2 = -32\ha^2$. Depending on how we approach this surface
we have one or two massless pions, see Fig. \ref{fig11}.a). This discrepancy reflects the fact that the transition from $II_\pm^{(\pi/2)}$ to the pion condensed phase $I$ is first order.
This is shown in Figs.~\ref{fig10}.d) and f) where the blue curves correspond to $\hmu^2 = \displaystyle{\lim_{\epsilon \to 0}(-32 \ha^2 -\epsilon )}$ while the green ones are the one for $\hmu^2 = \displaystyle{\lim_{\epsilon \to 0}(-32 \ha^2 +\epsilon )}$. The latter limit yields two massless Goldstone bosons which again reflect the fact that two angles in the unitary matrix $U_0$ are not fixed, see also the Table~\ref{t3}.

\begin{table}[t!] \centering
\rotatebox{90}{
\begin{tabular}[c]{c||c|c|c}
 phase &  $\displaystyle 
VF_\pi^2m_{\pi,1}^2$ & $\displaystyle 
VF_\pi^2m_{\pi,2}^2$ & $\displaystyle 
VF_\pi^2m_{\pi,3}^2$ \\ 
\noalign{\vskip\doublerulesep\hrule height 2pt} 
$I^{(\omega=0)}$ & 
$0$
 & $\displaystyle \widehat{\mu}_{\rm I}^2$ & $\displaystyle \underset{\ }{\overset{\ }{\widehat{\mu}_{\rm I}^2+32\widehat{a}^2+\frac{4\widehat{m}^2(3\widehat{\mu}_{\rm I}^2-32\widehat{a}^2)}{(32\widehat{a}^2+\widehat{\mu}_{\rm I}^2)^2}}}$ \\ \hline
\hline  $III_\pm^{(\omega=0)}$ & $\displaystyle (\sqrt{2|\widehat{m}|-32\widehat{a}^2}-\widehat{\mu}_{\rm I})^2$ & $2|\widehat{m}|-32\widehat{a}^2$ & $\displaystyle \overset{\ }{(\sqrt{2|\widehat{m}|-32\widehat{a}^2}+\widehat{\mu}_{\rm I})^2}$  \\
  \hline  Aoki phase & $\displaystyle 0$ 
& $\displaystyle 0$ & $\displaystyle \overset{\ }{32\widehat{a}^2-\frac{\widehat{m}^2}{8\widehat{a}^2}}$ 
\\  
$\widehat{\mu}_{\rm I}=\omega=0$ &&& \\
 \noalign{\vskip\doublerulesep\hrule height 2pt} 
$I^{(\omega=\pi/2)}$ & 
$0$
 & $\displaystyle \widehat{\mu}_{\rm I}^2+32\widehat{a}^2$ & $\displaystyle  \underset{\ }{\overset{\ }{\widehat{\mu}_{\rm I}^2+\frac{12\widehat{m}^2}{\widehat{\mu}_{\rm I}^2}}}$ \\ \hline
 $II_\pm^{(\omega=\pi/2)}$ 
& $\displaystyle \overset{\ }{\left(\sqrt{32|\widehat{a}|^2}-\widehat{\mu}_{\rm I}\right)^2}$ 
& $\displaystyle \overset{\ }{32|\widehat{a}|^2-\frac{\widehat{m}^2}{8|\widehat{a}|^2}}$ 
& $\displaystyle \overset{\ }{\left(\sqrt{32|\widehat{a}|^2}+\widehat{\mu}_{\rm I}\right)^2}$  \\ 
\hline  $III_\pm^{(\omega=\pi/2)}$ & $\displaystyle \overset{\ }{\left(\sqrt{2|\widehat{m}|}-\widehat{\mu}_{\rm I}\right)^2}$ & $\displaystyle 2|\widehat{m}|+32\widehat{a}^2$ & $\displaystyle \overset{\ }{\left(\sqrt{2|\widehat{m}|}+\widehat{\mu}_{\rm I}\right)^2}$  \\
  \hline  Aoki-like phase $\displaystyle\overset{\ }{\left(\omega=\frac{\pi}{2}\right)}$& &&
\\  
$\displaystyle\widehat{\mu}_{\rm I}^2=-32\widehat{a}^2-\epsilon$ & $\displaystyle 0$ 
& $\displaystyle \overset{\ }{32|\widehat{a}|^2-\frac{\widehat{m}^2}{8|\widehat{a}|^2}}$ 
& $\displaystyle \overset{\ }{128|\widehat{a}|^2}$ \\ 
$\displaystyle\widehat{\mu}_{\rm I}^2=-32\widehat{a}^2+\epsilon$ & $\displaystyle0$ & $\displaystyle0$ & $\displaystyle  \overset{\ }{32|\widehat{a}|^2+\frac{3\widehat{m}^2}{8|\widehat{a}|^2}}$ 
\end{tabular}}
\caption{\label{t3} Functional dependence of the three pion masses at zero ($\omega=0$) and maximal ($\omega=\pi/2$) twist. Note that we consider real isospin chemical potential ($\widehat{\mu}_{\rm I}^2>0$) such that the phase $II_\pm^{\omega=0}$ does not appear. At finite twist the expressions are quite complicated, cf. Eqs.~\eref{detHcoeff} and \eref{dispersion-II}. Hence we omit them in this list. We underline that we can approach the Aoki-like phase at maximal twist from two sides ($\widehat{\mu}_{\rm I}=\displaystyle{\lim_{\epsilon\searrow0}(-32\widehat{a}^2\pm\epsilon)}$) yielding a jump in the pion masses due to the first order phase transition which happens at this phase. This cannot be done for the original Aoki phase at zero twist ($\omega=\widehat{\mu}_{\rm I}=0$)
 since $\mu_I$ has to be real and thus $\widehat{\mu}_{\rm I}^2\geq0$.
}
\end{table}

\subsection{Phase $I$\; ($4\widehat{\mu}^2\geq |\widehat{m}|\sin\omega>0$)}\label{sec4.1}

In this phase both $\varphi$ and $\vartheta_1$ are neither zero nor $ \pi/2$. Then
the $\U(1)$ symmetry of the microscopic theory is spontaneously broken resulting in
a massless Goldstone boson corresponding to the angle $\vartheta_2$. The saddle point solution 
 $U_0$ is a combination of all three classical pion fields ($\pi^0$ and $\pi^\pm$) because in the
second order Lagrangian~(\ref{L2}) all three modes are coupled. This results
in a somewhat complicated expression for the masses
\begin{eqnarray}\label{detHcoeff}
\widehat{m}_{\pi,2}=\sqrt{c_1-\sqrt{c_2}},\quad\widehat{m}_{\pi,3}=\sqrt{c_1+\sqrt{c_2}}
\end{eqnarray}
with
\begin{eqnarray}\label{coefficients}
\fl c_1&=&\widehat{\mu}_{\rm I}^2\left(1+\frac{6\widehat{m}^2\sin^2\omega}{\widehat{\mu}_{\rm I}^4}+\frac{6\widehat{m}^2\cos^2\omega}{(32\widehat{a}^2+\widehat{\mu}_{\rm I}^2)^2}\right) +16\widehat{a}^2\left(1-\frac{4\widehat{m}^2\cos^2\omega}{(32\widehat{a}^2+\widehat{\mu}_{\rm I}^2)^2}\right)\geq \sqrt{c_2}\geq0,\nonumber\\
\fl c_2&=& (16\widehat{a}^2)^2\left(1-\frac{4\widehat{m}^2\cos^2\omega}{(32\widehat{a}^2+\widehat{\mu}_{\rm I}^2)^2}\right)^2+\frac{9}{4}\widehat{\mu}_{\rm I}^4\left(\frac{4\widehat{m}^2\cos^2\omega}{(32\widehat{a}^2+\widehat{\mu}_{\rm I}^2)^2}+\frac{4\widehat{m}^2\sin^2\omega}{\widehat{\mu}_{\rm I}^4}\right)^2\nonumber\\
\fl&&-48\widehat{a}^2\widehat{\mu}_{\rm I}^2\left(\frac{16\widehat{m}^4\cos^4\omega}{(32\widehat{a}^2+\widehat{\mu}_{\rm I}^2)^4}+\frac{4\widehat{m}^2\sin^2\omega}{\widehat{\mu}_{\rm I}^4}-\frac{4\widehat{m}^2\cos^2\omega(\widehat{\mu}_{\rm I}^4-4\widehat{m}^2\sin^2\omega)}{\widehat{\mu}_{\rm I}^4(32\widehat{a}^2+\widehat{\mu}_{\rm I}^2)^2}\right)\geq0.
\end{eqnarray}
 The third mass $\widehat{m}_{\pi,1}=0$ vanishes. In \ref{App_sec4.1} we show the details of the computation.

\subsection{Phase $II_\pm$ (\;$|\cos\vartheta_1|=1$ and $\omega\neq 0,\pi/2$)}\label{sec4.2}

In this phase the saddle point solution satisfies a reduced parameterization,
\be
U_0 = \cos\varphi + \imath \sin \varphi \tau_3.
\ee
From the second order chiral Lagrangian~(\ref{L2}) one can immediately read off 
 that the $\pi^0$ mode decouples from  the charged pions $\pi^\pm$.
Moreover, the second order terms in $\pi^0$ do not depend on the chemical potential $\widehat{\mu}_{\rm I}$.
The corresponding masses are worked out explicitly in \ref{App_sec4.2}. From Eq. (B.20) we obtain (using that $\alpha=cos\widetilde\varphi_0$) 
\be
m_{\pi^0}^2 = \frac{2|\Sigma m|}{F_\pi^2} \cos(\widetilde{\varphi}_0-\omega) 
       + \frac{32 a^2 C_2}{F_\pi^2} (\sin^2\varphi_0- \cos^2\varphi_0),
\ee
where $ \widetilde{\varphi}_0$ is determined by Eq.~\eref{sol-app}.

The chemical potential dependence of the charged pions is again given by relation~\eref{mass-relation}. However the mass at vanishing isospin chemical potential may differ from the one of the neutral pion, i.e. (see \ref{App_sec4.2})
\be
m_{\pi^{\pm}}(\mu_{\rm I}=0) =\frac{2|\Sigma m|}{F_\pi^2} \cos(\widetilde{\varphi}_0-\omega) 
       - \frac{32 a^2 C_2}{F_\pi^2}  \cos^2\varphi_0.
\ee
Additional details are worked out in \ref{App_sec4.2} and can be read off Figs.~\ref{fig11}, \ref{fig9}, \ref{fig10}, and Table~\ref{t3}.

\section{Summary}\label{sec5}

Starting from the leading order chiral Lagrangian in the $p$-regime we studied the phase diagram of 
two degenerate twisted mass Wilson fermions as a function
of the quark mass, the isospin chemical potential, the lattice spacing and the twist angle. 
This work extends previous studies of the phase diagram in the same parameter space but at
zero lattice spacing and vanishing twist~\cite{Kogut:2000ek,Son:2000xc} or at zero isospin chemical potential and finite twist and lattice spacing~\cite{Sharpe-Wu}. The phases are characterized by
order parameters, in particular by the mass dependent chiral condensate and by the neutral and charged pion condensate,
which enables us to distinguish the various phases. 
We consider both real and imaginary isospin chemical potential, and real and imaginary effective lattice
spacing (because of the sign of the low-energy constants). Since the saddle point approximation can be analyzed for parameters in the $\epsilon$-domain we can always stay well away from the Roberge-Weiss transition~\cite{RW} at imaginary isospin chemical potential.

At zero chemical potential and at zero twist we can distinguish, depending on the values of the low-energy constants,
two different possibilities in the approach to the chiral limit at non-zero lattice spacing. First,
a transition to the Aoki phase~\cite{Aokiclassic} with a non-zero neutral pion condensate which spontaneously breaks parity,
and second, the Sharpe-Singleton scenario~\cite{SharpeSingleton}, where when the quark mass crosses zero,  
we jump to a different minimum
separated by a potential barrier but with the same physics.

A non-zero isospin chemical potential
destroys the Aoki phase. Since the Aoki phase has two charged pions, a non-zero isospin chemical
potential immediately leads to pion condensation so that the partition function becomes singular at
$\mu_{\rm I} =0$. In the case of the Sharpe-Singleton scenario, the first order jump of the mass dependent chiral condensate  persists at imaginary effective lattice spacing until the chemical
potential reaches a critical value. Interestingly the Sharpe-Singleton scenario may extend to real effective lattice spacing where not the chiral condensate but the $\pi^0$ condensate may jump when the quark mass crosses the origin. Moreover, we are always in the normal phase for large quark mass. This phase is characterized
by a non-zero value of the mass dependent chiral condensate. At finite twist the same part of the chiral condensate is rotated into a $\pi^0$ condensate.

At non-zero twisted mass, the phase diagram greatly simplifies and only two phases are possible. 
Phase $I$ is connected to the pion condensed phase at zero lattice spacing when the chemical
potential is less than the pion mass. The other phase $II_\pm$ is continuously connected to
the ``normal" phase. The two phases are separated  by a second order phase transition.

We have also shown that in the considered parameter space, the phase diagram has a symmetry which
relates the mean field solution at a given twist angle $\omega$ to the complementary twist angle $\pi/2-\omega$. We underline that this symmetry is not a symmetry of the mass spectrum of the pseudo-Goldstone bosons. In particular, this implies
that the physical properties of the phase at maximum twist corresponding to the Aoki phase are different
from those of the Aoki phase.
Instead of two massless Goldstone bosons, we find either one or two massless Goldstone bosons at the critical
value of the isospin chemical potential depending on how the phase boundary is approached, $\widehat{\mu}_{\rm I}^2=\displaystyle{\lim_{\epsilon\searrow0}(-32\widehat{a}^2-\epsilon)}$ and $\widehat{\mu}_{\rm I}^2=\displaystyle{\lim_{\epsilon\searrow0}(-32\widehat{a}^2+\epsilon)}$, respectively. The pions in the phase $I$ do not have good isospin quantum numbers. Taking the isospin chemical potential away form its critical value does not result in pion condensation.

The analytical results presented in this paper are useful for lattice studies of QCD at non-zero isospin chemical potential with Wilson fermions with zero or maximum twist. In particular they will be helpful to determine the region of the parameter domain that is connected to the physical point. Interesting extensions of this study would be to QCD with more than two flavors~\cite{Baron:2010bv}, to lift the degeneracy in the quark mass as studied in \cite{HorShar1, HorShar2, HorShar3} for $\widehat{\mu}_{\rm I}^2=0$, and to QCD-like theories like two color QCD as it was recently analytically discussed by three of the authors~\cite{KVZ2, KVZmoriond}.

\section*{Acknowledgements}

This work was supported by  U.S. DOE Grant No. DE-FAG-88FR40388 (JV), the
Humboldt Foundation (MK, SZ), CRC 701: {\it Spectral Structures and Topological Methods in Mathematics} of the Deutsche Forschungsgemeinschaft,
the {\sl Sapere Aude} program of The Danish Council for 
Independent Research (KS) and the Centre National de la Recherche Scientifique (SZ).

\appendix

\section{Derivation of the Phase Diagram}\label{app1}

In this Appendix we determine the solutions of the mean field equations. Different types of solutions
are discussed separately in \ref{app1.1} and \ref{app1.2}. The case of zero twist is worked out in
\ref{app1.3} and the case of maximum twist in \ref{app1.4}.

In the thermodynamic limit we have to minimize Eq.~\eref{Lagrangian}
with respect to the variables $\varphi\in[0,\pi]$ and $\vartheta_1\in[0,\pi[$.
Before starting the calculation of the saddle point solutions let us discuss the Lagrangian in more detail. 
First, the Lagrangian does not depend on  the third angle $\vartheta_2$ parameterizing the group $\SU(2)$ but this angle may be fixed by the source terms via spontaneous symmetry breaking. Second, the minima of the Lagrangian have to satisfy the following two conditions
\begin{equation}\label{inequalities}
 \widehat{m}\cos\varphi\geq0\quad{\rm and}\quad \widehat{m}\cos\vartheta_1\leq0.
\end{equation}
This fixes the signs of the trigonometric functions of the angles $\varphi$ and $\vartheta_1$ which are $\sign\cos\varphi=-\sign\cos\vartheta_1=\sign\widehat{m}$ while $\sin\varphi=\sin\vartheta_1\geq0$ because $\varphi,\vartheta_1\in[0,\pi]$. For this choice the first two terms of Eq.~\eref{chiLagrcoor} are negative which would be positive when choosing other signs. The other terms in  Eq.~\eref{chiLagrcoor} remain the same when switching the sign of the trigonometric functions. The third thing we want to point out is that the Lagrangian~\eref{Lagrangian} is a quadratic function of the two parameters $x=\cos\varphi$ and $y=\sin\varphi\cos\vartheta_1$ on the centered unit disk. Therefore we have either a minimum inside the disk ($x^2+y^2<1$ equivalent to $|\cos\vartheta_1|<1$) or on the boundary ($x^2+y^2=1$ implying $\cos\vartheta_1=-\sign\widehat{m}$).

\subsection{Inside the unit disk ($|\cos\vartheta_1| < 1$)}\label{app1.1}

The global extremum of the polynomial in the variables $x=\cos\varphi$ and $y=\sin\varphi\cos\vartheta_1$ is
\begin{eqnarray}\label{extremum}
 x=\frac{2\widehat{m}\cos\omega}{32\widehat{a}^2+\widehat{\mu}_{\rm I}^2}\quad{\rm and}\quad y=\frac{2\widehat{m}\sin\omega}{\widehat{\mu}_{\rm I}^2}.
\end{eqnarray}
However it is only a minimum inside the unit disk $x^2+y^2<1$ (and thus a global minimum) if
\begin{equation}
0\leq2|\widehat{m}|\cos\omega<32\widehat{a}^2+\widehat{\mu}_{\rm I}^2\quad{\rm and}\quad 0\leq2|\widehat{m}|\sin\omega<\widehat{\mu}_{\rm I}^2.
\end{equation}
Additionally we have
\be 
1 > \sqrt{1- \left(\frac{2\whm \cos\omega}{32\wha^2+\whmu^2}\right )^2}
> \frac {2|\whm|\sin\omega}{\whmu^2} \ge 0
\label{parm-I}
\ee
which encodes the fact that we are inside the disc.
We denote the phase corresponding to this situation by $I$.

The angles take the values
\begin{eqnarray}\label{angles-P1}
 \varphi^{(I)}={\rm arccos}\left(\frac{2\widehat{m}\cos\omega}{32\widehat{a}^2+ \widehat{\mu}_{\rm I}^2}\right),\quad \vartheta_1^{(I)}={\rm arccos}\left(-\frac{2\widehat{m}\cos\omega}{\widehat{\mu}_{\rm I}^2\sin\varphi^{(I)}}\right)
\end{eqnarray} 
such that the group element $U\in\SU(2)$ is given by
\begin{eqnarray}\label{matrix-P1}
 \fl U_0^{(I)}&=&\frac{2\widehat{m}\cos\omega}{32\widehat{a}^2+ \widehat{\mu}_{\rm I}^2} \eins_2\\
 \fl&&\hspace*{-1cm}+\imath \left[\begin{array}{cc} \displaystyle-\frac{2\widehat{m}\sin\omega}{\widehat{\mu}_{\rm I}^2} & \displaystyle e^{\imath\vartheta_2}\sqrt{1-\frac{4\widehat{m}^2\cos^2\omega}{(32\widehat{a}^2+\widehat{\mu}_{\rm I}^2)^2}-\frac{4\widehat{m}^2\sin^2\omega}{\widehat{\mu}_{\rm I}^4}} \\ \displaystyle e^{-\imath\vartheta_2}\sqrt{1-\frac{4\widehat{m}^2\cos^2\omega}{(32\widehat{a}^2+\widehat{\mu}_{\rm I}^2)^2}-\frac{4\widehat{m}^2\sin^2\omega}{\widehat{\mu}_{\rm I}^4}} & \displaystyle\frac{2\widehat{m}\sin\omega}{\widehat{\mu}_{\rm I}^2}\end{array}\right]\nonumber
\end{eqnarray}
with $\vartheta_2\in[0,2\pi]$. The free energy is then equal to
\begin{eqnarray}\label{Lagrangian-P1}
 V\mathcal{L}^{(I)}_{\rm 0}=-\frac{\widehat{\mu}_{\rm I}^2}{2}-\frac{2\widehat{m}^2\sin^2\omega}{\widehat{\mu}_{\rm I}^2}-\frac{2\widehat{m}^2\cos^2\omega}{32\widehat{a}^2+ \widehat{\mu}_{\rm I}^2}.
\end{eqnarray}
Again we want to underline that the angle $\vartheta_2$ can be fixed by the source terms $j_\pm$.

\subsection{On the unit circle ($|\cos\vartheta_1|=1$) and $\omega\neq 0,\pi/2$}\label{app1.2}

The extrema on the boundary $(x,y)=(\cos\varphi,\sin\varphi)=(\sign\widehat{m}\cos\widetilde{\varphi},\sin\widetilde{\varphi})$ are given by the equation
\begin{eqnarray}\label{transcendent}
16\widehat{a}^2\cos\widetilde{\varphi}\sin\widetilde{\varphi}-|\widehat{m}|\sin\omega\cos\widetilde{\varphi}-|\widehat{m}|\cos\omega\sin\widetilde{\varphi}=0
\end{eqnarray}
which is the derivative of the chiral Lagrangian 
\begin{eqnarray}\label{Lagrangian-simp2b}
 V\mathcal{L}_{\rm 0}=16\widehat{a}^2\cos^2\varphi-2|\widehat{m}|\sin\omega\sin\widetilde{\varphi}-2|\widehat{m}|\cos\omega\cos\widetilde{\varphi}.
\end{eqnarray}
Note that this case is independent of the chemical potential such that the solution has a ``Silver-blaze-property".

 Let $\omega\neq 0,\pi/2$. Then the solution of the transcendental equation can be compactly written as an integral,
\begin{eqnarray}\label{sol-app}
 \fl\cos\widetilde{\varphi}=F\left(\frac{|\widehat{m}|}{16\widehat{a}^2}\right)=\int_0^1\Theta\left(|\widehat{m}|(\sqrt{1-y^2}\cos\omega-y\sin\omega)-16\widehat{a}^2y\sqrt{1-y^2}\right)dy,
\end{eqnarray}
where $\Theta$ is the Heaviside step function. One can easily show that Eq.~\eref{sol-app} is the solution of Eq.~\eref{transcendent} via integration by parts and taking into account that there is only one solution in the interval $\cos\widetilde{\varphi}\in]0,1[$.
Thus the angles freeze out at
\begin{eqnarray}
 \fl \varphi^{(II)}&=&\arccos\left[\sign{\widehat{m}}\int_0^1\Theta\left(|\widehat{m}|(\sqrt{1-y^2}\cos\omega-y\sin\omega)-16\widehat{a}^2y\sqrt{1-y^2}\right)dy\right],\nonumber\\
 \fl\vartheta_1^{(II)}&=&{\rm arccos}\left(-\sign\widehat{m}\right),\label{angles-P3}
\end{eqnarray}
and the corresponding unitary matrix is
\begin{eqnarray}\label{matrix-P3}
  \fl U^{(II)}&=&\sign\widehat{m}\int_0^1\Theta\left(|\widehat{m}|(\sqrt{1-y^2}\cos\omega-y\sin\omega)-16\widehat{a}^2y\sqrt{1-y^2}\right)dy \eins_2\\
  \fl&&-\imath\sign\widehat{m}\sqrt{1-\left[\int_0^1\Theta\left(|\widehat{m}|(\sqrt{1-y^2}\cos\omega-y\sin\omega)-16\widehat{a}^2y\sqrt{1-y^2}\right)dy\right]^2}\tau_3.\nonumber
\end{eqnarray}
This solution corresponds to the phase labelled by ``$II_\pm$'' where the subscript denotes the sign of the quark mass.
The free energy becomes
\begin{eqnarray}\label{Lagrangian-P3}
 V\mathcal{L}^{(II)}_{\rm 0}= 8\widehat{a}^2\cos 2\widetilde{\varphi}^{(II)}-2|\widehat{m}|\cos(\widetilde{\varphi}^{(II)}-\omega)+ 8\widehat{a}^2.
\end{eqnarray}

The complete region to find  the phase $II_\pm$ is given by
\begin{eqnarray}
\fl \left (\widehat{\mu}_{\rm I}^2\geq2|\widehat{m}|\sin\omega\geq0
\ {\rm and}\  \frac {2|\whm|\sin\omega}{\whmu^2} >\sqrt{1- \left(\frac{2\whm \cos\omega}{32\wha^2+\whmu^2}\right )^2}> 0 \right ) 
\ {\rm or}\  \left ( 2|\widehat{m}|\sin\omega> \widehat{\mu}_{\rm I}^2\right ).\nonumber\\ \fl\label{region-P2b}
\end{eqnarray}

Collecting everything we summarize that there are two phases, $I$ and $II_\pm$, at finite twist angle $\omega\neq0,\pi/2$.

\subsection{The limit of zero twist ($\omega\to0$)}\label{app1.3}

Let us now consider the limit of the two  phases when setting the twist angle $\omega$ to zero. The first phase $I$ will be in the region
\begin{equation}\label{region-P1-no}
 \widehat{\mu}_{\rm I}^2>0\quad {\rm and}\quad 0<2|\widehat{m}|\leq32\widehat{a}^2+\widehat{\mu}_{\rm I}^2
\end{equation}
with the unitary matrix
\begin{eqnarray}\label{matrix-P1-no}
  U^{(I)}|_{\omega=0}&=&\frac{2\widehat{m}}{32\widehat{a}^2+ \widehat{\mu}_{\rm I}^2} \eins_2+\imath\sqrt{1-\frac{4\widehat{m}^2}{(32\widehat{a}^2+\widehat{\mu}_{\rm I}^2)^2}} \left[\begin{array}{cc} \displaystyle 0 & \displaystyle e^{\imath\vartheta_2} \\ \displaystyle e^{-\imath\vartheta_2} & \displaystyle 0 \end{array}\right],
\end{eqnarray}
and the free energy
\begin{eqnarray}\label{Lagrangian-P1-no}
 V\mathcal{L}_{\rm 0}^{(I)}|_{\omega=0}=-\frac{\widehat{\mu}_{\rm I}^2}{2}-\frac{2\widehat{m}^2}{32\widehat{a}^2+ \widehat{\mu}_{\rm I}^2}.
\end{eqnarray}
This phase meets the $\widehat{\mu}_{\rm I}=0$ plane where the Aoki-phase is.

What happens with the phase $II_\pm$? In this case the solution of the
transcendental equation (\ref{transcendent}) breaks up into two branches,
\be
\sin \varphi = 0 \quad {\rm or} \quad \cos \varphi = \frac \whm{16 \wha^2}.
\ee
The first branch corresponds to the region
\begin{equation}\label{region-P3-no}
  \left(\widehat{\mu}_{\rm I}^2>0\ {\rm and}\ 2|\widehat{m}|> 32\widehat{a}^2+\widehat{\mu}_{\rm I}^2\right)\ {\rm or}\ \left(0> \widehat{\mu}_{\rm I}^2\ {\rm and}\ |\widehat{m}|>16\widehat{a}^2\right),
\end{equation}
and the second branch is found in the region
\begin{equation}\label{region-P2-no}
 \widehat{\mu}_{\rm I}^2<0\quad {\rm and}\quad 16\widehat{a}^2>|\widehat{m}|>0.
\end{equation}
 We always have the solution $\widetilde{\varphi}=0$
 which only dominates if we do not reach the solution 
$\widetilde{\varphi}={\rm arccos}(|\widehat{m}|/16\widehat{a}^2)$ 
only occurring in the region~\eref{region-P2-no}. 
In both regions the correct solution is given by Eq.~\eref{sol-app}.

 The first region~\eref{region-P3-no} yields the unitary matrix
\begin{eqnarray}\label{matrix-P3-no}
  U^{(III)}|_{\omega=0}&=&\sign\widehat{m} \eins_2
\end{eqnarray}
and the free energy is given by
\begin{eqnarray}\label{Lagrangian-P3-no}
 V\mathcal{L}^{(III)}_{\rm 0}|_{\omega=0}=16\widehat{a}^2-2|\widehat{m}|.
\end{eqnarray}
Note that the phase denoted by $III_\pm^{(\omega=0)}$ (corresponding 
to $\sign\widehat{m}=\pm1$) is continuously connected to the phase $I$ implying a second order phase transition.
 
 In the second region~\eref{region-P2-no} the unitary matrix becomes
\begin{eqnarray}\label{matrix-P2-no}
  U^{(II)}|_{\omega=0}&=&\frac{\widehat{m}}{16\widehat{a}^2} \eins_2-\imath\sign\widehat{m}\sqrt{1-\left(\frac{\widehat{m}}{16\widehat{a}^2}\right)^2}\tau_3
\end{eqnarray}
and the free energy is given by
\begin{eqnarray}\label{Lagrangian-P2-no}
 V\mathcal{L}^{(II)}_{\rm 0}|_{\omega=0}=-\frac{\widehat{m}^2}{16\widehat{a}^2}.
\end{eqnarray}
 At vanishing isospin chemical potential the phase  $II_\pm^{(\omega=0)}$ meets the phase $I$ at the Aoki 
phase which only exists in the $(\widehat{\mu}_{\rm I}^2=0)$-plane. 
Starting from a finite twist angle the phase  $II_\pm$ is continuously approached from the phase $I$. This is not true anymore for $\omega=0$ where it is a first order phase transition. To see this, one has to take the limit $\widehat{\mu}_{\rm I}^2\to0$ in Eqs.~\eref{matrix-P1-no} and \eref{matrix-P2-no}.
 
We want to underline that $\sign\widehat{m}$ has to be replaced by $\sign j_0$ in Eq.~\eref{matrix-P2-no} when we do not approach this phase from a finite twist but at zero twist with the source term $j_0$. Thus the order of the limits $\omega\to0$ and $j_0\to0$ is crucial in this phase to determine the correct saddle point equation. 
 
The Aoki phase only exists at $\widehat{\mu}_{\rm I}=0$ and is quite particular since the angle $\vartheta_1$ also drops out while $\sin\varphi\neq0$ such that the unitary matrix still depends on this angle
\begin{eqnarray}\label{matrix-Aoki}
 \fl U^{\rm (Aoki)}|_{\omega=0}&=&\frac{\widehat{m}}{16\widehat{a}^2} \eins_2+\imath\sqrt{1-\left(\frac{\widehat{m}}{16\widehat{a}^2}\right)^2}\left[\begin{array}{cc} \displaystyle \cos\vartheta_1 & \displaystyle e^{\imath\vartheta_2}\sin\vartheta_1 \\ \displaystyle e^{-\imath\vartheta_2}\sin\vartheta_1 & \displaystyle -\cos\vartheta_1 \end{array}\right]
\end{eqnarray}
with the free energy
\begin{eqnarray}\label{Lagrangian-Aoki}
 V\mathcal{L}^{\rm (Aoki)}_{\rm 0}|_{\omega=0}=-\frac{\widehat{m}^2}{16\widehat{a}^2}.
\end{eqnarray}
As a function of $\widehat{\mu}_{\rm I}$ it connects the phases
 $I$ and $III_\pm^{(\omega=0)}$ by a first order phase transition. Finite pion condensates can be created via the three source terms $j_k$, $k=1,2,3$.

\subsection{The limit of maximal twist ($\omega\to\pi/2$)}\label{app1.4}

At maximal twisted mass the phase $I$ yields a unitary matrix
\begin{eqnarray}\label{matrix-P1-max}
 U^{(I)}|_{\omega=\pi/2}&=&\imath \left[\begin{array}{cc} \displaystyle-\frac{2\widehat{m}}{\widehat{\mu}_{\rm I}^2} & \displaystyle e^{\imath\vartheta_2}\sqrt{1-\frac{4\widehat{m}^2}{\widehat{\mu}_{\rm I}^4}} \\ \displaystyle e^{-\imath\vartheta_2}\sqrt{1-\frac{4\widehat{m}^2}{\widehat{\mu}_{\rm I}^4}} & \displaystyle\frac{2\widehat{m}}{\widehat{\mu}_{\rm I}^2}\end{array}\right]
\end{eqnarray}
and the free energy
\begin{eqnarray}\label{Lagrangian-P1-max}
 V\mathcal{L}^{(I)}_{\rm 0}|_{\omega=\pi/2}=-\frac{\widehat{\mu}_{\rm I}^2}{2}-\frac{2\widehat{m}^2}{\widehat{\mu}_{\rm I}^2}
\end{eqnarray}
in the region
\begin{equation}\label{region-P1-max}
 \widehat{\mu}_{\rm I}^2\geq2|\widehat{m}|>0\quad {\rm and}\quad 32\widehat{a}^2+ \widehat{\mu}_{\rm I}^2>0.
\end{equation}

In the phase $II_\pm$ the equation for $\varphi$ again splits into two branches for
$\omega \to \pi/2$ (see Fig. \ref{fig2c}). One branch with $\cos \varphi=\sign\widehat{m}$
exists for $ |\whm| > -16\wha^2$ and the other branch with
 $\sin \varphi= -|\whm|/16\wha^2$
is the minimum for $ 0<|\whm| < -16\wha^2$.
Taking the limit $\omega \to \pi/2$ of the regions (\ref{region-P2b}) we thus obtain the following four regions
\begin{equation}\label{region-P2-max}
 \fl\left(\widehat{\mu}_{\rm I}^2\leq0\ {\rm and}\ -16\widehat{a}^2>|\widehat{m}|>0\right)\quad {\rm or}\quad \left(-16\widehat{a}^2>|\widehat{m}|>0\ {\rm and}\ -32\widehat{a}^2>\widehat{\mu}_{\rm I}^2>0\right)
\end{equation}
and
\begin{equation}\label{region-P4-max}
 \fl\left(\widehat{\mu}_{\rm I}^2\leq0\ {\rm and}\ |\widehat{m}|>-16\widehat{a}^2\right)\quad {\rm or}\quad \left(2|\widehat{m}|>\widehat{\mu}_{\rm I}^2>0\ {\rm and}\ |\widehat{m}|>-16\widehat{a}^2\right).
\end{equation}
 Both possibilities are covered by the solution~\eref{sol-app}.

When approaching the limit of maximal twist from a twist angle $\omega<\pi/2$ in the phase $II_\pm$ we only smoothly reach the unitary matrix
\begin{eqnarray}\label{matrix-P2-max}
  U^{(II)}|_{\omega=\pi/2}&=&\sign\widehat{m}\sqrt{1-\left(\frac{\widehat{m}}{16\widehat{a}^2}\right)^2} \eins_2+\imath\frac{\widehat{m}}{16\widehat{a}^2}\tau_3,
\end{eqnarray}
with the free energy
\begin{eqnarray}\label{Lagrangian-P2-max}
 V\mathcal{L}^{(II)}_{\rm 0}|_{\omega=\pi/2}=\frac{\widehat{m}^2}{16\widehat{a}^2}+16\widehat{a}^2
\end{eqnarray}
in the region~\eref{region-P2-max} if the quark mass $\widehat{m}$ remains finite. However the other phase (denoted by $III_\pm^{(\omega=\pi/2)}$) in the region~\eref{region-P4-max}  is approached by a second order phase transition. In this phase the unitary matrix is given by
\begin{eqnarray}\label{matrix-P4-max}
  U^{(III)}|_{\omega=\pi/2}&=&-\imath\sign\widehat{m}\tau_3,
\end{eqnarray}
and the free energy is equal to
\begin{eqnarray}\label{Lagrangian-P4-max}
 V\mathcal{L}^{(III)}_{\rm 0}|_{\omega=\pi/2}=-2|\widehat{m}|.
\end{eqnarray}
Again the labelling reflects the sign of the mass $\sign\widehat{m}=\pm1$.

Similar to the case of vanishing twist we can find a different vacuum polarization in the phase $II_\pm^{(\omega=\pi/2)}$ when interchanging the order of the limit $\omega\to\pi/2$ and $m_{\rm v}\to0$. In the case that we do not regularize with a finite twist but with the source $m_{\rm v}$ we have to replace $\sign\widehat{m}$ by $\sign m_{\rm v}$ in Eq.~\eref{matrix-P2-max}. 

Due to the symmetry~\eref{symmetry-U} the Aoki phase at $\omega=0$ has a counterpart at $\omega=\pi/2$ which yields the saddle point
\begin{eqnarray}\label{matrix-Aoki-no}
 \fl U^{\rm (Aoki)}|_{\omega=\pi/2}&=&\imath\frac{\widehat{m}}{16\widehat{a}^2} \tau_3+\sqrt{1-\left(\frac{\widehat{m}}{16\widehat{a}^2}\right)^2}\left[\begin{array}{cc} \displaystyle \cos\vartheta_1 & \displaystyle\imath e^{\imath\widetilde{\vartheta}_2}\sin\vartheta_1 \\ \displaystyle\imath e^{-\imath\widetilde{\vartheta}_2}\sin\vartheta_1 & \displaystyle \cos\vartheta_1 \end{array}\right]
\end{eqnarray}
and the free energy
\begin{eqnarray}\label{Lagrangian-Aoki-no}
 V\mathcal{L}^{\rm (Aoki)}_{\rm 0}|_{\omega=\pi/2}=\frac{\widehat{m}^2}{16\widehat{a}^2}+16\widehat{a}^2.
\end{eqnarray}
We already absorbed the phase shift resulting from the symmetry transformation~\eref{symmetry-U} in the angle $\widetilde{\vartheta}_2\in[-\pi,\pi]$. The phases $I$ and $II_{\pm}^{(\omega=\pi/2)}$ meet each other at this Aoki-like phase at $\widehat{\mu}_{\rm I}^2=-32\widehat{a}^2>0$. The transition from $I$ to $II_{\pm}^{(\omega=\pi/2)}$ is discontinuous, cf. Eqs.~\eref{matrix-P1-max} and \eref{matrix-P2-max}, and is thus a first order phase transition.

We find a finite charged pion condensate and chiral condensate in the Aoki-like phase at maximal twist when switching on the source terms $j_\pm$ and $m_{\rm v}$, respectively.  Also a finite twist $\omega\to\pi/2$ can align the chiral condensate.

\section{Computation of the pion masses}\label{MassDer}

To compute the pion masses we employ the vector notation of the Pauli matrices
\begin{eqnarray}\label{Paulivector}
\underline{\tau}=\left[\begin{array}{c} \tau_1 \\ \tau_2 \\ \tau_3 \end{array}\right].
\end{eqnarray}
Then any element in the Lie algebra ${\rm su}(2)$ can be written as
\begin{eqnarray}\label{vecdecomp}
 A=A^j\tau_j=\langle\underline{A},\underline{\tau}\rangle\in {\rm su}(2),\ 
{\rm with}\ \underline{A}=\left[\begin{array}{c} A^1 \\ A^2 \\ A^3\end{array}\right],
\end{eqnarray}
where $\langle.,.\rangle$ is the Euclidean scalar product in $\mathbb{R}^3$, and $|.|$ is the modulus of the corresponding norm. Note that the norm itself will not be a map to non-negative real numbers if the vector is complex. Therefore we need the modulus. The Pauli matrices themselves are given by
\begin{eqnarray}\label{Paulimatr}
 \tau_j=\langle\underline{e}_j,\underline{\tau}\rangle\in {\rm su}(2).
\end{eqnarray}
The chosen notation fulfills the following rules
\begin{eqnarray}
 AB&=&\langle\underline{A},\underline{\tau}\rangle\langle\underline{B},\underline{\tau}\rangle=\langle\underline{A},\underline{B}\rangle\eins_2+\imath\langle\underline{A}\times\underline{B},\underline{\tau}\rangle,\label{rule1}\\
 \tr A&=&\tr\langle\underline{A},\underline{\tau}\rangle=0,\label{rule2}\\
 \tr \tau_j A&=&\tr \langle\underline{e}_j,\underline{\tau}\rangle\langle\underline{A},\underline{\tau}\rangle=2\langle\underline{e}_j,\underline{A}\rangle=2A_j,\label{rule3}
\end{eqnarray}
for all $A,B\in {\rm su}(2)$.

With the help of this  notation we decompose the matrix
\begin{eqnarray}\label{u0-dec}
 U_0=\alpha\eins_2+\imath\langle\underline{\beta},\underline{\tau}\rangle,\  U_0^\dagger=\alpha\eins_2-\imath\langle\underline{\beta},\underline{\tau}\rangle,
\end{eqnarray}
where the components $\alpha,\beta^j\in[-1,1]$ with $\alpha^2+|\underline{\beta}|^2=1$ can be read off from Eq.~\eref{parametSU(2)}. The pion fields are
\begin{eqnarray}\label{pi-dec}
 \Pi=\Pi^j\tau_j=\langle\underline{\Pi},\underline{\tau}\rangle,\ 
 \Pi^\dagger=(\Pi^j)^*\tau_j=\langle\underline{\Pi}^*,\underline{\tau}\rangle.
\end{eqnarray}
In this notation  the six terms in 
the chiral Lagrangian $\mathcal{L}_2$, see Eq.~\eref{L2} read
\begin{eqnarray}
 \fl \tr\Pi\Pi^\dagger&=&2|\underline{\Pi}|^2,\label{terms}\\
 \fl \tr\Pi[\Pi^\dagger U_0^\dagger,\tau_3]_-U_0&=&-4\imath\langle\underline{\Pi}^*\times\underline{\Pi},\left(\alpha^2\underline{e}_3+\alpha\underline{\beta}\times\underline{e}_3+\langle\underline{\beta},\underline{e}_3\rangle\underline{\beta}\right)\rangle,\nonumber\\
 \fl \tr[\Pi,\tau_3]_-[\Pi^\dagger,U_0^\dagger\tau_3 U_0]_-&=&8(|\underline{\beta}|^2-\alpha^2-2\langle\underline{\beta},\underline{e}_3\rangle^2)|\underline{\Pi}|^2+8(\alpha^2-|\underline{\beta}|^2)|\langle\underline{\Pi},\underline{e}_3\rangle|^2\nonumber\\
 \fl&&+16\alpha\langle\underline{\Pi}^*,\underline{e}_3\rangle\langle\underline{\Pi},\underline{\beta}\times\underline{e}_3\rangle+16\langle\underline{\beta},\underline{e}_3\rangle\langle\underline{\Pi}^*,\underline{e}_3\rangle\langle\underline{\Pi},\underline{\beta}\rangle,\nonumber\\
 \fl \tr(e^{\imath\omega\tau_3}U_0+U_0^\dagger e^{-\imath\omega\tau_3})\Pi\Pi^\dagger&=&4(\alpha\cos\omega-\langle\underline{\beta},\underline{e}_3\rangle\sin\omega)|\underline{\Pi}|^2,\nonumber\\
 \fl |\tr(U_0-U_0^\dagger)\Pi|^2&=&16|\langle\underline{\Pi},\underline{\beta}\rangle|^2,\nonumber\\
 \fl \tr(U_0+U_0^\dagger)\Pi\Pi^\dagger\tr(U_0+U_0^\dagger)&=&16\alpha^2|\underline{\Pi}|^2.\nonumber
\end{eqnarray}
The third term is generally complex and only becomes real after summing over all energies $E$ and momenta $p$.
Note that all terms are rotation invariant in the $\underline{e}_1$--\,$\underline{e}_2$ plane so that we can 
rotate the vector $\underline{\beta}$ to $\underline{\beta}'=\beta_\bot\underline{e}_1+\beta_3\underline{e}_3$
 with $\beta_\bot^2=\beta_1^2+\beta_2^2$.
 
The aim is to find the dispersion relations from the Lagrangian $\mathcal{L}_2$, in particular we have to 
find the energies $\widehat{E}=E\sqrt{V}F_\pi\geq0$ for which the pion propagator diverges. Given the
bilinear form in the pion fields 
\begin{eqnarray}\label{bilinear}
 \mathcal{L}_2=\frac{1}{2VF_\pi^2}\sum_{\widehat{p},\widehat{E}>0}\underline{\Pi}^\dagger H(\widehat{E},\widehat{p})\underline{\Pi},
\end{eqnarray}
we have to find the energies for which the determinant of
\begin{eqnarray}
\fl H(\widehat{E},\widehat{p})&=&\left(\widehat{p}_k\widehat{p}^k-\widehat{E}^2+\widehat{\mu}_{\rm I}^2(2\beta_\bot^2-1)+2\widehat{m}(\alpha\cos\omega-\beta_3\sin\omega)-32\widehat{a}^2\alpha^2\right)\eins_3\nonumber\\
\fl&&\hspace*{-1.8cm}+\left[\begin{array}{ccc}  32\widehat{a}^2\beta_\bot^2 & 2\imath\widehat{\mu}_{\rm I}\widehat{E} (1-\beta_\bot^2) & ((\widehat{\mu}_{\rm I}^2+32\widehat{a}^2)\beta_3 +2\imath\alpha\widehat{\mu}_{\rm I}\widehat{E})\beta_\bot \\ -2\imath\widehat{\mu}_{\rm I}\widehat{E}  (1-\beta_\bot^2)  & 0 & (-\widehat{\mu}_{\rm I}^2\alpha+2\imath\widehat{\mu}_{\rm I}\widehat{E}\beta_3)\beta_\bot \\  ((\widehat{\mu}_{\rm I}^2+32\widehat{a}^2)\beta_3 -2\imath\alpha\widehat{\mu}_{\rm I}\widehat{E})\beta_\bot & (-\widehat{\mu}_{\rm I}^2\alpha-2\imath\widehat{\mu}_{\rm I}\widehat{E}\beta_3)\beta_\bot & \widehat{\mu}_{\rm I}^2(1-2\beta_\bot^2)+32\widehat{a}^2\beta_3^2 \end{array}\right]\nonumber\\
\fl&\equiv &H_0+2\imath\widehat{\mu}_{\rm I}\widehat{E} H_1+(\widehat{p}_k\widehat{p}^k-\widehat{E}^2)\eins_3\label{Hmatrix}
\end{eqnarray}
vanishes. Using that the real vector 
$\underline{u}^T=(\beta_3\beta_\bot,-\alpha\beta_\bot,(1-\beta_\bot^2))$ is in the kernel of 
the antisymmetric matrix $H_1$ the determinant of $H$ can be rewritten as
\begin{eqnarray}
 \fl\det H(\widehat{E},\widehat{p})&=&(\widehat{p}_k\widehat{p}^k-\widehat{E}^2)^3+\tr H_0(p_kp^k-\widehat{E}^2)^2-4|\underline{u}|^2\widehat{\mu}_{\rm I}^2\widehat{E}^2(\widehat{p}_k\widehat{p}^k-\widehat{E}^2)\nonumber\\
 \fl&&+ \frac{{\tr}^2H_0 -\tr H_0^2}{2}(\widehat{p}_k\widehat{p}^k-\widehat{E}^2)-4\langle\underline{u},H_0\underline{u}\rangle \widehat{\mu}_{\rm I}^2\widehat{E}^2+\det H_0.
\end{eqnarray}
 The matrix $H_0$ is the curvature of the potential and is therefore positive semi-definite at the saddle points. The norm of $\underline{u}$ is simply $\sqrt{1-\beta_\bot^2}$ while the other terms are quite lengthy.

Note in the
 phase $II_\pm$,  $II_\pm^{(\omega=0,\pi/2)}$ and $III_\pm^{(\omega=0,\pi/2)}$  the vector $\underline{\beta}\|  \underline{e}_3$ so that $\beta_\bot=0$, see \ref{App_sec4.2}. This results in a simplification of 
 the Lagrangian $\mathcal{L}_2$. 
For the phase $I$, a simplification is that we always have an exact zero mode, see \ref{App_sec4.1}.

\subsection{Phase $I$}\label{App_sec4.1}

In the phase $I$ we have $\beta_\bot\neq0$, and the parameters of $U_0$ are
given by
\begin{equation}\label{values_I}
 \alpha=\frac{2\widehat{m}\cos\omega}{32\widehat{a}^2+\widehat{\mu}_{\rm I}^2},\ \beta_3=-\frac{2\widehat{m}\sin\omega}{\widehat{\mu}_{\rm I}^2},\ \beta_\bot=\sqrt{1-\alpha^2-\beta_3^2}.
\end{equation}
With these values we can simplify $H$ to
\begin{eqnarray}
\fl H(\widehat{E},\widehat{p})&=&\left(\widehat{p}_k\widehat{p}^k-\widehat{E}^2\right)\eins_3\label{Hmatrix2}\\
\fl&&\hspace*{-1.8cm}+\left[\begin{array}{ccc}  (\widehat{\mu}_{\rm I}^2+32\widehat{a}^2)\beta_\bot^2 & 2\imath\widehat{\mu}_{\rm I}\widehat{E} (1-\beta_\bot^2) & ((\widehat{\mu}_{\rm I}^2+32\widehat{a}^2)\beta_3 +2\imath\alpha\widehat{\mu}_{\rm I}\widehat{E})\beta_\bot \\ -2\imath\widehat{\mu}_{\rm I}\widehat{E}  (1-\beta_\bot^2)  & \widehat{\mu}_{\rm I}^2\beta_\bot^2 & (-\widehat{\mu}_{\rm I}^2\alpha+2\imath\widehat{\mu}_{\rm I}\widehat{E}\beta_3)\beta_\bot \\  ((\widehat{\mu}_{\rm I}^2+32\widehat{a}^2)\beta_3 -2\imath\alpha\widehat{\mu}_{\rm I}\widehat{E})\beta_\bot & (-\widehat{\mu}_{\rm I}^2\alpha-2\imath\widehat{\mu}_{\rm I}\widehat{E}\beta_3)\beta_\bot & \widehat{\mu}_{\rm I}^2(1-\beta_\bot^2)+32\widehat{a}^2\beta_3^2 \end{array}\right].\nonumber
\end{eqnarray}
Then the determinant of $H_0$ vanishes, i.e. $\det H_0=0$. Therefore one pion mode is massless $\widehat m_{\pi,1}=\widehat E_{\pi,1}(\widehat{p}=0)=0$, see Eq. (\ref{Hmatrix}. The corresponding mode is given by
\be
\Pi_{\pi,1}\propto\
-\beta_3 \tau_1+\alpha\tau_2 +(1-\alpha^2-\beta_3^2)^{1/2} \tau_3.
\label{mode1}
\ee
At the phase boundary the coefficient of $\tau_3$ vanishes and
the mode $\Pi_{\pi,1}$ becomes a linear combination of the charged pion modes $\Pi_+$ and $\Pi_-$ with 
a mixing angle determined by the twist angle. At the phase boundary to phase
II, this mode joins the mode that  becomes massless starting from phase II. 

To find the other modes we have to solve the remaining quadratic equation which leads to the masses
\begin{eqnarray}
\widehat{m}_{\pi,2}=\sqrt{c_1-\sqrt{c_2}},\quad\widehat{m}_{\pi,3}=\sqrt{c_1+\sqrt{c_2}}
\end{eqnarray}
with
\begin{eqnarray}
 c_1&=&\frac{1}{2}(\tr H_0+4|\underline{u}|^2\widehat{\mu}_{\rm I}^2),\nonumber\\
 c_2&=&\frac{1}{4}(\tr H_0+4|\underline{u}|^2\widehat{\mu}_{\rm I}^2)^2-\frac{\tr^2 H_0-\tr H_0^2}{2}-4\langle\underline{u},H_0\underline{u}\rangle \widehat{\mu}_{\rm I}^2,
\end{eqnarray}
see Eq.~\eref{detHcoeff}.
The corresponding modes involve quite complicated expressions which we omit. 
Nonetheless let us emphasize that they mix all three pion modes $\pi^0$ and $\pi^{\pm}$.
At the phase boundary they become the $\pi^0$ mode 
and the massive charged pion. 

\subsection{Phases $II_\pm$, $II_\pm^{(\omega=0,\pi/2)}$ and $III_\pm^{(\omega=0,\pi/2)}$}\label{App_sec4.2}

In the phase $II_\pm$ as well as $III_\pm^{(\omega=0,\pi/2)}$ we have $\beta_\bot=0$. Therefore the matrix~\eref{Hmatrix} simplifies to
\begin{eqnarray}\label{HmatrixII}
 H&=&\left(\widehat{p}_k\widehat{p}^k-\widehat{E}^2-\widehat{\mu}_{\rm I}^2+\widehat{m}(\alpha\cos\omega-\beta_3\sin\omega)-32\widehat{a}^2\alpha^2\right)\eins_3\\
&&+\left[\begin{array}{ccc}  0 & 2\imath\widehat{\mu}_{\rm I}\widehat{E} & 0 \\ -2\imath\widehat{\mu}_{\rm I}\widehat{E}  & 0 & 0 \\  0 & 0 & \widehat{\mu}_{\rm I}^2+32\widehat{a}^2\beta_3^2 \end{array}\right].\nonumber
\end{eqnarray}
From this expression we can simply read off the dispersion relations
\begin{eqnarray}\label{dispersionII}
\widehat{E}_{\pi^0}&=&\sqrt{\widehat{p}_k\widehat{p}^k+2\widehat{m}(\alpha\cos\omega-\beta_3\sin\omega)+32\widehat{a}^2(\beta_3^2-\alpha^2)},\\
\widehat{E}_{\pi^\pm}&=&\sqrt{\widehat{p}_k\widehat{p}^k+2\widehat{m}(\alpha\cos\omega-\beta_3\sin\omega)-32\widehat{a}^2\alpha^2}\pm\widehat{\mu}_{\rm I},\nonumber
\end{eqnarray}
corresponding to the pions $\pi^0=\tau_3\Pi_3$ and $\pi^\pm\propto\tau_\mp\Pi_\pm$, respectively. Recall the relations $\alpha=\sign\widehat{m}|\alpha|$ and $\beta_3=-\sign\widehat{m}\sqrt{1-\alpha^2}$. Then the pion masses defined as the rest energy $\widehat{E}(\widehat{p}=0)=\widehat{m}_\pi$ is
\begin{eqnarray}\label{dispersion-II}
\widehat{m}_{\pi^0}&=&\sqrt{2|\widehat{m}|(|\alpha|\cos\omega+\sqrt{1-\alpha^2}\sin\omega)+32\widehat{a}^2(1-2\alpha^2)},\\
\widehat{m}_{\pi^\pm}&=&\sqrt{2|\widehat{m}|(|\alpha|\cos\omega+\sqrt{1-\alpha^2}\sin\omega)-32\widehat{a}^2\alpha^2}\pm\widehat{\mu}_{\rm I}.\nonumber
\end{eqnarray}
At the phase boundary with the phase $I$  one of the charged pions becomes
massless and becomes degenerate with the massless pion from phase I.

For the  particular phases one has to plug  the corresponding values for $\alpha$  into these expressions. Note again that only for $\widehat{\mu}_{\rm I}^2>0$ the masses remain positive semi-definite. For  $\widehat{\mu}_{\rm I}^2<0$ the masses become complex which is quite natural when considering the way the chemical potential affects the propagation of quarks.

\section*{Bibliography}



\end{document}